\documentclass[letter,11pt]{article}

\usepackage{jheppub}
\usepackage{hyperref}
\usepackage{graphicx}
\usepackage{amsmath}
\usepackage{amssymb} 
\usepackage[normalem]{ulem}

\usepackage{xspace}
\usepackage{slashed}
\usepackage[utf8]{inputenc}
\usepackage[T1]{fontenc}

\newcommand{\nn}{\nonumber}

\newcommand{\Pythia}{\textsc{Pythia}\xspace}
\newcommand{\Vincia}{\textsc{Vincia}\xspace}

\DeclareRobustCommand{\Sec}[1]{Sec.~\ref{#1}}
\DeclareRobustCommand{\Secs}[2]{Secs.~\ref{#1} and \ref{#2}}

\DeclareRobustCommand{\Fig}[1]{Fig.~\ref{#1}}
\DeclareRobustCommand{\Figs}[2]{Figs.~\ref{#1} and \ref{#2}}

\DeclareRobustCommand{\Eq}[1]{Eq.~(\ref{#1})}
\DeclareRobustCommand{\Eqs}[2]{Eqs.~(\ref{#1}) and (\ref{#2})}

\def\eqn#1{Eq.~(\ref{eq:#1})}
\def\eqns#1#2{Eqs.~(\ref{eq:#1}) and~(\ref{eq:#2})}

\newcommand{\be}{\begin{equation}}
\newcommand{\ee}{\end{equation}}
\newcommand{\bea}{\begin{eqnarray}}
\newcommand{\eea}{\end{eqnarray}}

\newcommand{\zcuta}{z_{{\rm cut}\,1}}
\newcommand{\zcutb}{z_{{\rm cut}\,2}}

\newcommand{\Qcuti}{Q_{{\rm cut}\,i}}
\newcommand{\jet}{{J}}
\newcommand{\cs}{{cs}}
\newcommand{\csa}{{cs1}}
\newcommand{\csb}{{cs2}}
\newcommand{\csi}{{csi}}
\newcommand{\gs}{{gs}}
\newcommand{\gsa}{{gs1}}
\newcommand{\gsb}{{gs2}}

\newcommand{\DSC}{{D_{Ci}}}
\newcommand{\tDSC}{{\tilde D_{Ci}}}

\preprint{MIT-CTP 5034}

\title{Collinear Drop}

\author{Yang-Ting Chien and Iain W. Stewart}

\affiliation{Center for Theoretical Physics, Massachusetts Institute of Technology, Cambridge, MA 02139, USA}

\emailAdd{ytchien@mit.edu,iains@mit.edu}

\abstract{
We introduce collinear drop jet substructure observables, which are unaffected by contributions from collinear radiation, and systematically probe soft radiation within jets. These observables can be designed to be either sensitive or insensitive to process-dependent soft radiation originating from outside the jet. Such collinear drop observables can be exploited as variables to distinguish quark, gluon, and color neutral initiated jets, for testing predictions for perturbative soft radiation in Monte Carlo simulations, for assessing models and universality for hadronization corrections, for examining the efficiency of pileup subtraction methods, and for any other application that leaves an imprint on soft radiation. We discuss examples of collinear drop observables that are based both on clustering and on jet shapes. Using the soft-collinear effective theory we derive factorization expressions for collinear drop observables from QCD jets, and carry out a resummation of logarithmically enhanced contributions at next-to-leading-logarithmic order.
We also identify an infinite class of collinear drop observables for which the leading double logarithms are absent. 
}

\keywords{QCD, Colliders, Jets, Jet Substructure}

\begin{document}
\maketitle

\section{Introduction}
\label{sec:intro}

Jets are collimated sprays of particles observed in high energy colliders. They emerge from energetic quarks and gluons produced in a hard collision, which are then converted into final state particles through parton splitting and hadronization. The parton shower approximation is built primarily around the logarithmic enhancement from collinear splitting and has been very successful for understanding and modeling jets. It is at the heart of a number of Monte Carlo event generators which are able to reproduce a significant amount of experimental data at various energy scales \cite{Buckley:2011ms,Sjostrand:2007gs,Sjostrand:2006za,Fischer:2016vfv,Gleisberg:2008ta,Bahr:2008pv,Bellm:2015jjp,Nagy:2014mqa}, after tuning of their hadronization models. Although the agreement between these simulations and data is not always perfect, and fails in some cases, these showers are important for experimental calibration and are usually the default method for making comparison to new measurements. They are also important as baselines for studying the dynamics and utility of jet substructure observables~\cite{Abdesselam:2010pt,Altheimer:2012mn,Altheimer:2013yza,Adams:2015hiv,Larkoski:2017jix,Andrews:2018jcm,Dasgupta:2018nvj}.  A program has also been developed to increase the accuracy of parton showers, through improved treatment of kinematic regions and matching to fixed-order calculations, see for example~\cite{Frixione:2002ik,Nason:2004rx,Bauer:2008qh,Platzer:2012bs,Lonnblad:2012ix,Alioli:2012fc,Hoeche:2012yf,Alioli:2015toa}. 
An important ingredient for testing such improvements is to define new types of observables that are sensitive to different regions of phase space, or which test aspects of the shower beyond the leading collinear approximation. 

Another driving force for making progress in understanding hard collisions has been systematically improvable field theoretic methods for making predictions for jet data.  This includes both methods based on Soft Collinear Effective Theory (SCET)~\cite{Bauer:2000ew,Bauer:2000yr,Bauer:2001ct,Bauer:2001yt,Bauer:2002nz} and coherent branching~\cite{Catani:1990rr,Catani:1991kz,Banfi:2004yd}, as well as state of the art fixed-order $pp$ collision calculations with final state jets, see for example~\cite{Boughezal:2015dra,Caola:2015wna,Chen:2016zka,Boughezal:2016dtm}. 
For such calculations our imperfect understanding of soft radiation, hadronization, and underlying event are now often limiting factors in various theoretical predictions. See Refs.~\cite{Korchemsky:1994is,Dokshitzer:1995zt,Salam:2001bd,Lee:2006nr,Cacciari:2009dp,Abbate:2010xh,Mateu:2012nk,Jouttenus:2013hs,Stewart:2014nna} for analytic work on predicting these types of soft corrections for jets. Thus testing analytic field theoretic methods for predicting soft radiation sensitive observables have also now become a priority.

The field of jet substructure was developed to systematically study and explore 
the dynamics of radiation inside jets. Typically, soft radiation has been viewed as a contaminant to be eliminated in order to improve the reconstruction of jet observables. This is accomplished by using jet grooming procedures \cite{Butterworth:2008iy,Ellis:2009me,Krohn:2009th,Dasgupta:2013ihk,Larkoski:2014wba} to suppress soft contributions to jet observables by systematically removing soft and wide-angle particles within the jet. This leads to groomed observables that are much less sensitive to the dynamics of any processes occurring outside of the jet, such as initial and final state soft radiation from other jets, underlying event, and pileup. This jet grooming is also  motivated by obtaining improved precision to search for new physics.  Often jet substructure observables can be strongly affected by jet grooming, since it may change their leading logarithmic structure~\cite{Dasgupta:2013ihk,Dasgupta:2013via}.  An example is the jet mass, which is strongly modified by the removal of peripheral soft radiation.  Another possible approach to truncating soft radiation is to use  jet shapes with angular weights~\cite{Berger:2003iw,Almeida:2008yp,Hornig:2016ahz} which suppress the contribution from wide-angle radiation.  In both cases one removes soft and wide-angle radiation contributions by effectively introducing an energy and angular cutoff, so that it is predominantly energetic collinear radiation that is retained. 

In this paper we introduce the ``collinear drop'' class of jet substructure observables to do precisely the opposite, retaining components of the soft radiation for detailed study, while removing collinear radiation. We show that such collinear drop observables can be constructed from approaches mimicking both the jet grooming and jet shape approaches. 
The goal here is to consider jet observables that are sensitive to physics in various soft phase space regions. This makes collinear drop observables ideal for studying perturbative soft dynamics, hadronization, underlying event, and pileup in proton-proton collisions. We show that if one wishes to study only soft radiation related to the jet itself, that collinear drop observables can easily facilitate this using the same techniques as in jet grooming.
Collinear drop observables can also be used to study the jet quenching mechanism and medium evolution in heavy ion collisions, which are known to be sensitive to jet information that can be probed with jet substructure~\cite{Chien:2018dfn}. Besides elucidating the soft regime of QCD, collinear drop observables are also useful for studying the color radiation pattern of the particle initiating a jet. This makes them useful for boosted particle tagging, an application that we intend to explore in more detail elsewhere. 

To illustrate the idea behind collinear drop observables we will consider two main examples.  As our first example we exploit the soft drop jet grooming algorithm~\cite{Larkoski:2014wba} (which generalizes the minimal mass drop algorithm~\cite{Dasgupta:2013ihk}). We consider the difference between two soft drop masses, $m_{{\rm SD}_{1}}$ and $m_{{\rm SD}_{2}}$, defined using two different choices for the soft drop parameters,
\begin{align}  \label{eq:delta_m2_def0}
   \Delta m^2 = m_{{\rm SD}_1}^2 - m_{{\rm SD}_2}^2\;.
\end{align}
We choose the parameters so that the ${\rm SD}_{2}$ grooming is more aggressive than that of ${\rm SD}_{1}$, implying that the particles remaining in the ${\rm SD}_{2}$ jet are a subset of those in the ${\rm SD}_{1}$ jet, and that $\Delta m^2\ge 0$. 
$\Delta m^2$ probes a jet region that is free from the energetic collinear radiation contained within the ${\rm SD}_{2}$ jet, thus making it a collinear drop observable. Furthermore, the choice of parameters in ${\rm SD}_{1}$ controls the initial jet to which we have applied this collinear drop procedure. If we wish to study underlying event or pileup contamination in the jet, then we can turn off the ${\rm SD}_{1}$ grooming so that $m_{{\rm SD}_1}^2\to m_J^2$, the full jet mass observable. On the other hand, if we wish to study soft radiation associated to the dynamics of the jet itself, then we can carry out grooming through the choice of ${\rm SD}_{1}$ to ensure that $\Delta m^2$ has reduced sensitivity to soft radiation originating from outside the jet.

As our second example we consider a class of jet shapes that we refer to as ``flattened angularities'',
\begin{align} \label{eq:tauw}
  \tau_\omega = \sum_{i\in{\rm jet}} z_i~\omega(\theta_i,\theta_0)\;,
\end{align}
where $z_i$ are energy or $p_T$ fractions for each particle $i$ in the jet, and $\omega(\theta,\theta_0)$ is an angular weight function, with $\theta$ measured relative to the jet axis. For $pp$ jets this would be $\omega(\Delta R,\theta_0)$, where $\Delta R$ is the usual rapidity-azimuthal distance measure to the jet axis.  We take the definition of flattened angularities to imply that for a chosen angular parameter $\theta_0>0$ the function $\omega(\theta,\theta_0)$ either vanishes identically or is exponentially suppressed for a finite region around the jet axis, which we denote by $\omega(\theta \le \theta_0,\theta_0) \simeq 0$.  By choosing the angle $\theta_0$ to contain the vast majority of the collinear radiation, we obtain a collinear drop observable. If desired, one can also define $\omega(\theta,\theta_0)$ in a manner that suppresses wide-angle soft radiation to obtain an analog of the jet-grooming present in our $\Delta m^2$ example. 
This flattened angularity  gives a collinear drop jet shape observable that does not require the jet reclustering that occurs in soft drop.

We intend to use $\Delta m^2$ and $\tau_\omega$ to illustrate the general principles behind collinear drop  as a new class of jet substructure observables, though it should be clear that one can construct many other examples beyond those considered here.  For instance a simple generalization would be to consider differences of other observables besides the jet mass in \eqn{delta_m2_def0}, like transverse momenta or angularities of particles in the groomed jets, or to use a different choice of jet groomer.

The rest of the paper is organized as follows. In \Sec{sec:obs} we describe general strategies for constructing collinear drop observables and discuss the two examples of $\Delta m^2$ and $\tau_\omega$ in more details. We also discuss examples of observables that are not collinear drop observables, by virtue of only having power-law suppressed contributions from collinear radiation rather than having a stronger veto on the contribution from these particles. 
In \Sec{sec:fact} we review the SCET factorization theorem for the soft drop jet mass cross section, discuss differences between the $e^+e^-$ and $pp$ collider cases, and develop appropriate scale choices that implement the groomed to ungroomed transition. We also develop scale variations that respect the jet mass transition and endpoint, and test the resulting uncertainty bands at next-to-leading-logarithmic order (NLL). 
In \Sec{sec:CD} we make perturbative predictions for the $\Delta m^2$ collinear drop observable for QCD jets. In particular we derive a factorization theorem for $\Delta m^2$ using SCET, and use it to provide analytic resummed partonic predictions at next-to-leading-logarithmic (NLL) order. We also discuss potential groomed to ungroomed transitions and the adjustable collinear drop spectrum's endpoint, and develop scale choices and scale variations that respect these constraints. In addition we explore the general features of collinear drop distributions when we vary grooming parameters at NLL order, and test our method for estimating perturbative uncertainties at this order. Here the resummation of logarithms arises both from the hierarchies involving the observable, $\Delta m^2 \ll Q^2$, as well as other hierarchies related to removing collinear and soft particles.  
In \Sec{sec:MC} we study collinear drop observables with \Pythia and \Vincia Monte Carlo simulations, including re-testing the parameter dependence and making explicit comparisons with the NLL SCET results. We also demonstrate the utility of using collinear drop to study hadronization in observables with little sensitivity to underlying event. On the flip side we show that other collinear drop observables have enhanced sensitivity to underlying event, and hence can be used to test models intended to describe it.  In all cases further light will be shed on these tests by confrontation with experimental data. 
Finally, as a second type of collinear drop observable we briefly analyze MC simulations for a $\tau_\omega$ example that we refer to as the annulus energy fraction. 
In \Sec{sec:conclusion} we conclude and give an outlook of the use of collinear drop observables in soft QCD.

\section{Collinear Drop Observables}
\label{sec:obs}

The goal of collinear drop is to specify observables that are sensitive to soft radiation within jets, while eliminating contributions from energetic collinear radiation that is collimated with the jet axis. Using light-cone coordinates we can write momenta components of any four vector $p^\mu$ as $p=(n\cdot p, \bar n\cdot p, p_\perp)$ where in four-component notation $n=(1,\hat n_J)$ and $\bar n=(1,-\hat n_J)$ are light-like vectors involving the jet axis unit vector $\hat n_J$.  Using light-cone components, collinear radiation can be defined as particle having momenta scaling as $p_n\sim Q(\lambda^2,1,\lambda)$. Here $Q=2 E_{\rm jet}$ with $E_{\rm jet}$ the jet energy, so each collinear particle carries a non-negligible fraction of the jets energy, and the small parameter $\lambda \ll 1$ determines how collimated the radiation is with the jet axis. For such collinear radiation the contribution to a collinear drop observable $\Delta\text{CD}$ should either vanish or be exponentially suppressed
\begin{align} \label{eq:CDcondition}
  \Delta\text{CD}[p] \simeq 0  \quad \text{for} \quad 
  \frac{|\vec p_{\perp}|}{p^0} \simeq \sin(\theta) < \lambda_0  \ll 1
  \,,
\end{align}
where $\lambda_0$ sets an angular cutoff scale for the polar angle $\theta$ measured relative to the jet-axis.\footnote{Note that this condition only implicitly depends on the fact that the eliminated collinear particles have a large energy, $\bar n\cdot p\sim Q$, through the pre-determined jet-axis which defines $\vec p_\perp$ and $\theta$.} For jet algorithm based observables the  particle with momentum $p$ could be a subjet or contained in a subjet.

Two examples of observables satisfying \eqn{CDcondition} have already been given in \eqns{delta_m2_def0}{tauw}, and we elaborate on these examples below in \Secs{sec:cdrop_grooming}{sec:cdrop_chop}, respectively.  Then in \Sec{sec:cdrop_not} we contrast this with examples that suppress the contribution from collinear radiation, but which do not fully qualify as collinear drop observables.

%On the other hand, soft contributions from initial and final state radiation, underlying events and pileup require substantial non-perturbative modeling. One can systematically study these contributions by introducing energy and angular scales to regulate the soft particles. In the following we will first use two sets of grooming procedures to separately constrain collinear and soft radiation. We will then discuss how angularities can be extended along this direction. It is also straightforward to construct collinear drop observables in the telescoping deconstruction framework.

\subsection{Collinear Drop from Jet Grooming}
\label{sec:cdrop_grooming}

In jet grooming, the constituents of a jet are reconsidered in order to remove soft wide-angle particles, many of which arise from processes like underlying event, hadronization, and pileup that contaminate the partonic description of the jet. This grooming effectively introduces an additional energy or angular cutoff scale that determines what radiation is removed.  These algorithms are designed to retain collinear radiation, and a smaller subset of soft radiation, which are then used to define the groomed jet observable.  A simple way to obtain a sample of particles on which to define a collinear drop observable is to use the complement, namely to define the observable using the subjets/particles that were removed by the jet grooming. 

%A jet grooming procedure typically introduces a low energy scale to disentangle contributions from radiation below that scale. The scale should not be too low such that complicated soft contributions can be effectively suppressed. On the other hand, the scale should be low enough such that one does not remove too much radiation and alter signal jets one wants to study dramatically. 

%Groomed jet observables then highlight contributions from collinear radiation, while soft and non-perturbative contributions are expected to be constrained. Note that a jet grooming procedure separates the jet constituents into two categories where one is removed from the jet and the other one remaining in the jet. Then jet observables can be constructed using the groomed jet constituents, and the removed particles do not directly contribute to the jet observable values.

As a concrete example, we use the soft drop grooming procedure \cite{Larkoski:2014wba} with two different degrees of grooming. Given a jet reconstructed with radius $R$ using any algorithm (such as anti-$k_t$ \cite{Cacciari:2008gp}), we recluster the jet using the Cambridge/Aachen (C/A) algorithm \cite{Dokshitzer:1997in,Wobisch:1998wt} to obtain an angular ordered branching tree. We then traverse the tree starting from the largest angles, making pairwise comparisons that remove the softer subjet branch until 
the soft drop condition is satisfied:
\begin{align}\label{eq:SD}
  &  \frac{\min(p_{T_i},p_{T_j})}{p_{T_i}+p_{T_j}} > z_{\rm cut} \biggl(\frac{\Delta R_{ij}}{R_0}\biggr)^{\beta}
  && \hspace{-1cm} \text{for }  pp\text{ collisions} 
   \,,\nn \\ 
 &  \frac{\min(E_{i},E_{j})}{E_i+E_j} > z_{\rm cut} \biggl( \sqrt{2} 
   \frac{\sin(\theta_{ij}/2)}{\sin(R_0^{ee}/2)}\biggr)^{\beta}
   && \hspace{-1cm} \text{for } e^+e^-\text{ collisions.}  
\end{align}
For the $pp$ case, $p_{T_i}$ and $p_{T_j}$ are the transverse momenta of the two branches, and $\Delta R_{ij}$ is the longitudinally boost invariant distance between the two branches in the plane of the rapidity ($y$) and azimuth angle ($\phi$),
\begin{align} \label{eq:DeltaR}
 \Delta R_{ij}^2 = 2\cosh(y_i-y_j)-2\cos(\phi_i-\phi_j)
   \simeq (y_i-y_j)^2 + (\phi_i-\phi_j)^2 \,,
\end{align}
where the last approximation is valid in the small angle limit, and is sometimes simply adopted for the definition of $\Delta R_{ij}^2$. 
For $e^+e^-$ collisions we instead use the energies $E_i$ and $E_j$, and the geometric angular distance $\theta_{ij}$ between particles. The parameters $R_0$ (or $R_0^{ee}$) set a reference angular scale in the soft drop condition, and are usually taken to be equal to the initial jet radius $R$. In \eqn{SD} $z_{\rm cut}$ is a dimensionless soft drop parameter which sets an upper momentum cutoff for the removal of soft branches, and the parameter $\beta$ provides a weight factor which for $\beta>0$ makes the cutoff stronger for branches separated by a wider  angle.
 
In the small angle limit the physics of the soft drop constraint is universal between the $pp$ and $e^+e^-$ cases. Approximating $\cosh \eta \simeq \cosh \eta_J$ where $\eta_J$ is the jet's pseudo-rapidity, we have $\Delta R = \theta \cosh\eta_J + {\cal O}(\theta^2)$, and the ratios on the LHS of \eqn{SD} are also both equal to a common parameter $z_{ij}$. The soft drop condition therefore becomes
\begin{align} \label{eq:ztcut}
  z_{ij} > \, \tilde z_{\rm cut}\: \theta_{ij}^\beta
\end{align} 
where following Ref.~\cite{Hoang:2018SDNP} we have defined a parameter $\tilde z_{\rm cut}$ that differs for $pp$ and $e^+e^-$ collisions. In particular, $\tilde z_{\rm cut} = z_{\rm cut} (\sqrt{2}\sin\big(R_0^{ee}/2)\big)^{-\beta}$ for  $e^+e^-$, and 
 $\tilde z_{\rm cut}=z_{\rm cut} (\cosh\eta_J/R_0)^\beta$ for $pp$.
An additional definition that will be useful later on is
\begin{align}
  Q_{\rm cut} \equiv 2^\beta\, \tilde z_{\rm cut}\, Q \,.
\end{align}

\begin{figure}[t!]
\centering
    \includegraphics[height=3.6cm]{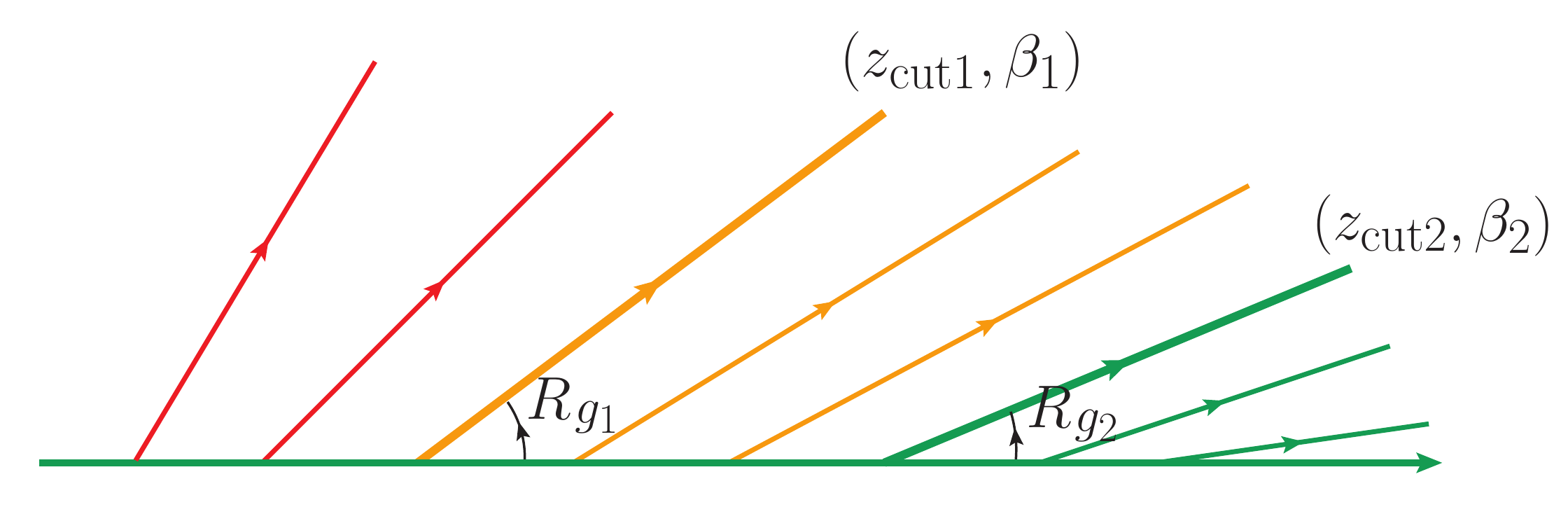}
    \caption{Illustration of the particles kept in the collinear drop sample, displaying for simplicity a set of angular-ordered emissions from a single branch. The soft-drop parameters ${\rm SD}_1 =(z_{{\rm cut}\,1},\beta_1)$ determine what soft wide-angle red particles are dropped, while the soft-drop parameters ${\rm SD}_2$ enforce collinear drop by determining which green collinear particles are dropped. The collinear-drop observable is then defined on the remaining orange particles, roughly contained between the two groomed jet radii $R_{g_1}$  and $R_{g_2}$.}
\label{fig:cd}
\end{figure}

To setup an adjustable sample of soft particles on which to define a collinear drop observable, we consider an initial jet that has been groomed with soft drop parameters ${\rm SD}_1 = (z_{{\rm cut}\,1},\beta_1)$ and then we remove all particles that are kept by a stronger soft drop grooming given by parameters ${\rm SD}_2 = (z_{{\rm cut}\,2},\beta_2)$. Intuitively this implies taking  $z_{{\rm cut}\,1} \le z_{{\rm cut}\,2}$ and $\beta_1 \geq \beta_2$, such that ${\rm SD}_2$ grooms the jet more aggressively than ${\rm SD}_1$. Technically we only require that the ${\rm SD}_2$ jet constituents are a subset of the ${\rm SD}_1$ jet constituents,
\begin{equation}\label{eq:SDs}
    \{{\rm jet}_{{\rm SD}_2}\} \subset \{{\rm jet}_{{\rm SD}_1}\}\;.
\end{equation}
The sample used to define collinear drop observables is then taken to be the particles which are groomed away by ${\rm SD}_2$ but not by ${\rm SD}_1$, i.e., in the complement set $\{{\rm jet}_{{\rm SD}_1}\} \setminus \{{\rm jet}_{{\rm SD}_2}\}$. \Fig{fig:cd} gives an illustration of this with two soft drop settings.  Note that $\{{\rm jet}_{{\rm SD}_2}\}$ contains the energetic collinear radiation, and removing these particles is the crucial ingredient for collinear drop. In contrast, if so desired, the parameters of ${\rm SD}_1$ can be relaxed so that $\{{\rm jet}_{{\rm SD}_1}\}$ is the full jet, which enables a better probe of underlying event and pileup.  In contrast, choosing a non-trivial ${\rm SD}_1$ enables collinear drop to primarily probe soft radiation associated to the jet. Thus we see that this definition of a collinear drop observable can be adjusted depending on the type of soft radiation one wants to look at.  

Given this setup we can then directly define a collinear drop observable $O_{\rm CD}$ using only particles from the complement set
\begin{align}  \label{eq:OCD1}
  O_{\rm CD} = O\big[ \{{\rm jet}_{{\rm SD}_1}\} \setminus \{{\rm jet}_{{\rm SD}_2}\} \big]
 \,,
\end{align} 
or alternatively by considering the difference of groomed jet observables each defined by one of the sets of grooming parameters,
\begin{align}  \label{eq:OCD2}
    O_{\rm CD} =  O_{{\rm SD}_1} - O_{{\rm SD}_2} \;.
\end{align}
The results from using \eqns{OCD1}{OCD2} will agree for observables $O$ that are linear in their contributions from constituents, which is true of many observables of interest. For observables that are not even approximately linear, one should use only particles in the complement set as in \eqn{OCD1}. 
 
As an explicit example of the above construction we consider the collinear drop jet mass, $\Delta m^2$, which can be defined as
\begin{equation}\label{eq:delta_m2_def}
    \Delta m^2 = m_{{\rm SD}_1}^2 - m_{{\rm SD}_2}^2\;.
\end{equation}
Here $m_{{\rm SD}_i}$ is the groomed jet mass with the soft drop condition ${\rm SD}_i$,
\begin{equation}\label{eq:m_def}
    m_{{\rm SD}_i}^2 = p_{{\rm SD}_i}^2\;,~~~{\rm where}~~p_{{\rm SD}_i}^\mu = \sum_{j\in {\rm jet}_{{\rm SD}_i}} p_j^\mu\;.
\end{equation}
Note that when there is no cause for confusion we will simply use $m_J$ for the soft drop jet mass, but like we do here, we will use the alternate notation $m_{{\rm SD}_i}$ if we want to specify the soft drop parameter set $i$ from which the jet mass is derived.
Defining $\Delta p^\mu = p_{{\rm SD}_1}^\mu - p_{{\rm SD}_2}^\mu$ we have
\begin{equation}\label{eq:delta_m2_phys}
    \Delta m^2 = 2p_{{\rm SD}_2}\cdot \Delta p + (\Delta p)^2
  = 2p_{{\rm SD}_2}\cdot \Delta p +\ldots 
  = Q \, n \cdot \Delta p + \ldots \;,
\end{equation}
where the ellipses denote contributions power-suppressed by $m_{{\rm SD}_i}/Q\ll 1$. Thus we see that in the region of interest, where the jet mass is much smaller than the energy of the jet, $\Delta m^2$ is to a very good approximation a linear observable, and \eqns{OCD1}{OCD2} both lead to the same leading description.  Here $\Delta m^2$ probes the lightcone projection of the particle momenta $n \cdot \Delta p$ for those particles which are roughly between the two groomed jet radii $R_{g_1}$ and $R_{g_2}$, but which technically are dynamically determined on a jet-by-jet basis. 

In later sections we will use Monte Carlo to explore the physics that $\Delta m^2$ can be used to probe, and show that analytic resummed expressions can be obtained for the $d\sigma/d\Delta m^2$ cross section using factorization in SCET.

\subsection{Collinear Drop from Jet Shapes}
\label{sec:cdrop_chop}

In our construction of collinear drop observables in \Sec{sec:cdrop_grooming}, the definition intrinsically relied on a clustering algorithm that is inherent in the jet grooming. Another way of defining a collinear drop observable is with a so-called jet shape, defined by directly summing over observed final state particles in a pre-determined jet with rapidity $y_{\rm jet}$ and azimuthal angle $\phi_{\rm jet}$. The precise values for this jet axis depend on the algorithm used to determine the jets.\footnote{This jet axis can be chosen as a conventional jet axis like anti-$k_T$ or a soft-recoil free axis \cite{Larkoski:2014uqa} such as the Winner-Take-All axis \cite{Bertolini:2013iqa}.}

With a single sum over particles we can define the following jet shapes
\begin{align}  \label{eq:CDshape1}
& \text{ for } e^+e^- \text{ collisions:}
 &\tau_\omega & = \sum_{i\in{\rm jet}} z_i~\omega(\theta_i,\theta_0) \,,
  && \text{ where } z_i= \frac{E_i}{E_{\rm jet}}
  \,,\\
& \text{ for } pp \text{ collisions:}
 &\tau_\omega &= \sum_{i\in{\rm jet}} z_i~\omega(\Delta R_i,\theta_0) \,,
  && \text{ where } z_i= \frac{p_{Ti}}{p_T^{\rm jet}}
  \,,\nn
\end{align}
where the function $\omega(\theta,\theta_0)\ge 0$ is an angular weight factor depending on a fixed parameter $\theta_0$. Here $\theta_i$ is the angular distance of particle $i$ to the jet-axis, and $\Delta R_i$ is the angular distance from the jet-axis in the rapidity-azimuthal plane, defined as in \eqn{DeltaR} but with $y_j\to y_{\rm jet}$ and $\phi_j\to \phi_{\rm jet}$. The definitions in \eqn{CDshape1} are generalizations of the classic angularity jet shapes \cite{Berger:2003iw,Almeida:2008yp}. To ensure these are collinear drop jet shapes we demand, for some angular distance parameter $\theta_0$ within which the majority of collinear particles are contained, that 
\begin{align}  \label{eq:omegaCD}
  \omega(\theta \le \theta_0,\theta_0) \simeq 0  \,.
\end{align}
Here $\simeq 0$ could be an exact equality, or indicate that the contribution from this region is exponentially suppressed relative to the dominant contributions.  
To ensure collinear safety we take linear dependence on $z_i$ in \eqn{CDshape1}, and we can impose the condition that $\omega(\theta,\theta_0)$ is continuous as $\theta\to \theta_0$ from above. Since soft emissions have $z_i\to 0$ in the soft limit, $\tau_\omega$ is always infrared safe. 

We can also define collinear drop observables which involve correlations between two or more particles in the jet. For example, as a collinear drop extension of the 2-point energy correlation function~\cite{Larkoski:2013eya} we can define
\begin{align} \label{eq:CDshape2}
  e_2^{(\beta)\rm CD} = \sum_{\stackrel{i,j\in{\rm jet}}{i<j}}  z_i\, z_j\, \theta_{ij}^\beta \, \omega(\theta_i,\theta_0) \omega(\theta_j,\theta_0)
  \,,
\end{align}
with similar extensions for higher point energy correlation functions. 
Here the extra multiplicative factors of $\omega(\theta_i,\theta_0)$ ensure that only comparisons that do not involve collinear particles give non-negligible contributions in the sum. 

The above construction still leaves considerable freedom in specifying the function $\omega(\theta,\theta_0)$ whose choice is needed to fully specify the collinear drop observable. One potentially desirable feature is to also induced a suppression for wide-angle soft particles near the jet boundary, in order to mimic some of the features of jet grooming in the jet shape variable. This can be accomplished by demanding that  $\omega(\theta,\theta_0)\simeq 0$ in a region of $\theta$ about $\theta= R$. 

For definiteness and our later analyses, we give a few examples for collinear drop jet shape observables by specifying $\omega(\theta,\theta_0)$. One simple example is a double-sided step function at radii $R_1$ and $R_2$, %\TODO{Discuss IR Safety of \eqn{theta-annulus}}
\begin{align}   \label{eq:theta-annulus}
\omega_a(\theta,R_1)
 &=\left\{
\begin{array}{lr}
	1 & ~~~~~R_1 < \theta < R_2\\
	0 & ~~~~~\mbox{otherwise}
\end{array}
\right. \,.
\end{align}
This observable is closely related to the classic observable of jet energy profile $\rho(r)=d\Psi(r)/dr$ averaged over a jet sample \cite{Ellis:1992qq,Seymour:1997kj,Li:2011hy,Chien:2014nsa,Cal:2019hjc}. Here we highlight the dependence on $R_1>0$ in the argument of this $\omega$ since choosing $R_1$ to contain the majority of the collinear radiation is what makes this a collinear drop observable. 
The resulting jet shape is equivalent to the momentum fraction $z$ of particles within the ring region $R_1 < \theta < R_2$,\footnote{The non-collinear drop case where $R_1 = 0$ is the jet shape distribution considered in \cite{Gallicchio:2012ez}.} and we will refer to the corresponding $\tau_{\omega_a}$ as the ``annulus energy fraction''. If we take $R_2>R$ then there is no suppression for particles near the jet boundary, whereas for $R_2<R$ we remove a subset of the wide-angle soft radiation. Here $R_1$ and $R_2$ play a similar role to the groomed soft drop radii $R_{g2}$ and $R_{g1}$ of our example in \Sec{sec:cdrop_grooming}. We will consider Monte Carlo simulations and analytic resummation results for the annulus energy fraction in the later sections. Another example is the gaussian angularity or ``gaussianity''\footnote{We thank Christopher Lee for this suggestion.} $\tau_{\omega_g}$ that is obtained using,
\be
\omega_{g}(\theta,r-2\sigma)
= e^{-(\theta-r)^2/2\sigma^2}\;.
\ee
Here the weight function has its dominant support around the angular region $r-\sigma < \theta < r+\sigma$, and we can choose the angular distance $\theta_0=r-2\sigma>0$ to ensure that collinear particles from small angles give only exponentially suppressed contributions. For $r+2\sigma \lesssim R$ this $\omega_g$ choice also give exponentially suppressed contributions for soft particles near the jet boundary. Yet another possibility for defining a collinear drop observable would be to retain exponential suppression for collinear particles, but make the suppression for wide angle soft particles polynomial by using
\be
\omega_e(\theta,r/10)
= (1-\theta/R)^\alpha\, e^{-r/\theta}\;.
\ee
These examples should make clear the method for constructing other possible collinear drop jet shapes, and that we have not attempted to provide an exhaustive list.

\subsection{Examples that are Not Collinear Drop Observables}
\label{sec:cdrop_not}

In this subsection we consider combinations of standard jet measurements which have the property that they suppress the contribution of collinear particles.  In particular, we wish to highlight some examples that at first glance appear to be similar to collinear drop observables, but which actually do not satisfy our definition because they still obtain non-trivial contributions from energetic collinear particles.

One example of an observable that changes the weight of collinear and soft particles are the conventional angularities~\cite{Berger:2003iw,Almeida:2008yp}, which include a angular weight indexed by a parameter $\alpha > 0$. For the jet shape angularity  they can be defined by
\begin{align}
\tau_\alpha &= \sum_{i\in{\rm jet}} z_i \biggl(\frac{\theta_i}{R}\biggr)^\alpha \,,
 && \tau_\alpha = \sum_{i\in{\rm jet}} z_i \biggl(\frac{\Delta R_i}{R}\biggr)^\alpha \,,
\end{align}
for $e^+e^-$ and $pp$ collisions respectively, with the same definitions for $\theta_i$ and $\Delta R_i$  as in \eqn{CDshape1}. For small $\tau_\alpha$ or for jets with small $R$, the angularity with $\alpha=2$ is the same as the jet mass, while $\alpha=1$ corresponds to a jet shape broadening. 
Note that, wide-angle radiation near the jet boundary with $\theta_i\approx R$ has the maximal angular weight, while the contribution from collinear radiation with $\theta_i\ll R$ is angularly suppressed. A larger $\alpha$ will suppress the collinear radiation and enhance the relative contribution of the wide-angle soft radiation. However this suppression is only polynomial with the power $\alpha$, and hence it is weaker than what we require in the definition of a collinear drop observable.

% Angularities define a broad class of jet observables which have different sensitivities to particles at different angular scales. To enjoy the infrared and collinear safety, the angularity $\tau_\alpha$ is defined conventionally as the transverse momentum fractions $z_i$ of particles power-law weighted by their angles $\theta_i$ to a pre-determined axis along $(y_{\rm jet},\phi_{\rm jet})$,

Motivated by the definition of collinear drop observables in \eqn{OCD2}, one might also consider the difference of two angularities as a potentially related observable. Taking $\beta > \alpha$ we let
\be
    \Delta\tau= \tau_\alpha - \tau_\beta = \sum_{i\in{\rm jet}} z_i \Big[\Big(\frac{\theta_i}{R}\Big)^\alpha - \Big(\frac{\theta_i}{R}\Big)^\beta\Big] \ge 0\;.
\ee
Here the angular weighting factor vanishes when $\theta_i\rightarrow 0$ and $\theta_i\rightarrow R$, which seems similar to our collinear drop observables. One can also determine that the contributions to $\Delta\tau$ peak at a finite angle,
\be
    \theta_{\rm peak} = \Big(\frac{\alpha}{\beta}\Big)^{\frac{1}{\beta-\alpha}}R < R
     \,.
\ee
However, for energetic collinear particles with $z_i\sim 1$ and $\theta_i\ll R$ we have
\be
    \Delta\tau\sim \Big(\frac{\theta_i}{R}\Big)^\alpha - \Big(\frac{\theta_i}{R}\Big)^\beta
    \sim \Big(\frac{\theta_i}{R}\Big)^\alpha\;,
\ee
so the angular weight is dominated by a power-law with the exponent $\alpha$. Thus $\Delta \tau$ is again not a collinear drop observable. Note that it also suppresses wide-angle soft radiation linearly as $\sim (\beta-\alpha)(1-\theta/R)$ for $\theta\to R$.

%Therefore
%\be
%    \theta_i\approx \Delta\tau^{1/\alpha}R\;,
%\ee
%with the similar scaling as the collinear mode for $\tau_\alpha$. We see that the subtraction operation in this case does not drop contributions from collinear particles. 

%On the other hand, it suggests a way to suppress wide-angle radiation alternative to jet grooming. This suppresses wide-angle radiation beyond $\theta_0$ and functions similarly as jet grooming.  For $\theta_i\approx \theta_0$,
%\be
%    \Delta\tau\approx z_i \Big[\Big(\frac{\alpha}{\beta}\Big)^{\frac{\alpha}{\beta-\alpha}} - \Big(\frac{\alpha}{\beta}\Big)^{\frac{\beta}{\beta-\alpha}}\Big]
%    = z_i~C(\alpha,\beta)\;.
%\ee
%The relevant soft mode has the similar momentum scaling as the one for angularities up to an overall, constant normalization factor $C(\alpha,\beta)$ which can be absorbed by redefining the observable. The characteristic energy scale for the soft mode is $E_J\Delta\tau$.

The issue with the difference of the two angularities is simply that they do not give the same weight to collinear particles, which therefore do not cancel out in $\Delta\tau$. When considering classic event shapes in $e^+e^-$ collisions it is known that for thrust $\tau=1-T$ \cite{PhysRevLett.39.1587} and $C/6$, where $C$ is the $C$-parameter~\cite{Parisi:1978eg,Donoghue:1979vi}, have the same resummation formula  up to next-to-leading logarithmic order~\cite{Catani:1998sf}. The difference 
\begin{align}
    \Delta_{\tau C} = \tau - \frac{C}{6}\;,
\end{align}
therefore seems like a potential candidate for a collinear drop observable. In terms of a sum over all particles in the $e^+e^-$ event we can write
\begin{align}
    \tau &= \frac{1}{Q} \sum_{j} p_j^\perp \min (e^{\eta_j},e^{-\eta_j})\;,
    && \frac{C}{6}=\frac{1}{Q} \sum_{j} p_j^\perp \frac{1}{e^{\eta_j}+e^{-\eta_j}}\;.
\end{align}
Here $p_j^\perp$ and $\eta_j$ are the particle transverse momentum and pseudo-rapidity defined with respect to the thrust axis. To see whether this qualifies as a collinear drop observable we consider the limit where particles are collinear to the thrust axis, $\theta_j \ll 1$, where $p_j^\perp \simeq z_j\theta_j$ and we have $e^{-|\eta_j|}\approx \theta_j/2$ therefore,
\begin{align}
    \Delta_{\tau C} = \sum_j p_j^\perp  \Big(e^{-|\eta_j|} - \frac{1}{e^{|\eta_j|}+e^{-|\eta_j|}}\Big)\approx \sum_j z_j \theta_j  e^{-3|\eta_j|}
    \approx\frac{1}{8}\sum_j z_j \theta_j^4\;,
\end{align}
while each of $\tau$ and $C/6$ alone behave as $\sum_j z_j \theta_j^2$ in the collinear limit. Thus we see that the observable $\Delta_{\tau C}$ suppresses the collinear contribution by increasing the power of the angular exponent by two, but since this behavior is still polynomial in $\theta_j$, this $\Delta_{\tau C}$ is not a collinear drop observable. 

%The examples considered in this section had power law suppression of the collinear region, as opposed to the collinear drop criteria of having either zero or an exponentially suppressed contribution from collinear particles. This difference has observable consequences for jet distributions, as we will highlight in our numerical analysis in \Sec{sec:MC}.  \TODO{Confirm that this is true, strengthen or soften.}

%
%To suppress wide-angle radiation, we introduce an angular scale $r$ beyond which the weight function $\omega(\theta)$ decreases,
%\be
%\omega_2(\theta)
%=\left\{
%\begin{array}{lr}
%	(\theta/r)^\alpha (r/R)^\alpha & ~~~~~\theta < r\\
%	(r/\theta)^\beta (r/R)^\alpha & ~~~~~\theta > r
%\end{array}
%\right.
%\ee
%where $\alpha,\beta > 0$. In fact, this observable is related to soft drop angularity with soft drop parameters $(z_{\rm cut}, \beta)$. However, all the particles within jets contribute to this new observable whereas soft drop angularity removes particles from jets. 

\subsection{Comparison of Phase Space with Soft Drop and Collinear Drop}

For our analytic QCD based analysis we will focus on the collinear drop observable $\Delta m^2$ from \eqn{delta_m2_def}.
In the following sections we will derive a factorization formula for $d\sigma/d\Delta m^2$ using SCET, and use it to resum logarithmically enhanced terms from the hierarchies 
\begin{align}  \label{eq:dm2_hierarchies}
  & \Delta m^2 \ll Q^2 \,,
  && \frac{\theta_i}{2}\simeq \biggl( \frac{\Delta m^2}{ Q \Qcuti  } \biggr)^{\frac{1}{2+\beta_i}} \ll \frac{R}{2} \,,
\end{align}
where $i=1,2$. Since collinear radiation has been dropped, the first condition in \eqn{dm2_hierarchies} ensures that only soft radiation contributes to $\Delta m^2$. The second condition with $i=1,2$ ensures that the angle of soft radiation contributing to $\Delta m^2$ is always parametrically smaller than the jet radius $R$.  For this reason we refer to them as collinear-soft radiation contributions.  We will also assume that the soft drop grooming is removing soft radiation, and the collinear drop grooming is removing all collinear radiation, which requires
\begin{align}  
& z_{{\rm cut}\,1} \ll 1  \,, 
& z_{{\rm cut}\,2} \ll 1 
\,.
\end{align}

\begin{figure}[t!]
	\hspace{-1.1cm} 
	\includegraphics[width=0.56\textwidth]{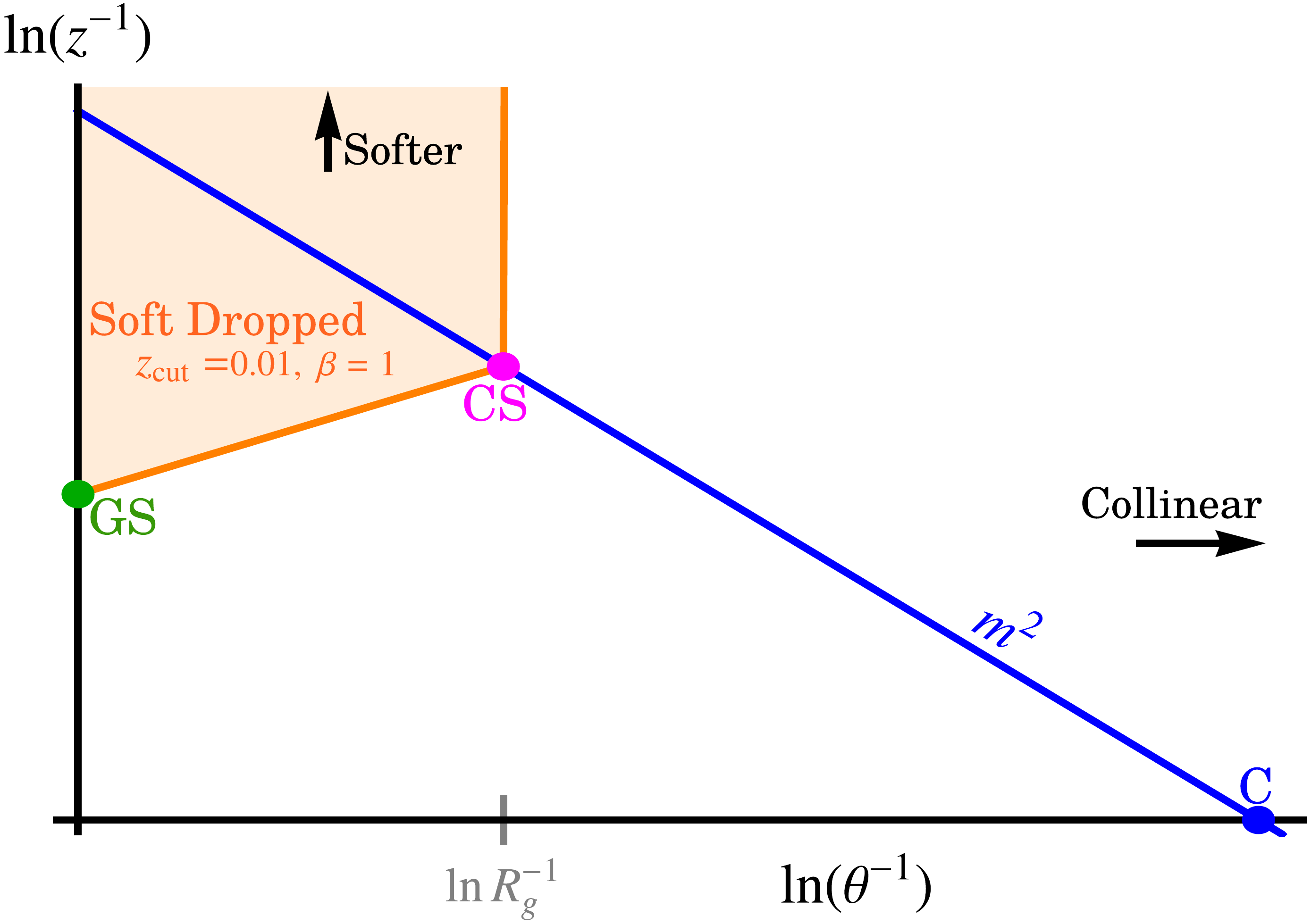}
	\includegraphics[width=0.53\textwidth]{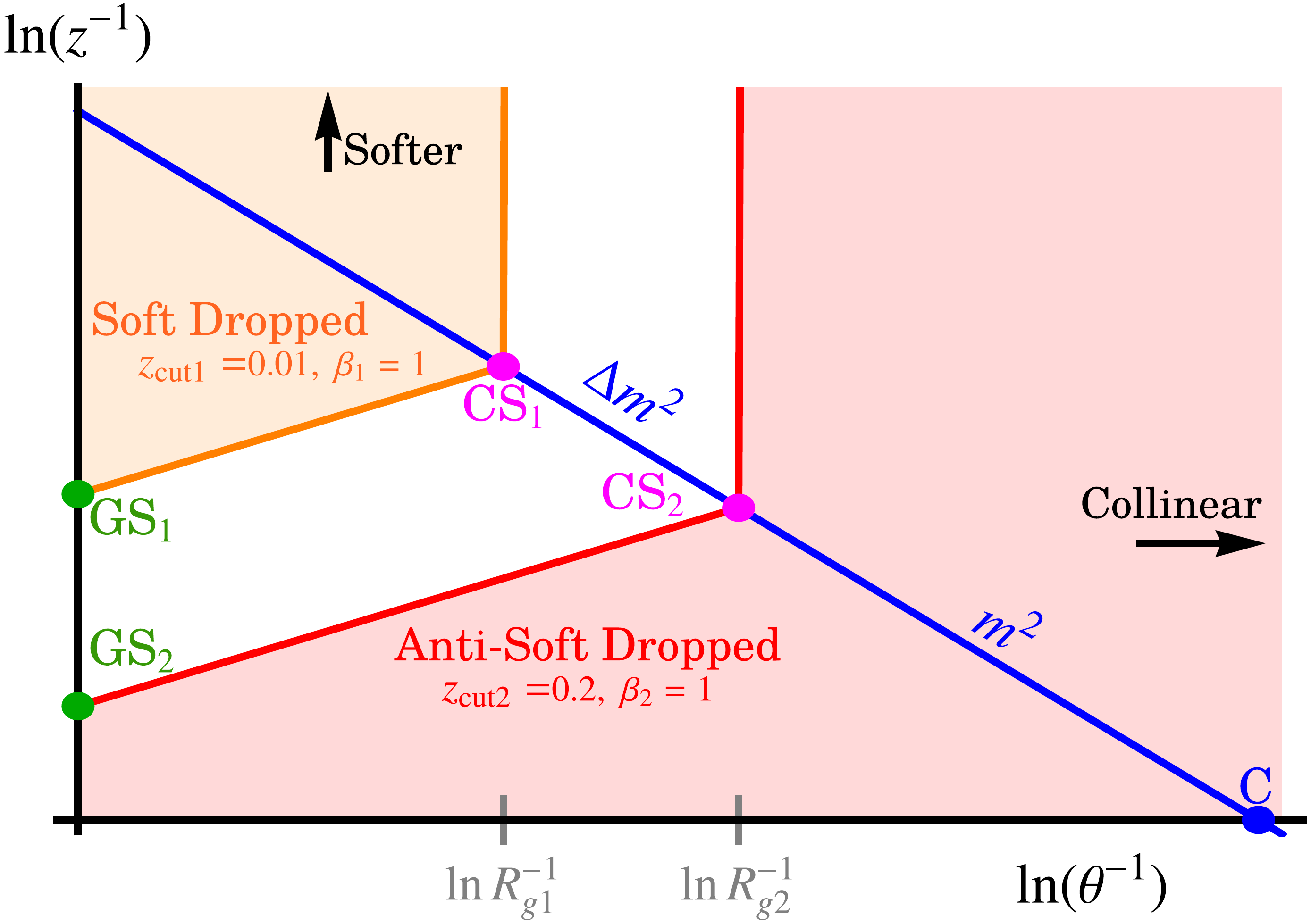}

	\caption{Phase space regions in the plane of energy fraction ($z$) and polar angle from the jet axis ($\theta$), where the white regions are those that are kept. 
		The left panel shows the result for soft drop where the orange shaded region is eliminated.  The right panel shows the result for collinear drop where in addition the red shaded region is eliminated. In SCET the 
		relevant degrees of freedom are collinear (C) modes, collinear-soft (CS) modes, and global soft (GS) modes, shown by solid dots. 
	}
	\label{fig:mode0}
\end{figure}

An illustration of the phase space of the radiation contributing to the soft-drop jet mass and our collinear drop jet mass observable is shown in \Fig{fig:mode0}. Here $z$ and $\theta$ are the energy fraction and angle relative to the jet energy and jet axis, respectively. The blue solid line indicates the scaling associated to the measurement, and the open white region is the phase space that contributes after one or both of the soft-drop and collinear drop conditions are applied.  The solid circles in these panels correspond to modes in SCET which we will discuss in more detail below.  The panel illustrate that, depending on the choice of parameters there can be hierarchies between the collinear drop and soft-drop constraints, such as:
\begin{align} \label{eq:SDhier}
 &  \zcuta \ll \zcutb \,,
 &  \frac{\theta_2}{2}\simeq \biggl(\frac{\Delta m^2}{ Q Q_{{\rm cut}\,2} }\biggr)^{\frac{1}{2+\beta_2}} 
 \ll \biggl( \frac{\Delta m^2}{ Q Q_{{\rm cut}\,1} } \biggr)^{\frac{1}{2+\beta_1}} \simeq  \frac{\theta_1}{2}  
  \,.
\end{align}
The hierarchies in \eqn{SDhier} require additional resummation, which we also carry out using our factorization based resummation approach. For simplicity we will restrict our results to NLL order\footnote{Note that since our results are presented for generic choices for $\beta_{1,2}$ we refer to them as NLL. For the special case of $\beta=0$ there is not a double logarithmic series in the groomed $m_J^2$ (or in $\Delta m^2$)~\cite{Dasgupta:2013ihk}, and hence these NLL terms are actually the leading-logs.}, though the factorization formula we have derived can be used for resummation at higher orders, and makes the procedure for this systematic. Indeed, in $e^+e^-$ collisions several event shape observables have been resummed at next-to-next-to-next-to-leading logarithmic accuracy \cite{Becher:2008cf,Abbate:2010xh,Chien:2010kc,Hoang:2014wka} using SCET based techniques.

%\section{Analytic Predictions for Collinear Drop $\Delta m^2$ }

\section{Soft Drop Factorization and the Groomed-Ungroomed Transition}
\label{sec:fact}

Since the collinear drop observable $\Delta m^2$ is defined using soft-drop jet masses, we first summarize in \Sec{sec:sdreview} key ingredients of the factorization of soft-drop jet mass~\cite{Frye:2016okc,Frye:2016aiz}, which are relevant for our factorization of $\Delta m^2$. Extensions required for small-$R$ resummation of jet masses were studied in Refs.~\cite{Kolodrubetz:2016dzb,Idilbi:2016hoa,Dasgupta:2016ktv,Marzani:2017mva,Baron:2018nfz}, and we discuss the analogous extension for soft-drop jet mass in \Sec{sec:sdforpp}.  Then in \Sec{sec:transition} we discuss how to handle the mass region where the transition from soft drop being effective to ineffective happens, developing corresponding profile scales, since to the best of our knowledge this has not yet been done in the SCET framework. 

\subsection{Review of Soft Drop Modes and Factorization for $e^+e^-$}
\label{sec:sdreview}

In this section we review the modes used to carry out resummation for soft drop in SCET following the analysis of Ref.~\cite{Frye:2016aiz} for hemisphere jets in $e^+e^-$ with $m_J^2/Q^2 \ll z_{\rm cut}\ll 1$. For our discussion we  specialize to the soft drop groomed jet-mass observable, $m_J$.  

Modes in SCET can be characterized by the scaling of their momentum components $(p^+,p^-,p_\perp)$ $ =(n\cdot p,\bar n\cdot p, p_\perp)$, which here are defined relative to the jet axis $\hat n_J$ by using the light-like vectors $n=(1,\hat n_J)$ and $\bar n = (1,-\hat n_J)$. 
The measurement of the soft-drop jet mass $m_J$ determines the relevant energetic collinear modes with momenta
\be
    p_c \sim \Big(\frac{m_J^2}{Q},Q,m_J\Big) = Q (\lambda^2,1,\lambda)\;,
\ee
where $\lambda = m_J/Q\ll 1$ is the power-counting parameter. $Q=p_J^-\simeq 2E_J$ is the center-of-mass energy of the $e^+e^-$ collision and $E_J$ is the jet energy. 
Together with the soft-drop condition with parameters $(z_{\rm cut},\beta)$, the relevant collinear-soft mode~\cite{Frye:2016aiz} emerges by solving the following system of constraints,
\begin{align}
    Qz \Big(\frac{\theta}{2}\Big)^2 \approx p_{\cs}^+\approx \frac{m_J^2}{Q}\;,~~~~~~z\approx \tilde z_{\rm cut}\, \theta^\beta
  \,,
\end{align}
where $z=E/E_J \simeq  p^-/p_J^-$ is the energy fraction, $\theta$ is the polar angle relative to the jet-axis, and we have taken $\theta\ll 1$. The parameter $\tilde z_{\rm cut} \propto z_{\rm cut}$ was defined in \eqn{ztcut} for $e^+e^-$ collisions.  
Therefore the collinear-soft mode has the following momentum scaling,
\begin{align}  \label{eq:pcsscaling}
    p_{\cs} &\sim \frac{m_J^2}{Q\zeta_{\cs}}
    \Big( \zeta_{\cs}, \frac{1}{\zeta_{\cs}},1\Big)
  \ \,, \qquad\qquad 
\end{align}
where we have made use of the shorthands
\begin{align}
 \zeta_{\cs} & \equiv \biggl( \frac{m_J^2}{QQ_{\rm cut}}\biggr)^{\frac{1}{2+\beta}} 
  \,,
 &&  Q_{\rm cut} \equiv 2^\beta\, \tilde z_{\rm cut}\, Q \,.
\end{align}
Note that this corresponds to a characteristic energy scale of $E_{\cs} \sim m_J^2/(2Q\zeta_{\cs}^2)=Q_{\rm cut} \zeta_{\cs}^\beta/2$ and angle $\theta_{\cs} \sim 2 \zeta_{\cs}$.\footnote{Taking $\eta_J=0$ for $pp$ collisions we have $\tilde z_{\rm cut}= z_{\rm cut}/R_0^\beta$, and \eqn{pcsscaling} becomes 
\begin{align}
p_{cs} & \sim \Big(\frac{m_J^2}{E_J},\:
 E_J z_{\rm cut}\Big(\frac{m_J}{E_JR_0\sqrt{z_{\rm cut}}}\Big)^{\frac{2\beta}{2+\beta}},\:
 m_J\sqrt{z_{\rm cut}}\Big(\frac{m_J}{E_JR_0\sqrt{z_{\rm cut}}}\Big)^{\frac{\beta}{2+\beta}}\Big) 
 \,,
\end{align}
which then agrees with Ref.~\cite{Frye:2016aiz}. Note that the combination $Q\zeta_{cs}$ is independent of $\eta_J$ since factors of $\cosh\eta_J$ cancel. 
}

The collinear and collinear-soft modes contribute to the soft-drop jet mass at leading power, and the distribution has the following factorized form,
\begin{align}
    \frac{d\sigma}{d m_J^2}=\sum_{i=q,g}
      N_i(\Phi_J,\tilde z_{\rm cut},\beta,\mu) 
   \: P^{\rm SD}_i(m_J^2,Q,\tilde z_{\rm cut},\beta,\mu)\;,
\end{align}
where $\Phi_J=\{E_J,\theta_J\}$ encodes the jet energy and angle (or $p_T$ and $\eta_J$ for a jet from a $pp$ collision). The perturbative $m_J$ spectrum is determined by
\begin{align} \label{eq:PSD}
   P^{\rm SD}_i(m_J^2,Q,\tilde z_{\rm cut},\beta,\mu) 
   =Q_{\rm cut}^{\frac{1}{1+\beta}} \int\!\! ds\, dk^+\, 
    J_i(s,\mu)\: 
    S_{Ci} \Big(k^+ Q_{\rm cut}^{\frac{1}{1+\beta}},\beta,\mu \Big)\: 
    \delta\bigl(m_J^2-s-Q k^+\bigr)\;,
\end{align}
which is a convolution of the inclusive jet function $J_i$ and the collinear-soft function $S_{Ci}$, which describe collinear and collinear-soft contributions to the groomed jet mass respectively.\footnote{In order to make manifest the functional dependence for $S_{Ci}$ derived in Ref.~\cite{Frye:2016aiz}, our notation for $S_{Ci}$ follows Ref.~\cite{Hoang:2018SDNP}.} The perturbative function $P_i^{\rm SD}$ has mass dimension $-2$.  The index $i=q,g$ labels the parton initiating the jet as either a quark or a gluon. The function $N_i$ encodes the process dependence, including for example a hard function for the hard scattering process, global soft function $S_{Gi}$, and proton parton distribution functions in the case of $pp$ collisions. The hard function describes the hard scattering process producing the energetic quark or gluon $i$ which initiates the jet. The global soft function describes how soft radiation within the jet is removed by the soft drop procedure, with the relevant global soft mode scaling as
\begin{align} \label{eq:pgsSDee}
    p_{\gs}\sim Q \tilde z_{\rm cut} (1,1,1) \;,
\end{align}
which does not depend on the measurement of the jet mass $m_J$. $N_i$ also encodes information of the radiation outside jets that affects the jet cross section. Note that the factor $N_i$ has multiple characteristic energy scales which depend on $z_{\rm cut}$, $\beta$ and $R$, which themselves could require resummation, but does not depend on the jet mass. The factorization scale dependence in the perturbative calculations of $N_i(\mu)$ and $P^{\rm SD}_i(\mu)$ cancels and the physical cross section is independent of the scale $\mu$.

Because of the convolution form of the factorized expression, it is convenient to study it in Laplace space. For any momentum space function $f(s)$ we define the Laplace transform $\tilde f(y)$ and its inverse by 
\begin{align} \label{eq:laplace}
 \tilde f(y) &= \int_0^\infty\!\!\! ds\,
  e^{-y e^{-\gamma_E} s } f(s)
  \;,
 && f(s) 
   = e^{-\gamma_E} \int_{c-i\infty}^{c+i\infty} \frac{dy}{2\pi i}\:
   e^{y e^{-\gamma_E} s} \, \tilde f(y)
 \,,
\end{align}
where we include the factor of $e^{-\gamma_E}$ when defining $y$ to simplify later equations.  With this transformation \eqn{PSD} is converted to a product form,
\begin{align}
    \tilde P_i^{\rm SD}(y,Q,\tilde z_{\rm cut},\beta,\mu) 
   &= \tilde J_i(y,\mu) \: 
     \tilde S_{Ci}\Bigl(Q Q_{\rm cut}^{\frac{-1}{1+\beta}}y,\beta,\mu\Bigr)\
   \\
   &= \widetilde J_i\Bigl( \ln\frac{1}{y\mu^2},\alpha_s(\mu) \Bigr) \:
      \widetilde S_{Ci} \Biggl( \ln\frac{Q_{\rm cut}^{\frac{1}{1+\beta}}}{yQ\mu^{\frac{2+\beta}{1+\beta}}}, \beta,\alpha_s(\mu) \Biggr)
  \,, \nn
\end{align}
where the Laplace space $\tilde J_i$, and $\tilde S_{Ci}$ are  dimensionless functions, and in the last line we have defined forms whose first arguments are given by the only logarithms that can appear. 
They satisfy multiplicative renomalization group (RG) equations,
\begin{align} \label{eq:RGEJiSCi}
  \frac{d}{d\ln \mu}\tilde J_i(y,\mu)
    &= \biggl[-2 \Gamma^i_{\rm cusp}(\alpha_s)\ln\frac{1}{y\mu^2}+\gamma^{J_i}(\alpha_s)\biggr]
   \tilde J_i(y,\mu)\;,
  \\
  \frac{d }{d\ln \mu}\tilde S_{Ci}\Bigl(Q Q_{\rm cut}^{\frac{-1}{1+\beta}} y,\beta,\mu\Bigr)
     &= \Biggl[ 2\Gamma^i_{\rm cusp}(\alpha_s) \ln\frac{Q_{\rm cut}^{\frac{1}{1+\beta}}}{\mu^{\frac{2+\beta}{1+\beta}} Q y }
     +\gamma^{S_{Ci}}(\alpha_s)\Biggr] \tilde S_{Ci}\Bigl(Q Q_{\rm cut}^{\frac{-1}{1+\beta}} y,\beta,\mu\Bigr)
  \;.
  \nn 
\end{align}
Here  $\Gamma^i_{\rm cusp}(\alpha_s)$ is the cusp anomalous dimension, which obeys Casmir scaling up to 3-loops, $\Gamma^i_{\rm cusp}(\alpha_s) = C_i \Gamma_{\rm cusp}(\alpha_s)$, where $C_q = C_F=4/3$ and $C_g = C_A=3$, and up to two-loops
\begin{align} \label{eq:Gcusp}
  \Gamma_{\rm cusp}(\alpha_s) = 4 \Big(\frac{\alpha_s}{4\pi}\Big) + 
  4 \Bigl[ \Big(\frac{67}{9}-\frac{\pi^2}{3}\Big) C_A - \frac{20}{9}T_F n_f
  \Bigr] \Big(\frac{\alpha_s}{4\pi}\Big)^2 \,,
\end{align}
where $T_F=1/2$ and $n_f$ is the number of active quark flavors.
Note that the cusp anomalous dimension term for $\tilde J_i$ and $\tilde S_{Ci}$ each depend only on the dimensionless combination of their two arguments, as expected.  The product appearing in $\tilde S_{C_i}$ can also be written in terms of the jet energy as
\begin{align}
  Q Q_{\rm cut}^{\frac{-1}{1+\beta}} &=  
    E_J^{\frac{\beta}{1+\beta}} \tilde z_{\rm cut}^{\frac{-1}{1+\beta}}
   \,.
\end{align} 

The $\mu$ dependence of the product $\tilde J_i \tilde S_{Ci}$ is canceled by the $\mu$ dependence of $N_i$, ensuring that the cross section is $\mu$ independent. The RGE for $N_i$ is also multiplicative,  
\begin{align}  \label{eq:RGENi}
  \frac{d}{d\ln \mu} N_i(\Phi_J,\tilde z_{\rm cut},\beta,\mu)
    &= \biggl[-2 \Gamma^i_{\rm cusp}(\alpha_s) \ln \frac{\mu^{\frac{\beta}{1+\beta}} Q_{\rm cut}^{\frac{1}{1+\beta}}}{Q}+\gamma^{N_i}(\alpha_s)\biggr] 
    N_i(\Phi_J,R,\tilde z_{\rm cut},\beta,\mu)
     \,,
\end{align}
where $\gamma^{N_i}(\alpha_s) + \gamma^{J_i}(\alpha_s)+\gamma^{S_{Ci}}(\alpha_s)=0$, and the sum of $\Gamma_{\rm cusp}^i$ terms in \eqns{RGEJiSCi}{RGENi} also vanishes.  The anomalous dimensions for $N_i$ in \eqn{RGENi} is independent of $y$, and has contributions from two scales that can be seen by writing
\begin{align} 
  \ln \frac{\mu^{\frac{\beta}{1+\beta}} Q_{\rm cut}^{\frac{1}{1+\beta}}}{Q}
  &=
\frac{1}{1+\beta} \ln \frac{\mu^\beta }{Q^{1+\beta} Q_{\rm cut}^{-1}}
  = \ln \frac{\mu}{Q} - \frac{1}{1+\beta} \ln \frac{\mu}{Q_{\rm cut}}
  \,.
\end{align}
Here the first term in the last equality comes from the hard function in $N_i$ and the second term from the global soft function $S_{Gi}$.

With $R\sim 1$ the fixed-order calculations of the jet, collinear-soft, global-soft, and hard functions determine the corresponding momentum space scales $\mu_J$, $\mu_{\cs}$, $\mu_{\gs}$, and $\mu_h$ where they have no large logarithms.  These are~\cite{Frye:2016aiz}
\begin{align} \label{eq:canscales}
\mu_{\jet} &= m_J \,,
& \mu_{\cs} &= \bigg( \frac{m_J^2}{Q} \bigg)^{\frac{1+\beta}{2+\beta}} Q_{\rm cut}^{\frac{1}{2+\beta}} 
 \,,
& \mu_{\gs} &= Q_{\rm cut} \,,
& \mu_h & = Q \,.
\end{align}
Note that the scale for the collinear-soft function, $\mu_{\cs}$, is a geometric average of the ultrasoft scale $m_J^2/Q$ that is present for jets without any grooming, and the scale $Q_{\rm cut}$. The canonical scales also satisfy the relation
\begin{align} \label{eq:canscalerelation}
  \mu_{cs}^{\frac{2+\beta}{1+\beta}}\: \mu_h = \mu_{\jet}^2 \: \mu_{\gs}^{\frac{1}{1+\beta}} \,.
\end{align}
Using RG equations we can evolve the jet and collinear-soft functions from their characteristic energy scales $\mu_j$ and $\mu_{\cs}$ to the factorization scale $\mu$, and the RG evolution will resum the logarithms of scale ratios. Note that when $\beta=0$, $\mu_\cs$ depends linearly on $m_J$ therefore the scale ratio $\mu_j/\mu_\cs$ is independent of $m_J$. This implies that the $m_J$ distribution only consists of single logarithms, as originally discussed in \cite{Dasgupta:2013ihk}. 

The  resummed expression of the soft-drop jet function is obtained by evolving the jet and soft-collinear functions from their natural scales ($\mu_j$ and $\mu_\cs$ respectively) up to the global-soft scale $\mu_\gs$. This gives 
\begin{align} \label{eq:sigSDresum}
    \frac{d\sigma}{d m_J^2}=\sum_{i=q,g}
      N_i(\Phi_J,\tilde z_{\rm cut},\beta,\mu_h,\mu_\gs) 
   \: P^{\rm SD}_i(m_J^2,Q,\tilde z_{\rm cut},\beta,\mu_\gs)\;,
\end{align}
where the normalization factor $N_i$ now also contains a resummation of large logarithms between $\mu_h$ and $\mu_\gs$ (which will not concern us here), and the resummation of large logarithms that modify the $m_J$ spectrum are all contained in the resummed result for $P_i^{\rm SD}$. Its resummed expression is 
\begin{align} \label{eq:PSDresum}
 & \hspace{-0.2cm}
P^{\rm SD}_i(m^2,Q,\tilde z_{\rm cut},\beta,\mu_\gs) 
  \\
    &= \exp\biggl[4C_i K(\mu_{\jet},\mu_\gs)-\frac{2(2+\beta)}{(1+\beta)}C_i K(\mu_{\cs},\mu_\gs) \biggr] \:
   \Biggl(\frac{\mu_\jet^{2}\, Q_{\rm cut}^{\frac{1}{1+\beta} } }
         {\mu^{\frac{2+\beta}{1+\beta}}_{\cs} Q }
       \Biggr)^{  2C_i \, \omega(\mu_\cs,\mu_\gs)} 
  \nn\\
    & \ \times 
   \exp\bigl[ \omega_{J_i}(\mu_\jet,\mu_\gs) 
    + \omega_{S_{Ci}}(\mu_{\cs},\mu_\gs)\bigr] 
  \nn\\ 
 & \ \times
     \widetilde J_i\bigl(\partial_\eta,\alpha_s(\mu_\jet) \bigr) \,
     \widetilde S_{Ci} \Biggl(\partial_\eta+\ln\bigg(\frac{\mu_\jet^2 Q_{\rm cut}^{\frac{1}{1+\beta}}}{\mu^{\frac{2+\beta}{1+\beta}}_{\cs} Q}\bigg) ,\beta,\alpha_s(\mu_{\cs})\Biggr)
   \frac{e^{-\gamma_E\eta}}{\Gamma(\eta)}\,
   \frac{1}{m^2}\Big(\frac{m^2}{\mu_\jet^2}\Big)^{\eta}
   \:\Bigg|_{\eta=2 C_i\, \omega(\mu_{\cs},\mu_\jet)}
  . \nn
\end{align}
Here $\partial_\eta = \partial/\partial\eta$ enter in a polynomial fashion through fixed order terms in the functions $\widetilde J_i$ and $\widetilde S_{Ci}$. 
The functions $K(\mu_1,\mu_2)$, $\omega(\mu_1,\mu_2)$,  $\omega_F(\mu_1,\mu_2)$ in the exponent are  RG evolution kernels, 
defined by
\begin{align}
    K(\mu_1,\mu_2) & = 
   \int_{\alpha_s(\mu_1)}^{\alpha_s(\mu_2)}  d\alpha
   \frac{\Gamma_{\rm cusp}(\alpha)}{\beta(\alpha)}
    \int_{\alpha_s(\mu_1)}^\alpha\frac{d\alpha'}{\beta(\alpha')}
    \;,
  \\
 \omega(\mu_1,\mu_2) & = \int_{\alpha_s(\mu_1)}^{\alpha_s(\mu_2)} 
      d\alpha\frac{\Gamma_{\rm cusp}(\alpha)}{\beta(\alpha)}
\,,
  & \omega_F(\mu_1,\mu_2) &= \int_{\alpha_s(\mu_1)}^{\alpha_s(\mu_2)} d\alpha\frac{\gamma^F(\alpha)}{\beta(\alpha)}
  \;. \nn
\end{align}
If we specialize to NLL order then the boundary conditions $\widetilde J_i$ and $\widetilde S_{Ci}$ can be set to $1$, the result for $\Gamma_{\rm cusp}(\alpha)$ is kept at 2-loops, and the result for the $\gamma^F(\alpha)$ terms are kept at 1-loop.

\subsection{Soft-Drop for $pp$ collisions with a Jet of Radius $R$}
\label{sec:sdforpp}

In this section we consider the generalization of the hemisphere $e^+e^-$ results from \Sec{sec:sdreview} to the case of $pp$ collisions with ungroomed jets of radius $R$. We will include also the case where $R/2$ is small, which is typically the case at the LHC and especially in heavy ion studies. 
Various parts of this generalization are straightforward. In particular for $pp$ collisions $Q= 2 E_J = 2 p_T\cosh(\eta_J)$, where $p_T$ and $\eta_J$ are the jet's transverse momentum and rapidity. Also we now use the $pp$ version of the soft-drop definition where $\tilde z_{\rm cut}= z_{\rm cut} (\cosh\eta_J/R_0)^\beta$ from \eqn{ztcut}. For the kinematic limit we are considering, the jet function $J_i$ is not modified relative to the $e^+e^-$ case since these modes never see the jet boundary. The hard function and other contributions to $N_i$ are modified for the $pp$ case, and in particular the relevant hard scale is $\mu_h = p_T R$. For a jet of radius $R$, the normalization function $N_i(\Phi_J,R,\tilde z_{\rm cut},\beta,\mu)$ also has explicit $R$ dependence because of the jet selection, so for $pp$ we have
\begin{align} \label{eq:factpp}
    \frac{d\sigma}{d m_J^2}=\sum_{i=q,g}
      N_i(\Phi_J,R,\tilde z_{\rm cut},\beta,\mu) 
   \: P^{\rm SD}_i(m_J^2,Q,\tilde z_{\rm cut},\beta,\mu)
   \,,
\end{align}
where $\Phi_J=\{p_T,\eta_J\}$ encodes the jet kinematics. We will discuss below the $R$ independence of $P^{\rm SD}_i$ for $pp$ in the soft-drop factorization region.

One important source of $R$ dependence is the in-jet global soft modes, whose scaling for $(p^+,p^-,p_\perp)$ is modified relative to \eqn{pgsSDee}.  To derive the modified scaling we set $\Delta R \simeq R$, and note that the polar angle $\theta$ relative to the jet axis $\hat n_J$ scales as $\theta \sim 2\sqrt{p^+/p^-} \sim R/\cosh\eta_J$. The overall energy scale is fixed by saturating the soft drop condition, 
yielding
\begin{align} \label{eq:pgsSDR}
  p_\gs &\sim Q z_{\rm cut}' \biggl( \frac{R^2}{4\cosh^2\eta_J},1,\frac{R}{2\cosh\eta_J} \biggr)
    \,,
  & z_{\rm cut}'& \equiv \tilde z_{\rm cut}  \biggl(\frac{R}{\cosh\eta_J}\biggr)^\beta
  = z_{\rm cut} \biggl(\frac{R}{R_0}\biggr)^\beta 
  \,.
\end{align}
This scaling relation drops ${\cal O}(1)$ factors associated to the deviation in the shape between a $\theta < R$ and $\Delta R < R$ jet, and also differs for a jet in an $e^+e^-$ collision.\footnote{For an $e^+e^-$ collision with a geometric jet of radius $R$ we instead have $p_{gs} \sim Q z_{\rm cut}' (R^2/4, 1, R/2)$ with
\begin{align}
z_{\rm cut}' \equiv  \tilde z_{\rm cut} \Bigl(2\sin\frac{R}{2} \Bigr)^\beta 
 = z_{\rm cut}\biggl(\frac{\sqrt{2} \sin(R/2)}{\sin(R_0^{ee}/2)} \biggr)^\beta
 \;,
\end{align}
which is equal to $\tilde z_{\rm cut} R^\beta$ in the small-$R$ limit. Again we require $z_{\rm cut}' \ll 1$ for the scale of the in-jet global soft mode to be parametrically smaller than the hard(-collinear) modes, $p_{gs}^2 \ll (Q R/2)^2$. 
}
Note that $p_{gs}^2 \sim (p_T R\, z_{\rm cut}')^2$ so we require $z_{\rm cut}'\ll 1$ to make the scale of the global soft mode distinct from the hard(-collinear) scale $\sim p_T R$. Here $p_T R$ is a hard-collinear scale for $R\ll 1$, and a hard scale for $R\sim 1$.  In the case of $R<R_0$, The condition $z_{\rm cut}'\ll 1$ holds when one requires that $z_{\rm cut} \ll 1$. However, in the case of $R>R_0$, a sufficiently large value of $\beta$ will break this condition so that $z_{\rm cut}'$ can go beyond $1/2$ which is the maximum value of the soft branch momentum fraction. In this case  the jet instead has a reduced radius $R_{\rm red}<R$ because the particles with their angle $\theta > R_{\rm red}$ are all dropped. 

One can consider the in-jet global-soft function which accounts for the cross section of dropped soft radiation and includes the part of the soft contributions in $N_i$ needed to cancel the $\mu$ scale dependence in $P_i$. For a single gluon radiation with $d=4-2\epsilon$ we have the bare 
result\footnote{The virtual contributions is scaleless and only contributes here by ensuring the proper interpretation of $1/\epsilon$ poles as UV.}
\begin{align}
 S_{Gi}(Q_{\rm cut},\beta,R,\epsilon) 
 &= 1+ \frac{ 4g^2 C_i\, \mu^{2\epsilon}e^{\epsilon\gamma_E}}
  {(2\pi)^d(4\pi)^\epsilon}
 \!\int\!\! \frac{d^dq\: 2\pi\delta^+(k^2)}{(q^+q^-)} 
 ~\overline\Theta_{\rm SD}^{(\gs)}~\Theta_{\rm alg}\;,
\end{align}
where $\delta^+(k^2) = \delta(k^2) \Theta(k^0)$ and
\begin{align} \label{eq:ThetaSD}
 \overline\Theta_{\rm SD}^{(\gs)} 
  &=\Theta\Bigl( Q \tilde z_{\rm cu t} (2q^+/q^0)^{\beta/2}-2q^0\Bigr) 
  = \Theta\Bigl( Q_{\rm cut}^{\frac{2}{2+\beta}}
      (q^+)^{\frac{\beta}{2+\beta}} - q^+-q^-\Bigr)
 \,, \nn\\
 \Theta_{\rm alg}& = \Theta\bigl(R^2-\Delta R^2\bigr)
  \,.
\end{align}   
For simplicity we will take $\Delta R^2 = \cosh^2\!\eta_J\, 4q^+/q^-$ which is strictly true in the $R/2\ll 1$ limit, noting that this also suffices to determine the appropriate scale for the global soft function even when $R/2\sim 1$.
This gives
\begin{align} \label{eq:SGppbare}
  S_{Gi}(Q_{\rm cut},\beta,R,\epsilon)
  &=1+ \frac{C_i\alpha_s(\mu)}{\pi}\frac{1}{1+\beta}\biggl( 
  \frac{1}{2\epsilon^2} + \frac{1}{\epsilon} \ln\frac{\mu}{Q_{\rm cut}'} 
  + \ln^2\frac{\mu}{Q_{\rm cut}'}  \biggr)+ \ldots \,,
\end{align}
where the ellipses denote terms that are not relevant to our discussion and
\begin{align} \label{eq:Qcutprime}
 Q_{\rm cut}' \equiv Q_{\rm cut} \biggl( \frac{R}{2\cosh\eta_J} \biggr)^{1+\beta}
  = p_T R\, z_{\rm cut}'  = p_T R\,z_{\rm cut} \Bigl( \frac{R}{R_0} \Bigr)^\beta \,.
\end{align}
This determines the appropriate result for the global soft scale to be $\mu_{gs} \simeq Q_{\rm cut}'$. Note that this $Q'_{\rm cut}$ is independent of $\eta_J$, and thus invariant to boosts along the beam axis. Also note that the $(1/\epsilon)\ln(\mu/Q_{\rm cut}')$ term in \eqn{SGppbare} induces a $\ln R/\epsilon$ term that is independent of $\beta$. This is only apparent because we have distinguished $R$ and $R_0$. 
 
The anomalous dimensions for $\tilde J_i$ and $\tilde S_{Ci}$ are not modified by the presence of the jet radius $R$, so RG consistency implies that $N_i$ must still satisfy \eqn{RGENi}.  The cusp term there involves the combination
\begin{align}  \label{eq:QQforpp}
  Q Q_{\rm cut}^{\frac{-1}{1+\beta}} 
 &= p_T \Big( p_T z_{\rm cut} R_0^{-\beta}  \Big)^{\frac{-1}{1+\beta}} 
  = p_T R \Big( p_T R z_{\rm cut} (R/R_0)^{\beta}  \Big)^{\frac{-1}{1+\beta}} 
   \,,
\end{align} 
which is both $\eta_J$ and $R$ independent. Since the contribution from the global soft function in \eqn{SGppbare} involves $Q_{\rm cut}'$ rather than 
$Q_{\rm cut}$, there must be an extra $R$ dependent contribution to the cusp contributions to $\mu d/d\mu N_i$.  This arises from contributions from outside the jet. 
For exclusive jet production~\cite{Ellis:2010rwa} this contribution is from the ``unmeasured soft function'' $S^{\rm unmeas}(R,\mu,\ldots)$, which is independent of the jet grooming but may depend on parameters for other parts of the event. With $R/2 \ll 1$ it involves the term
\begin{align}
S^{\rm unmeas}(R,\mu,\ldots) 
 &= 1+  \frac{C_i\alpha_s(\mu)}{\pi} \: \frac{1}{\epsilon}  
  \ln\Big( \frac{R}{2} \Big) + \ldots \,,
\end{align}
where the ellipses are finite ${\cal O}(\epsilon^0)$ terms or terms associated with other parts of the event. This cancels the $(1/\epsilon) \ln R$ term in the product $S_{Gi}S^{\rm unmeas}$ at ${\cal O}(\alpha_s)$, an contributes the appropriate term to give the $R$ independent anomalous dimension in \eqn{RGENi}.
For inclusive jet production with $R/2\ll 1$ the required contribution to $N_i$ occurs from a hard-collinear matching coefficient $H_{i\to i}(R p_T,\mu,\ldots)$~\cite{Kang:2017mda}, whose anomalous dimension now has the required $\ln R$ dependence to cancel that from \eqn{Qcutprime}. These cancellations between the global soft function and contributions from outside the jet are direct analogs of the cancellation of factors of $R$ in the final result in \eqn{QQforpp}.

Next let us discuss how the collinear-soft function should be generalized for $pp$. When $R\sim 1$ the scaling for the collinear-soft mode is identical to \eqn{pcsscaling}, since \eqn{dm2_hierarchies} implies that $\theta_{cs}\ll R$, and the collinear-soft function does not see the jet boundary. Thus the required function is still $S_{Ci} \Big(k^+ Q_{\rm cut}^{\frac{1}{1+\beta}},\beta,\mu \Big)$ or $\tilde S_{Ci}\Bigl(Q Q_{\rm cut}^{\frac{-1}{1+\beta}} y,\beta,\mu\Bigr)$ with the same anomalous dimension in \eqn{RGEJiSCi}. Th appropriate canonical scale for $\mu_{cs}$ is also still given by \eqn{canscales}. 
In fact we also have these same functions for the case $R\ll 1$, but now we must be more careful in determining the upper limit on $m_J$ for which this analysis in terms of collinear-soft and global-soft functions still holds.  To determine the limit, consider the one gluon emission calculation for  $S_{Ci}$, but including a jet boundary constraint $\Theta_{\rm alg.}$ for $R\ll 1$.  This yields the integral
\begin{align}
 \frac{ 2g^2 C_i\, \mu^{2\epsilon}e^{\epsilon\gamma_E}}
  {(2\pi)^d(4\pi)^\epsilon}
\int \frac{dq^+dq^-d\Omega_{d-2}}{(q^+q^-)^{1+\epsilon}}~(2\pi)
\delta(q^+-k^+) ~\Theta_{\rm SD}~\Theta_{\rm alg.}
\;,
% \big[\delta(q^+-k^+)-\delta(q^+)\big] 
\end{align}
Since the modes in $S_{C_i}$ are collinear-soft we must take $q^+\ll q^-$ for the $\Theta$ functions in \eqn{ThetaSD}, so the appropriate soft-drop and jet boundary constraints are
\begin{align}
    \Theta_{\rm SD} &= \Theta\bigg(q^-- Q_{\rm cut} \Bigl(\frac{q^+}{q^-}\Big)^{\beta/2}\biggr)
   \;,
     &\Theta_{\rm alg.} &= \Theta\Big(\frac{R^2}{4\cosh^2\eta_J}-\frac{q^+}{q^-}\Big)
   \;.
\end{align}
For fixed $q^+=k^+$ both of these constraints give a lower limit on $q^-$. Thus the range of validity is determined by having the $\Theta_{\rm SD}$ constraint be stronger than the $\Theta_{\rm alg.}$ constraint. This requires
\begin{align}
  k^+ < Q_{\rm cut} \Bigl( \frac{R}{2\cosh\eta_J}\Bigr)^{2+\beta}
%  = \frac{Q R^2}{4\cosh^2\eta_J} z_{\rm cut}' 
  = Q_{\rm cut}' \Bigl( \frac{R}{2\cosh\eta_J}\Bigr)
\,.
\end{align}
Setting $k^+ = m_J^2/Q$ this implies that the factorized description with a collinear-soft function is valid for
\begin{align} \label{eq:SDlimit}
  m_J < m_0 \equiv p_T R \sqrt{z_{\rm cut}'}  \,.
\end{align}
Above the mass value $m_0$ the soft drop grooming is no longer effective, and the factorization theorem transitions to the ungroomed jet mass result. We discuss this transition in detail in the next subsection.

In summary, at NLL with $m_J < m_0$ the same resummation formula for $P_i^{\rm SD}$ in \eqn{PSDresum} applies for the $pp$ case, except now the scale choices in \eqn{canscales} become 
\begin{align}  \label{eq:ppcanscales}
  \mu_\cs & = \bigg( \frac{m_J^2}{Q} \bigg)^{\frac{1+\beta}{2+\beta}} Q_{\rm cut}^{\frac{1}{2+\beta}} 
   = \bigg( \frac{m_J^2}{p_T R} \bigg)^{\frac{1+\beta}{2+\beta}} Q_{\rm cut}^{\prime\,\frac{1}{2+\beta}} 
    \,, 
  &  \mu_\gs & =  Q_{\rm cut}^\prime 
   = p_T R\,z_{\rm cut} \Bigl( \frac{R}{R_0} \Bigr)^\beta \,,
  \nn\\
    \mu_\jet &= m_J \,,  
  &  \mu_h & = p_T R 
   \,.
\end{align}
Once again these canonical scales obey the relation in \eqn{canscalerelation}. 
Note that the collinear-soft scale $\mu_\cs$ depends on $R_0$ but is independent of $R$. Also, all scales are independent of $\eta_J$, as is the combination $Q_{\rm cut}^{\frac{1}{1+\beta}}/Q=Q_{\rm cut}^{\prime\,\frac{1}{1+\beta}}/(p_T R)$ appearing explicitly in \eqn{PSDresum}.
The full result for $pp$ is also affected by changes to the calculation of $N_i$ which differs from the $e^+e^-$ case. 

For completeness we note that the analogous formulas to \eqn{ppcanscales} for a jet of radius $R$ in an $e^+e^-$ collision are
\begin{align}  \label{eq:eeRcanscales}
  \mu_\cs & = \bigg( \frac{m_J^2}{Q} \bigg)^{\frac{1+\beta}{2+\beta}} Q_{\rm cut}^{\frac{1}{2+\beta}} 
    \,, 
  &  \mu_\gs & =  Q_{\rm cut} \tan^{1+\beta}\Bigl(\frac{R}{2}\Bigr) \,,
  \nn\\
    \mu_\jet &= m_J \,,  
  &  \mu_h & = Q \tan\Bigl(\frac{R}{2}\Bigr)
   \,,
\end{align}
which reduce to the hemisphere case for $R=\pi/2$.

\subsection{Transition Between Groomed and Ungroomed Regions and Profiles}
\label{sec:transition}

\begin{figure}[t!]
\centering
    \includegraphics[height=7cm]{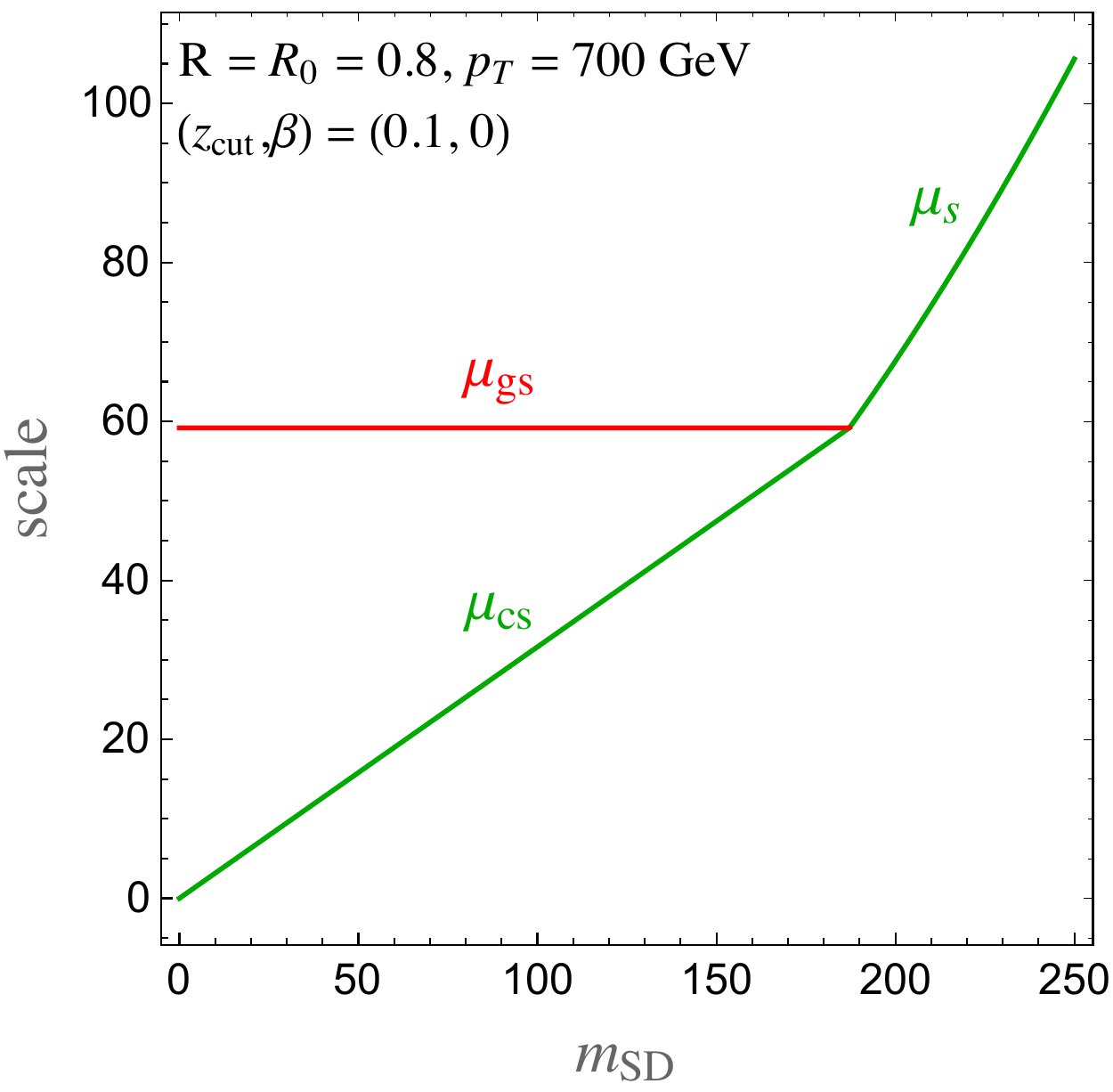}
    \caption{Canonical global soft scale $\mu_\gs$ and collinear soft scale $\mu_\cs$, which merge into the soft scale $\mu_s$ at the point $m_J = m_0$.  For $m_J > m_0$  the grooming  is no longer effective.}
\label{fig:scale}
\end{figure}

From \eqn{SDlimit} we saw that soft-drop for a jet of radius $R$ in $pp$ is no longer active when $m_J\ge m_0$. At $m_0$ the collinear-soft and global soft scales are equal,
\begin{align}
  \mu_{cs}(m_J^2=m_0^2) 
 &=\bigg( \frac{m_0^2}{Q} \bigg)^{\frac{1+\beta}{2+\beta}} Q_{\rm cut}^{\frac{1}{2+\beta}}  
  =\bigg( \frac{p_T R^2}{2 \cosh\eta_J } z_{\rm cut}'  \bigg)^{\frac{1+\beta}{2+\beta}}  \Bigl( p_T (2 \cosh\eta_J)^{1+\beta} z_{\rm cut}' R^{-\beta} \Bigr)^{\frac{1}{2+\beta}} 
 \nn\\
 &= p_T R\, z_{\rm cut}' = \mu_{gs}  \,.
\end{align}
Therefore the corresponding collinear-soft function ($S_{Ci}$) and global-soft function (inside $N_i$)  should be merged into a single soft function. For $R/2\sim 1$ the new relevant mode is (ultra)soft, while for $R/2\ll 1$ it is a different collinear-soft mode. In general the scaling for this mode is
\begin{align}
    p_s^\mu 
    \sim \frac{4m_{J}^2 \cosh^2\eta_J}{Q R^2} \Bigl(\frac{R^2}{4\cosh^2\eta_J},1,\frac{R}{2\cosh\eta_J} \Bigr)\;.
\end{align}
The canonical scale for this soft function is the standard (ultra)soft scale
$p_s^2\simeq \mu_s^2$ where $\mu_s = m_J^2/(p_T R)$. The description is continuous at $m_J=m_0$ since the value of $\mu_s$ is equal to the scale of the collinear-soft and global-soft functions at  $m_0$,
\begin{align} \label{eq:muSDungroom}
 \mu_s(m_0^2) = \mu_{cs}(m_0^2) = \mu_{gs}
 \,. 
\end{align}
The behavior of these scales and their merging is plotted in \Fig{fig:scale}. 

Thus we see that the soft drop factorization theorem must be smoothly transitioned to the ungroomed factorization theorem at $m_J=m_0=  p_T R \sqrt{z_{\rm cut}'}$. In our NLL resummed predictions this can be trivially accomplished by making the appropriate transition for the $\mu_{\cs}$ and $\mu_{\gs}$ scales. For the canonical scale choice we simply replace in \eqns{sigSDresum}{PSDresum} the scales as
\begin{align}
\mu_\cs \to  \mu_\cs(m_J^2)
& =\left\{
\begin{array}{lc}
   \Big( \frac{m_J^2}{Q} \Big)^{\frac{1+\beta}{2+\beta}} Q_{\rm cut}^{\frac{1}{2+\beta}}
%	m_J \sqrt{z_{\rm cut}}\Big(\frac{m_J}{E_JR_0\sqrt{z_{\rm cut}}}\Big)^{\frac{\beta}{2+\beta}}
   & ~~~~~m_J < m_0\\[5pt]
    \frac{m_J^2}{p_TR} & ~~~~~m_J \ge m_0
\end{array}
\right.\;, 
 \nn\\[5pt]
\mu_\gs \to \mu_\gs(m_J^2)
&=\left\{
\begin{array}{lc}
	p_T R\, z_{\rm cut}' & ~~~~~m_J < m_0\\[2pt]
    \frac{m_J^2}{p_TR} & ~~~~~m_J \gtrsim m_0
\end{array}
\right.\;,
\end{align}
At our NLL precision the endpoint of the (effectively ungroomed) jet mass spectrum is at $m_J=m_{\rm max}\equiv p_T R$, above which the cross section vanishes.  Our resummed distribution vanishes at this value due to the relation
\begin{align}\label{eq:muSDmax}
  \mu_s(m_{\rm max}^2) = \mu_J(m_{\rm max}^2) = \mu_h \,.
\end{align}
Note that for $z_{\rm cut}'\gtrsim1/2$, the transition does not happen and is superseded by the truncation at the reduced jet radius $m_J\lesssim p_TR_{\rm red}\sqrt{1/2}$.  In the next section we will provide theoretical predictions of soft-drop jet mass distributions. 

We will estimate the theoretical uncertainty by varying the scales $\mu_h$, $\mu_J(m_J^2)$, $\mu_{gs}(m_J^2)$ and $\mu_{cs}(m_J^2)$ in the resummation formula. This is done by using the method of profile functions in the SCET framework~\cite{Ligeti:2008ac,Abbate:2010xh}.  These scale variations are devised so that they always maintain the joining conditions in \Eqs{eq:muSDungroom}{eq:muSDmax}, and maintain the hierarchies between scales so that $\mu_h>\mu_{gs}>\mu_{cs}$ and $\mu_J>\mu_{cs}$.  In particular we determine the uncertainties at NLL by considering the following four variations
\begin{enumerate}
	\item Overall variation of all scales simultaneously up/down by a factor of two, so $\mu_i\to e_0  \mu_i$ with $e_0=1/2$ or $2$.
	\item Variation of the $\mu_{gs}$ and $\mu_{cs}$ scales by a multiplicative factor of $e_{s}=3/2$ or $2/3$ in the region $m_J\le m_0$, while simultaneously multiplying $\mu_s(m_J^2)$ for the region $m_J\ge m_0$ by a power $[m_J^2/(p_TR)^2]^{\ln e_s/\ln z_{\rm cut}}$ to maintain \eqns{muSDungroom}{muSDmax}.
	\item Variation of $\mu_J$ by a multiplicative trumpet factor of $\big[1+ e_J \big(1- \frac{m_J}{p_T R} \big)^2 \big]$ with $e_J=\pm 1/3$. 
	\item Variation of $\mu_{cs}$  by a multiplicative trumpet factor of $\big[1+ e_{cs} \big(1- \frac{m_J}{m_0} \big)^2 \Theta(m_0-m_J)\big]$ with $e_{cs}=\pm 1/3$. 
\end{enumerate}
We then compute the total uncertainty as simply the outer envelope of these variations.

\subsection{Monte Carlo and Partonic SCET Results for $m_J$}
\label{sec:SDcurves}

\begin{figure}[t!]
	\centering
	\includegraphics[width=0.49\textwidth]{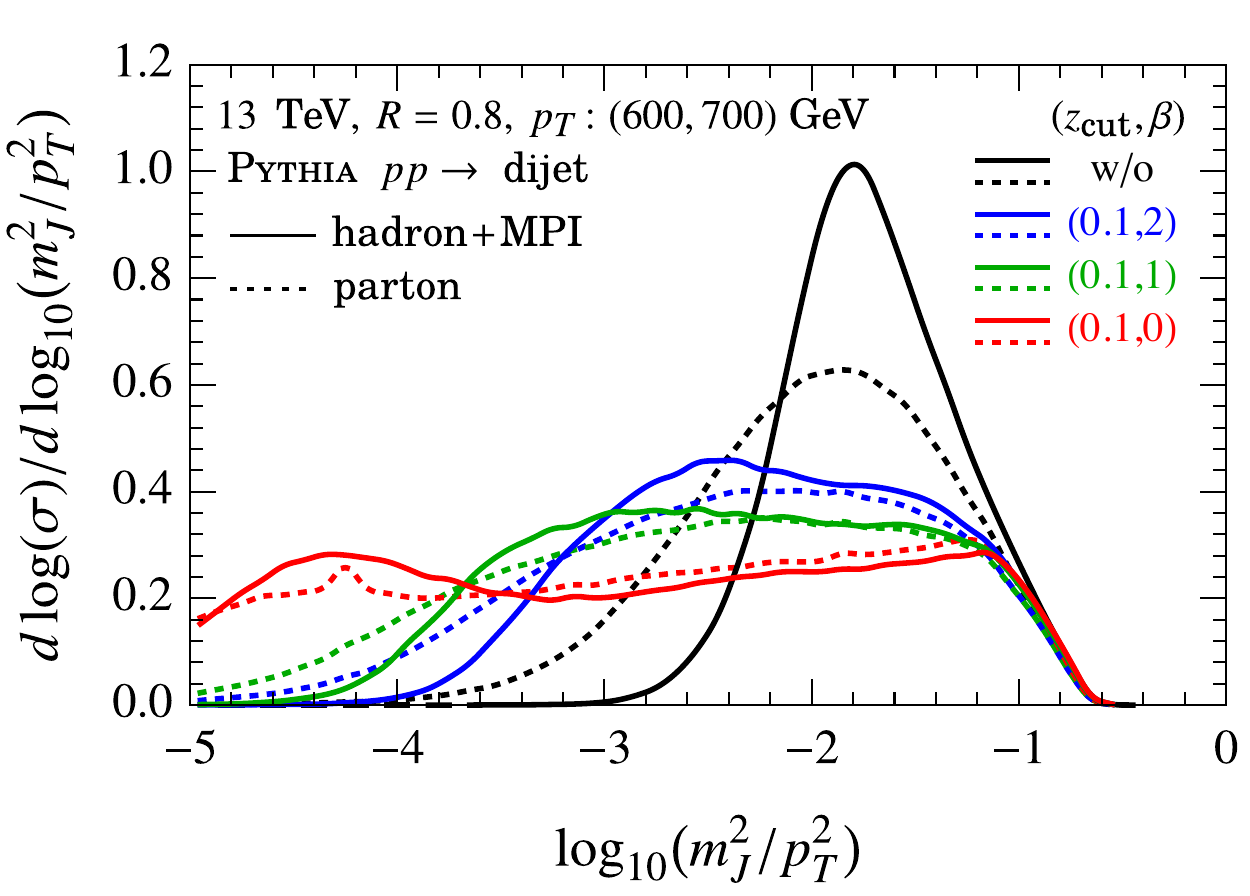}
	\includegraphics[width=0.49\textwidth]{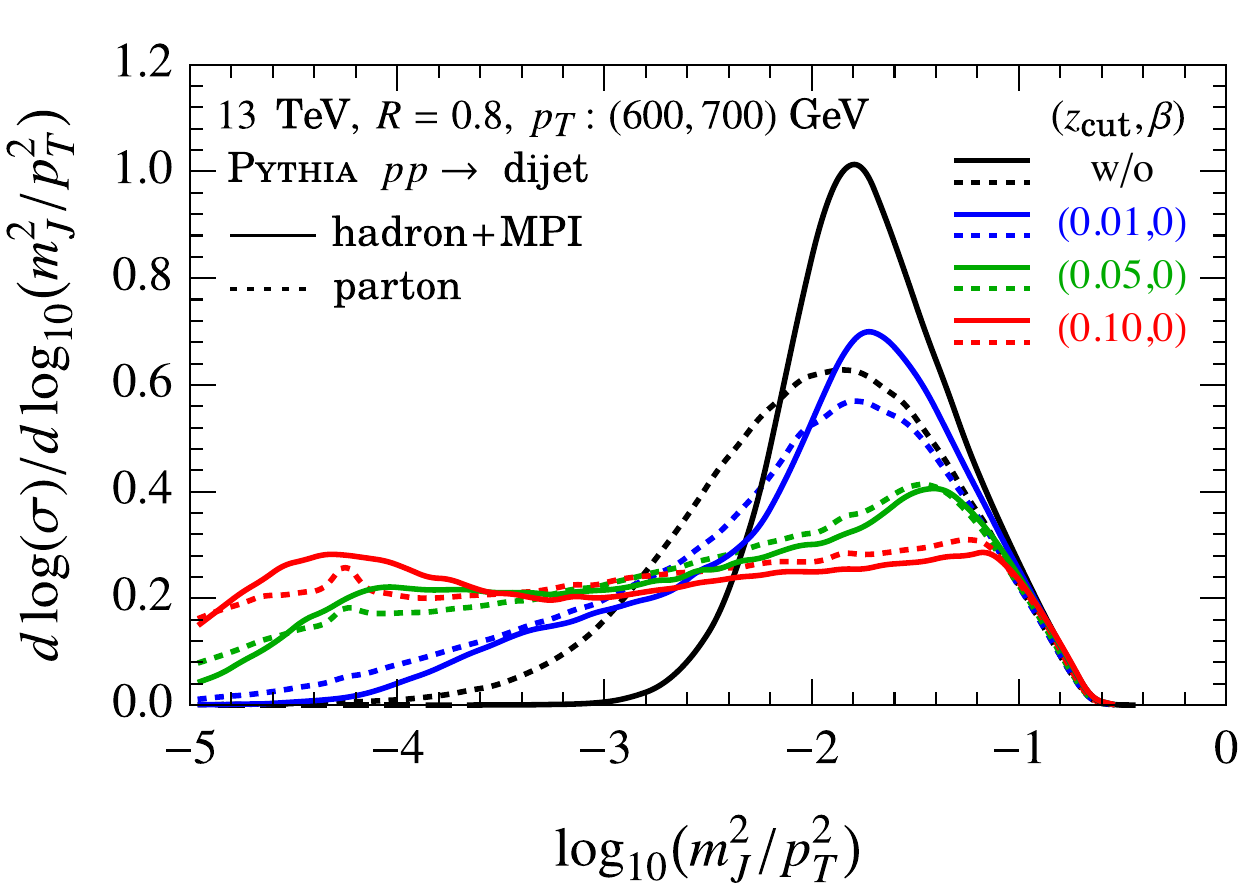}
\caption{Soft-Drop jet mass distributions from \Pythia simulations of $pp\to$ dijets at 13 TeV. The distributions are given at both parton level (dotted) and hadron level with MPI effects (solid). The jet mass distribution without grooming is also shown for comparison (black curves). For the soft drop curves the left panel fixes $z_{\rm cut}=0.1$ and varys $\beta=0,1,2$, while the right panel fixes $\beta=1$ and varies $z_{\rm cut}=0.01, 0.05, 0.1$.  }
	\label{fig:MC0_soft_drop}
\end{figure}

Having discussed soft drop groomed jet mass calculations, in this section we briefly discuss some features of the corresponding jet mass distributions~\cite{Dasgupta:2013ihk,Larkoski:2014wba,Marzani:2017kqd,Marzani:2017mva,Kang:2018jwa}. This will be useful for the purpose of drawing contrasts between the behavior of soft drop and collinear dropped jet mass for different choices of the grooming parameters.
Furthermore it will set a baseline of our discussion when we compare our theoretical NLL predictions to results from Monte Carlo simulations, which in this case can be compared also with data. We will also discuss the groomed to ungroomed region transition, which will have a direct analog in the collinear drop distributions. In all cases the jets are reconstructed using the anti-$k_t$ algorithm with $R_0=R=0.8$ in dijet events from 13 TeV proton-proton collisions, and we impose the jet $p_T$ cut of 600 GeV $< p_T<$ 700 GeV to select high $p_T$ jets.

Figure \ref{fig:MC0_soft_drop} shows the \Pythia simulation of soft-drop jet mass distributions with various soft-drop parameters. The left panel shows results with $z_{\rm cut}=0.1$ fixed, varying $\beta=0,1,2$,  which were also the values used in the ATLAS measurement~\cite{Aaboud:2017qwh}. 
Soft drop groomed jet mass measurements have also been made by CMS~\cite{Sirunyan:2018xdh}.
The right panel shows various values of $z_{\rm cut}$ with a fixed $\beta$, and both panels also include the ungroomed jet mass distribution for comparison. 
The dotted lines are purely partonic \Pythia results, while the solid lines include  hadronization and multi-parton interactions (MPI).
The curves are plotted using $\rho=\log_{10}(m_J^2/p_T^2)$ to better highlight the various regions of the distribution, and are normalized $d\log\sigma/d\rho = (1/ \sigma)(d\sigma/d\rho)$. Here $\sigma$ is chosen so the area is normalized to 1 in the plotted region. In general, stronger grooming pushes the distribution to smaller $m_J$ and flattens the peak in the displayed distributions.
From the left panel of \Fig{fig:MC0_soft_drop} we see that the groomed to ungroomed region transition occurs at $\log_{10}(R^2 z_{\rm cut})\approx -1.2$ for $z_{\rm cut} = 0.1$, independent of the $\beta$ as expected, while in the right panel one can observe that the transition value changes as we vary $z_{\rm cut}$. For $m_J>p_T R \sqrt{ z_{\rm cut}}$ soft drop becomes ineffective, and the groomed distributions transition to the ungroomed distribution in this region. For $m_J<p_T R \sqrt{ z_{\rm cut}}$, soft drop  removes wide-angle soft radiation and deforms the Sudakov peak by increasing the distribution in the smaller $\log_{10}(m_J^2/p_T^2)$ region. 
Generally hadronization and MPI increase the value of the jet mass and soft drop suppresses these effects. As one decreases the value of $\beta$ or increases $z_{\rm cut}$, soft drop removes more particles and results in a wider distribution further toward small jet mass region. One can also see that the region where hadronization and MPI effects are significant is further pushed toward the left with more aggressive grooming.
With a very small $z_{\rm cut}\simeq 0.01 $ the grooming transition can even pass the Sudakov peak so that few particles are removed by soft drop and the distribution is only distorted in the very small jet mass region.

%On the other hand, the right panel of Figure \ref{fig:MC0_soft_drop} shows the distributions with $\beta$ fixed at the value 0 and multiple $z_{\rm cut}$ values (0.01, 0.05 and 0.1), which correspond to groomed to ungroomed region transition at $\log_{10}(R^2 z_{\rm cut})\approx -2.2, -1.5, -1.2$, respectively. As one decreases the value of $z_{\rm cut}$, soft drop becomes less aggressive and the transition point moves further into the Sudakov peak region where nonperturbative effects are expected to be significant.  In the $z_{\rm cut} \rightarrow 0$ limit soft drop is smoothly turned off. Therefore the parameter $z_{\rm cut} $ can be used as a probe of the Sudakov peak region which has promising application in discriminating quark-initiated and gluon-initiated jets.
 
\begin{figure}[t!]
	\centering
	\includegraphics[width=0.48\textwidth]{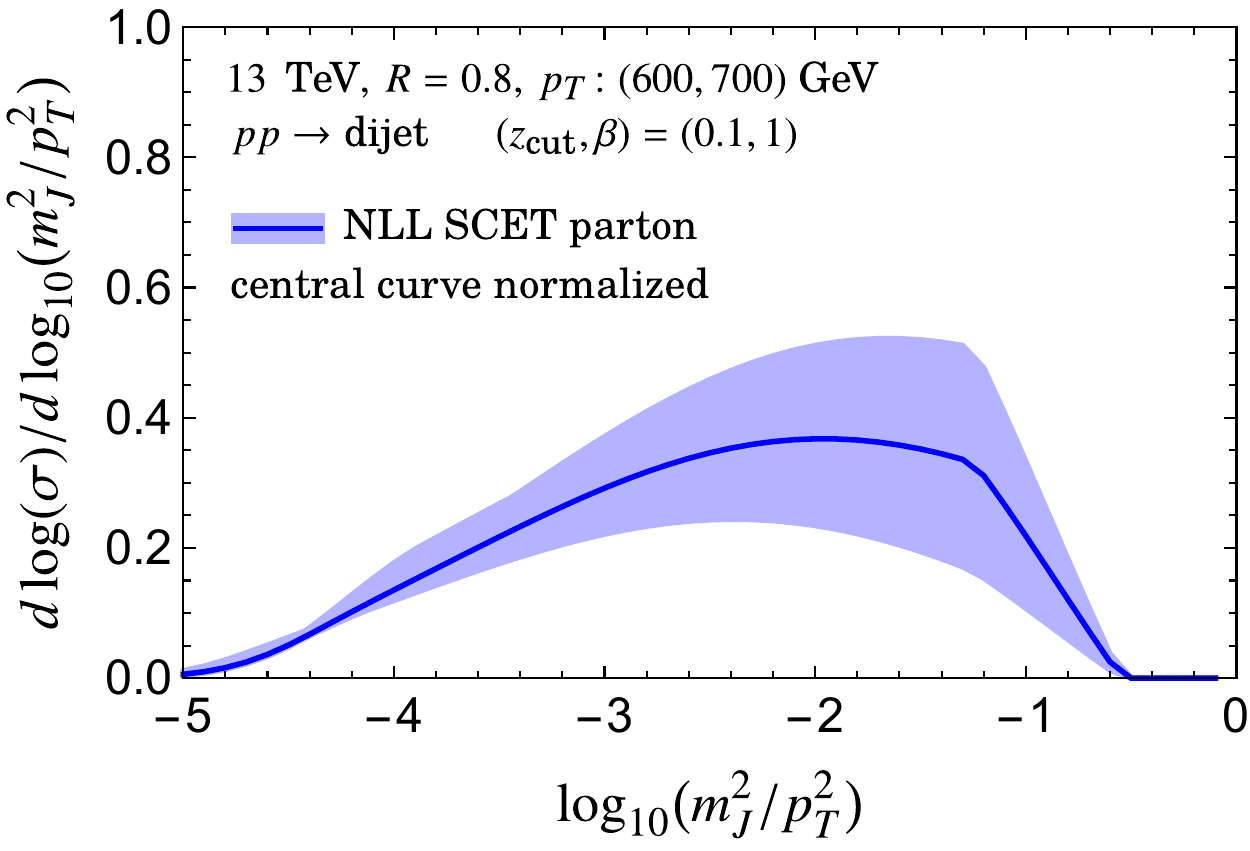}
	\includegraphics[width=0.48\textwidth]{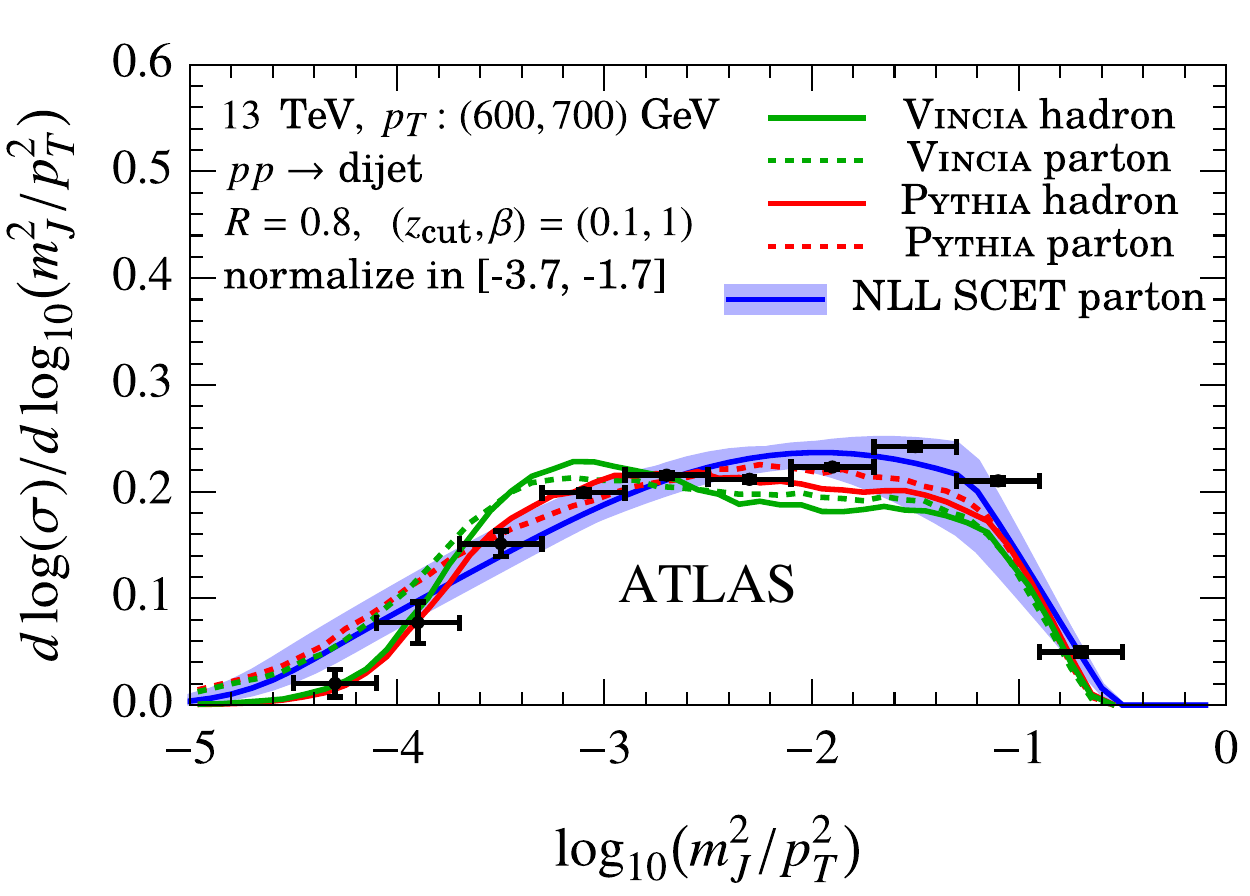}
	\includegraphics[width=0.48\textwidth]{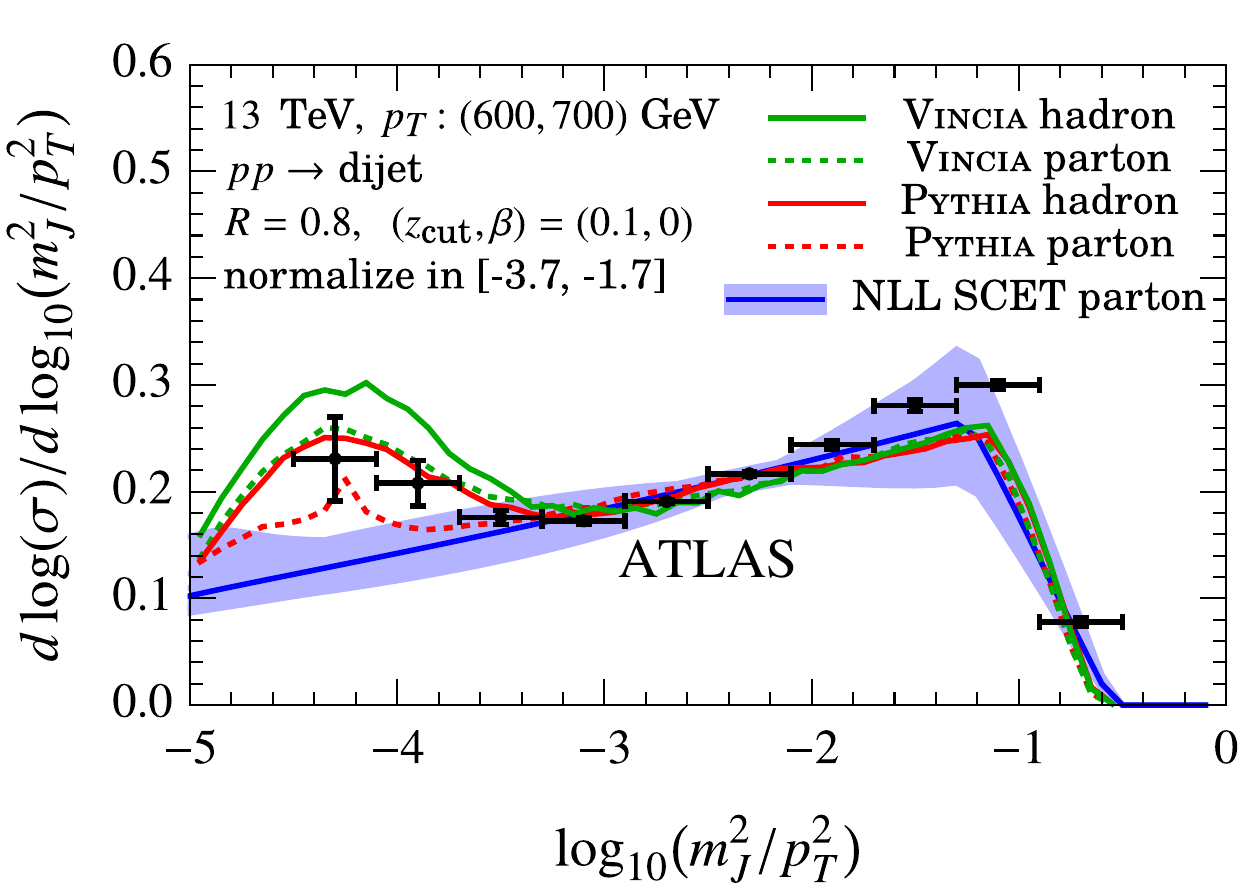}
	\includegraphics[width=0.48\textwidth]{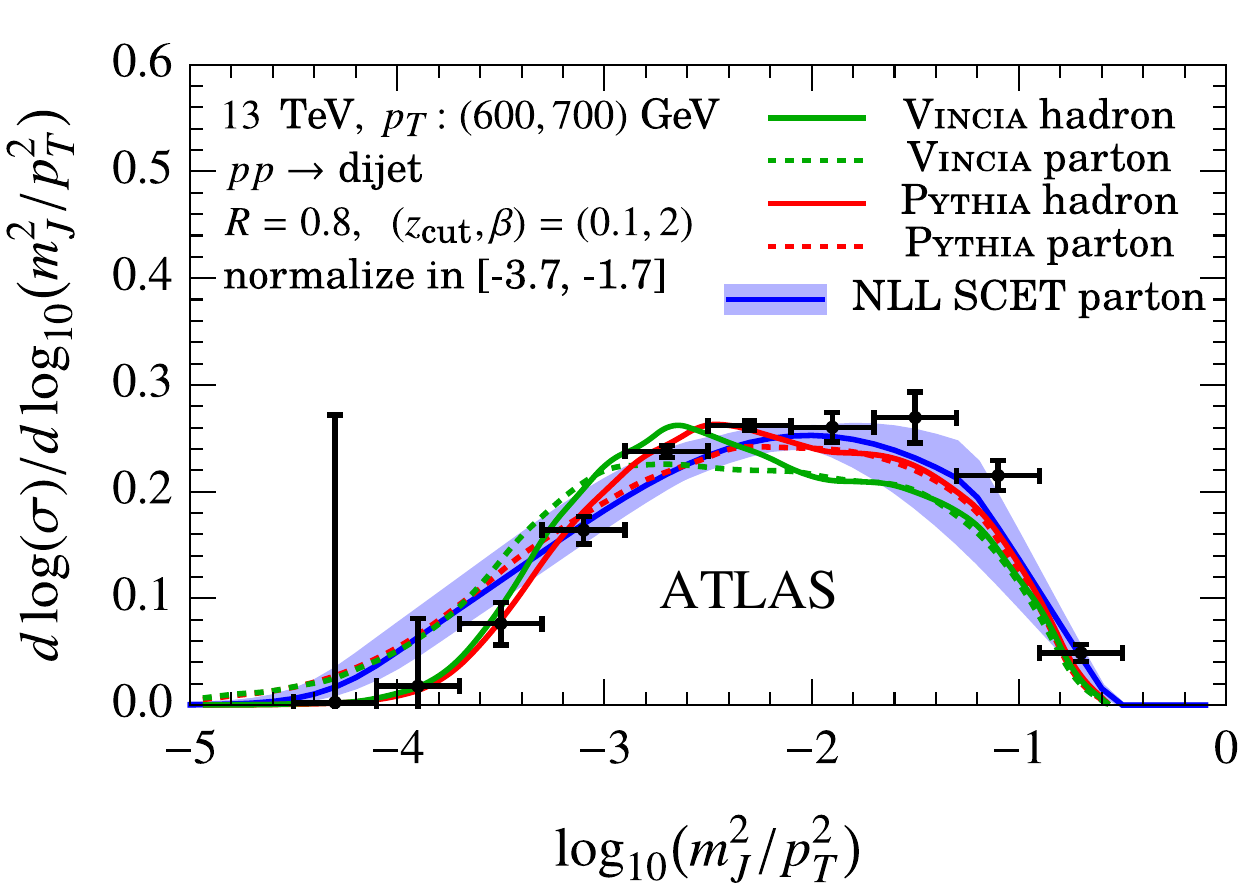} 
	\caption{Comparisons of soft drop jet mass distributions with $z_{\rm cut} = 0.1$ and $\beta=0, 1, 2$ in proton-proton collisions at 13 TeV. The solid blue curves represent the partonic NLL SCET results with uncertainty band estimated by scale variation. The top-left panel includes normalization uncertainty, whereas the remaining panels do not. Results are also shown for partonic and hadronic+MPI simulations from \Pythia and \Vincia, and compared to ATLAS data~\cite{Aaboud:2017qwh}.
	}
	\label{fig:ATLAS_soft_drop}
\end{figure}

Figure \ref{fig:ATLAS_soft_drop} shows results for the NLL partonic soft drop jet mass distribution with $z_{\rm cut}=0.1$. For the top left panel we take $\beta=1$ and display the uncertainty band obtained following the method described in \Sec{sec:transition}, while normalizing all variations to the central curve over the range shown. Since this panel includes the uncertainty in the normalization, the displayed variations are rather large at NLL. However, for a comparison to experimental data the cross sections are often normalized, making a comparison with only shape uncertainties more relevant. This is achieved for the NLL prediction in the top-right panel of \Fig{fig:ATLAS_soft_drop} by normalizing results within the range of $-3.7<\log_{10}(m_J^2/p_T^2)<-1.7$ as in the ATLAS measurement.  In particular, each of the scale variations used to estimate the perturbative are normalized in this fashion, prior to taking their envelope. This plot also shows for comparison partonic and hadronic  distributions for \Pythia and \Vincia simulations as well as ATLAS data from Ref.~\cite{Aaboud:2017qwh}. The lower panels show analogous results for $\beta=0$ and $\beta=2$ respectively. 

For $\beta=0$ all the NLL, \Pythia, and \Vincia curves somewhat undershoot the data in the region where $\log_{10}(m_J^2/p_T^2)\simeq -1.2$, but the NLL results are within our estimate for the perturbative uncertainties. In this region higher order fixed order perturbative corrections (included in the more detailed analyses in Refs.~\cite{Marzani:2017kqd,Marzani:2017mva,Kang:2018jwa}) 
are important. For small $\log_{10}(m_J^2/p_T^2)\lesssim -3.4$ one enters the region where nonperturbative hadronization corrections become ${\cal O}(1)$, as can be seen by the difference between partonic and hadronic simulation results. These differences are also visible at small $m_J$ in the $\beta=1,2$ panels.
For $\beta=1,2$ one can also see some difference between the \Pythia, \Vincia, and NLL SCET results at larger $m_J$ values. In both cases the central SCET partonic NLL curve being closer to that of \Pythia. We caution that no hadronization corrections have been included here in the SCET results, though such corrections have recently been rigorously characterized in Ref.~\cite{Hoang:2019ceu}.
In general we see that examining the NLL partonic SCET results enable us to see the bulk features of the soft drop jet mass spectrum, while not yet capturing the finer details entailed by inclusion of hadronization corrections and fixed order matching corrections.  The goal of our presentation of NLL results for collinear drop will be at a similar level, leaving more detailed analyses that reduce the theoretical uncertainties and include hadronization corrections to future work.

\section{Analytic Predictions for Collinear Drop $\Delta m^2$ }
\label{sec:CD}

In this section we carry out perturbative calculations for the collinear drop observable $\Delta m^2$, given by the difference of jet masses in \eqn{delta_m2_def} with soft-drop parameters $(\zcuta,\beta_1)$ and collinear drop parameters $(\zcutb,\beta_2)$. This allows us to carry out an all order resummation of large logarithms induced by the allowed soft radiation, and determine transition regions for this observable. 

\subsection{Collinear Drop from Soft Drop Grooming at ${\cal O}(\alpha_s)$} 

  To familiarize ourselves with $\Delta m^2$ consider the calculation of its distribution at ${\cal O}(\alpha_s)$. We consider only the region where $\Delta m \ll p R \sqrt{z_{{\rm cut}\,i}} \ll p R$ and $p$ is the initial parton momentum.  

\begin{figure}[t!]
\centering
    \includegraphics[height=7.5cm]{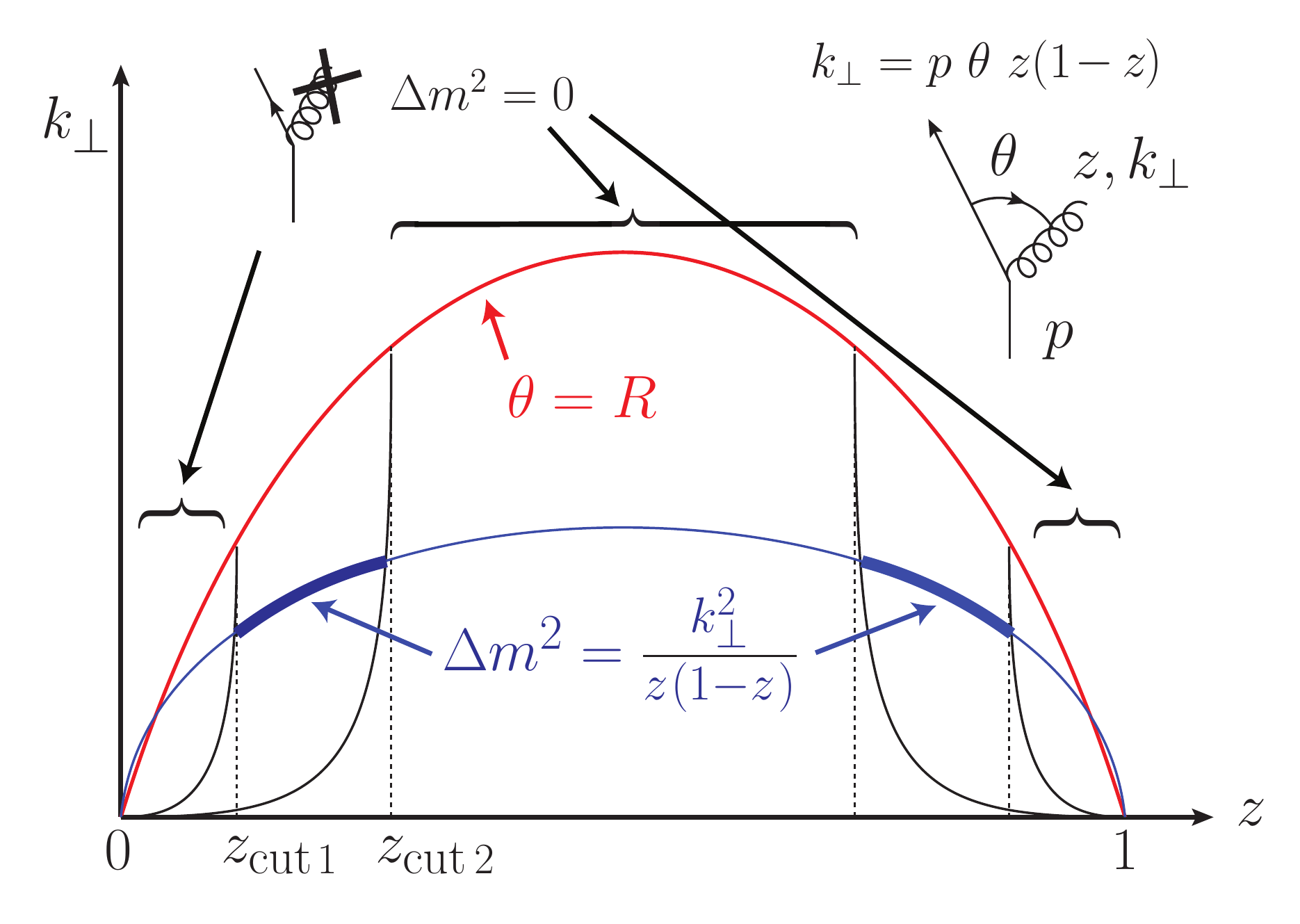}
    \caption{Single emission phase space regions that are kept and eliminated by collinear drop for the measurement of $\Delta m^2$.  The values of $z_{{\rm cut}\,1}$ and $z_{{\rm cut}\,2}$ are exaggerated for visibility.}
\label{fig:1loop}
\end{figure}

For this calculation we take $R_0=R$ and $\eta_J=0$ so that $p=p_T$, and use the Altarelli-Parisi splitting function ${\cal P}_{i\rightarrow j,k}(z)$, where the indices $i,j,k$ label the parton types in the $1\rightarrow 2$ splitting. This gives 
\begin{align}
    \frac{d\sigma_i^{(\alpha_s)}}{d\Delta m^2}
    = \sum_{j,k} \int dz \frac{dk_\perp}{k_\perp} \,
     {\cal P}_{i\rightarrow j,k}(z)\,
     \delta(\Delta m^2-\Delta m^2(z,k_\perp))~
     \Theta_{\rm CD}~\Theta_{\rm alg.}
   \,,
\end{align}
where $\Delta m^2(z,k_\perp) = k_\perp^2/[z(1-z)]$, and  the constraint imposed by the jet algorithm is given by $\Theta_{\rm alg.}=\Theta(R-\theta) = \Theta\big(p R\, \sqrt{z(1-z)}-\Delta m\big)$. Here $\theta = k_\perp /[p z(1-z)]=\Delta m/\big[p\sqrt{z(1-z)}\,\big]$, and the equalities involving $\Delta m$ use the relation imposed by the $\delta$-function.  The constraint $\Theta_{\rm CD}$ is the collinear drop condition which restricts the phase space to a soft region,
\begin{align}
	\Theta_{\rm CD}=
  \Theta\Big(\min (z,1-z)-z_{{\rm cut}\,1}
   \Big(\frac{\theta}{R}\Big)^{\beta_1}\Big) \:
  \Theta\Big( z_{{\rm cut}\,2} 
    \Big(\frac{\theta}{R}\Big)^{\beta_2}-\min (z,1-z)\Big)\;.
\end{align}
These constraints leave two strips in the phase space as shown in \Fig{fig:1loop}. Since $z_{{\rm cut}\,1} < z_{{\rm cut}\,2}\ll 1$, the allowed regions for $z$ either satisfy $z\ll 1$ or $(1-z)\ll 1$, implying that one of the two final state particles must be soft.

If $z\ll 1$ then the integration region is
\begin{align}
 \Big(\frac{\Delta m^2}{(pR)^2}\Big)^{\frac{\beta_1}{2+\beta_1}}
 \Big(z_{{\rm cut}\,1}\Big)^{\frac{2}{2+\beta_1}} 
 < z < 
 \Big(\frac{\Delta m^2}{(pR)^2}\Big)^{\frac{\beta_2}{2+\beta_2}}
 \Big(z_{{\rm cut}\,2}\Big)^{\frac{2}{2+\beta_2}}
\,.
%
%\Delta m^2(z,k_\perp)
%=\left\{
%\begin{array}{lc}
%	\frac{k^2_\perp}{z(1-z)} & ~~~~~\Big(\frac{\Delta m^2}{\mu_{J_R}^2}\Big)^{\frac{\beta_1}{2+\beta_1}}\Big(z_{{\rm cut}\,1}\Big)^{\frac{2}{2+\beta_1}} < z < \Big(\frac{\Delta m^2}{\mu_{J_R}^2}\Big)^{\frac{\beta_2}{2+\beta_2}}\Big(z_{{\rm cut}\,2}\Big)^{\frac{2}{2+\beta_2}}\\
%    \frac{k^2_\perp}{z(1-z)} & ~~~~~1-\Big(\frac{\Delta m^2}{\mu_{J_R}^2}\Big)^{\frac{\beta_2}{2+\beta_2}}\Big(z_{{\rm cut}\,2}\Big)^{\frac{2}{2+\beta_2}} < z < 1-\Big(\frac{\Delta m^2}{\mu_{J_R}^2}\Big)^{\frac{\beta_1}{2+\beta_1}}\Big(z_{{\rm cut}\,1}\Big)^{\frac{2}{2+\beta_1}} \\
%	0 & ~~~~~\mbox{otherwise}
%\end{array}
%\right.\;.
\end{align}
And for $(1-z)\ll 1$ there is the mirror image region obtained from $z\to 1-z$. Performing the integrals
the singular term at ${\cal O}(\alpha_s)$ involves a logarithm, 
\begin{align}
 \Delta m^2 \, \frac{d\sigma^{(\alpha_s)}}{d\Delta m^2}
   = \frac{\alpha_s(\mu)C_i}{\pi }
   \ln\Biggl[ \frac{z^{\frac{2}{2+\beta_2}}_{{\rm cut}\,2}}{z^{\frac{2}{2+\beta_1}}_{{\rm cut}\,1}}
   \biggl(\frac{\Delta m^2}{(p_TR)^2}
   \biggr)^{\frac{\beta_2}{2+\beta_2}-\frac{\beta_1}{2+\beta_1}}\Biggr]
    +{\cal O}\biggl[ \Big(\frac{\Delta m^2}{(p_TR)^2}\Big)^{\frac{\beta_i}{2+\beta_i}}\,
      z^{\frac{2}{2+\beta_i}}_{{\rm cut}\,i}\biggr] 
  \,.
\end{align}
The displayed term is the first term in the leading logarithmic series, while the terms not displayed are power suppressed in the limit we are considering as indicated.
We will use SCET to resum these logarithmically enhanced terms to all orders in $\alpha_s$, including terms up to the next-to-leading-logarithms. This includes at least all terms
$\Delta m^2 d\sigma/d\Delta m^2 \sim \sum_{k=1}^\infty [ \alpha_s^{k} L^{2k-1} + \alpha_s^{k} L^{2k-2} ]$ (with $L$ a generic large logarithm). Technically the resummation includes more terms since the counting and resummation are done by including the first two series of logarithms in the exponential in Fourier space. 

If we take $\beta_1=0$ and/or $\beta_2=0$ then the associated leading logarithmic singularity that depends on $\Delta m^2$ is removed, which is consistent with the behavior expected for the minimal-mass-drop limit of soft drop ($\beta_1=0$). Interestingly, there is also no double-logarithmic singularity at ${\cal O}(\alpha_s)$  in $\Delta m^2$ for $\beta_1=\beta_2$. We will demonstrate in \Sec{sec:b1isb2} that this absence of double logarithms persists to all orders in $\alpha_s$ for the leading logarithmic series.

\subsection{Factorization for Collinear Drop using Soft Drop Grooming}

\begin{figure}[t!]
	\includegraphics[width=0.49\textwidth]{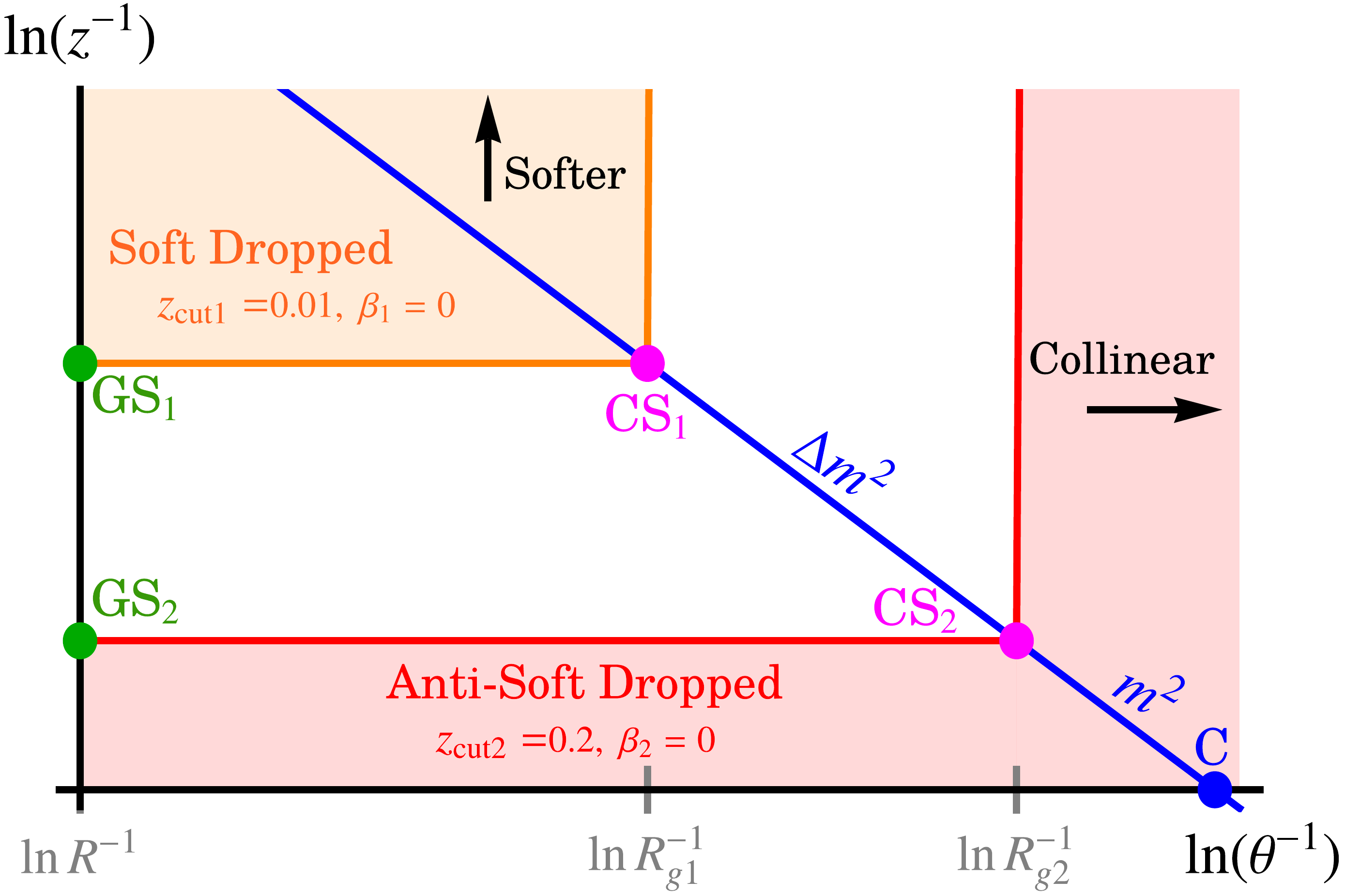}
	\includegraphics[width=0.49\textwidth]{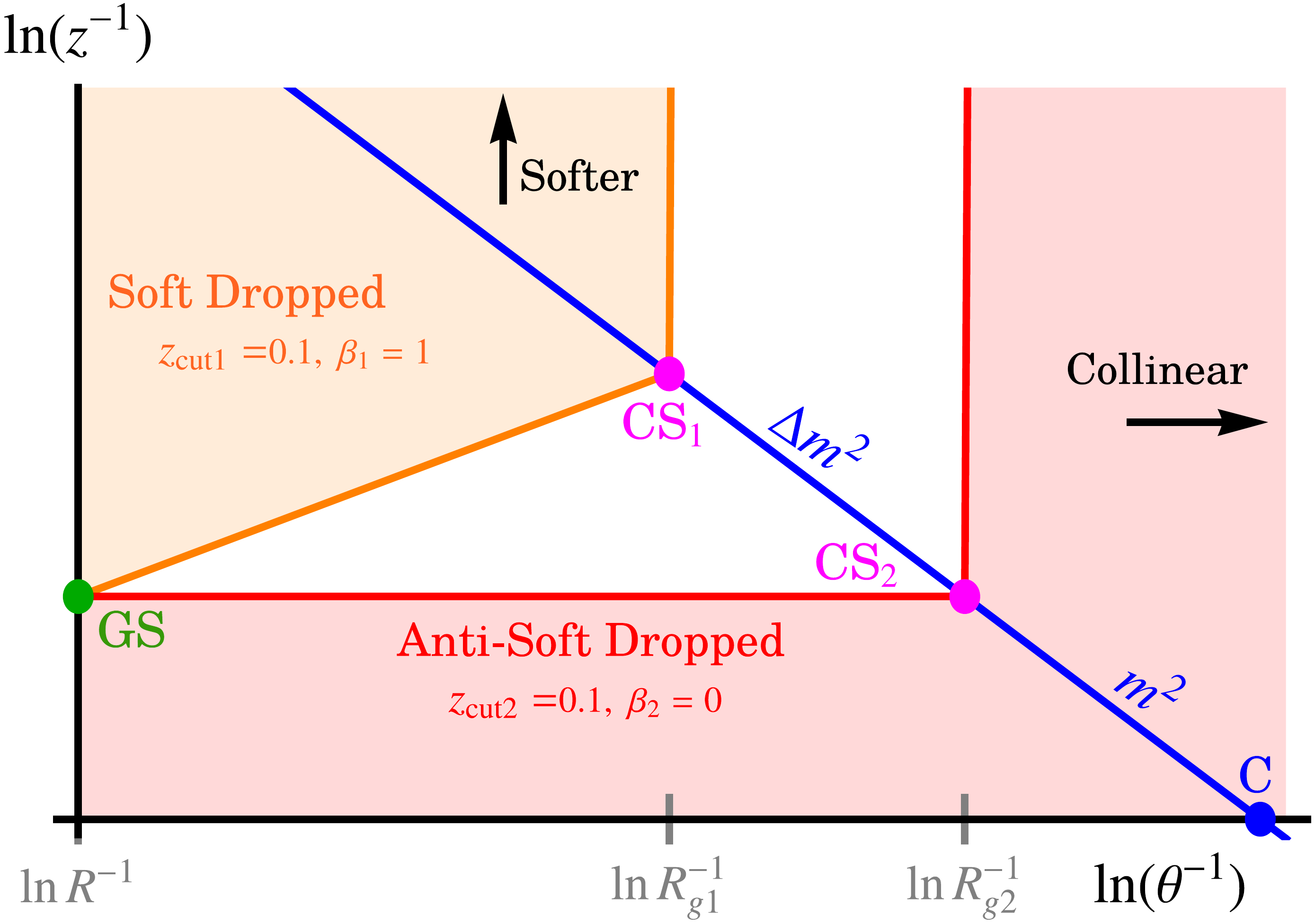}
	\\ 
	\includegraphics[width=0.49\textwidth]{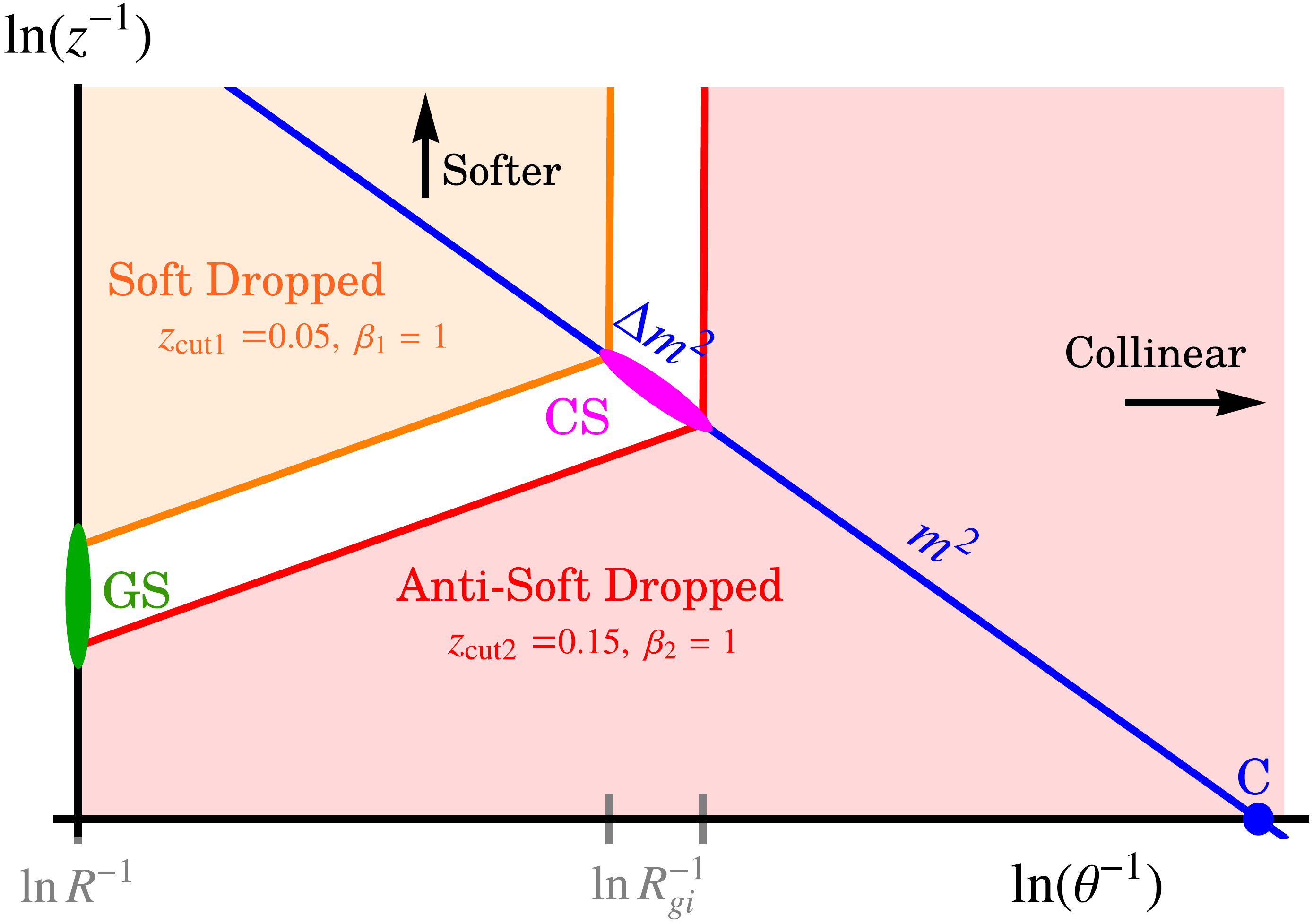}
	\includegraphics[width=0.49\textwidth]{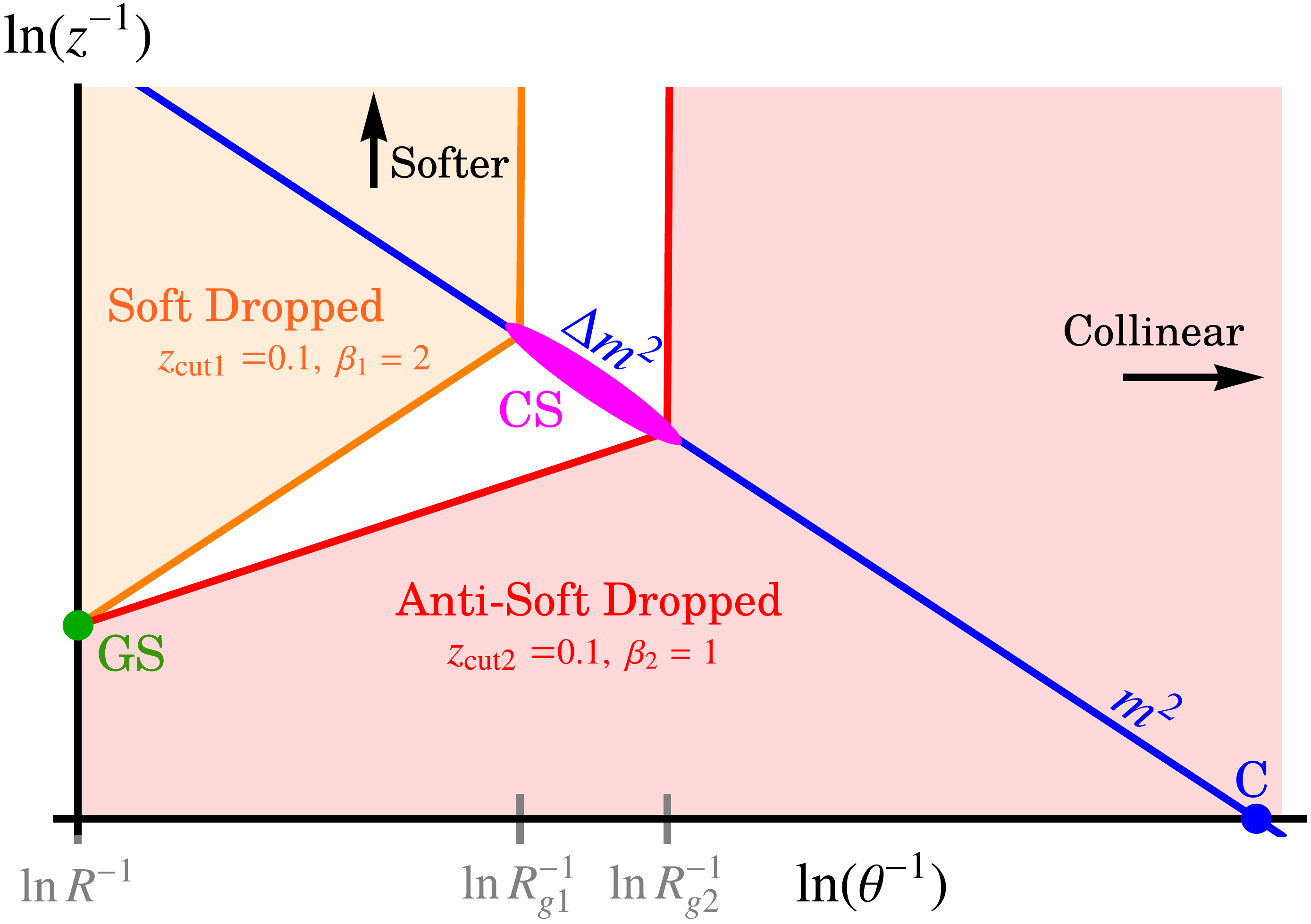}
	\caption{Regions kept by Collinear Drop in the plane of energy fraction ($z$) and polar angle from the jet axis ($\theta$).  Results are shown for four different values of $\{z_{{\rm cut}\,1},\beta_1\}$ and $\{z_{{\rm cut}\,2},\beta_2\}$, along with the corresponding modes needed for the SCET calculation. In the top two panels we have two collinear-soft (CS) modes and two global soft (GS) modes, while in the lower panels one or both of these pairs are combined into a single mode. 
	}
	\label{fig:mode1}
\end{figure}
 
Having summarized the major ingredients for soft drop jet mass calculations in SCET, we can now derive a factorization formula to perform resummation of large logarithms to all orders in $\alpha_s$  for $\Delta m^2$.  We carry out our analysis in the form relevant for $pp$ collisions and a jet of radius $R$. 
Our factorization analysis makes use of the SCET$_+$~\cite{Bauer:2011uc,Procura:2014cba,Larkoski:2015zka,Pietrulewicz:2016nwo} extension of SCET, which in our case includes two collinear-soft modes.

Since our collinear drop measurement also has a soft drop component, the dynamics of the jet being measured continue to factorize from the rest of the event, so analogous to \eqn{factpp} we have
\begin{align} \label{eq:CDfact}
    \frac{d\sigma}{d \Delta m^2}=\sum_{j=q,g}N_j^{\rm CD}(\Phi_J,R,\tilde z_{{\rm cut}\,i}, \beta_i,\mu) 
\:
  P^{\rm CD}_j(\Delta m^2,Q,\tilde z_{{\rm cut}\,i}, \beta_i ,\mu)
  \,.
\end{align}
Here $P_j^{\rm CD}$ determines the $\Delta m^2$ spectrum, while $N_j^{\rm CD}$ is a normalization factor, and we include a collinear drop superscript (CD) to indicate that both of these factors differ from the soft drop case. The functions $N_j^{\rm CD}$ and $P^{\rm CD}_j$ each depend on both $\{z_{{\rm cut}\,1},\beta_1\}$ and $\{z_{{\rm cut}\,2},\beta_2\}$, the former due to the contributions from two global soft modes, and the latter due to contributions from two collinear-soft modes.  In addition we still have $Q=2E_J$, and define
\begin{align}
 \tilde z_{{\rm cut}\,i} \equiv z_{{\rm cut}\,i} \, (\cosh\eta_J/R_0)^{\beta_i}
 \,.
\end{align}

The measurement of $\Delta m^2$ as well as the collinear drop condition impose the following constraints on the kinematics of the emissions,
\begin{align}
    z\theta^2\approx \frac{\Delta m^2}{E_J^2}\;,~~~~~~
  \tilde z_{{\rm cut}\,1}\: \theta^{\beta_1}\lesssim z
  \lesssim \tilde z_{{\rm cut}\,2}\: \theta^{\beta_2}\;.
\end{align}
These constraints are plotted in \Fig{fig:mode1} for several different choices of the $z_{{\rm cut}\,i}$ and $\beta_i$ parameters, taking $R=R_0$.  In these plots the intersection of the blue $\Delta m^2$ measurement line and the orange and red regions removed by collinear drop define collinear soft modes CS$_i$, whereas the intersection of the collinear drop constraints with the $\theta\simeq R$ line defines global soft modes GS$_i$. In all cases the collinear drop constraint involving $\{z_{{\rm cut}\,2},\beta_2\}$ removes the collinear region of phase space, including contributions from the collinear modes denoted by C.  

The soft drop mass measurement can be expressed as the insertion of a measurement function $\hat \delta$ that incorporates the jet reclustering and the collinear drop constraints. For a single emission
$\Theta_{{\rm SD}_1}$ imposes the constraint that we are below the orange boundary, and $\overline\Theta_{{\rm SD}_2}=1-\Theta_{{\rm SD}_2}$ ensures we are above the red boundary.  Therefore $\Theta^{\rm CD} = \Theta_{{\rm SD}1}\overline\Theta_{{\rm SD}2}=\Theta_{{\rm SD}1}-\Theta_{{\rm SD}2}$ selects the white unshaded regions in \Fig{fig:mode1}, so the measurement function is
\begin{align} \label{eq:deltaCD}
 \hat\delta = \delta\Bigl( \Delta m^2 -  \bigl[
    \Theta_{{\rm SD}_1} - \Theta_{{\rm SD}_2} \bigr] Q n\cdot \hat p \Bigr)
   \,.
\end{align}
For these $\Theta$ functions we can make approximations appropriate for collinear-soft radiation.
The global soft modes do not contribute to the $\Delta m^2$, and these modes capture contributions from the shaded regions to the event normalization. For a single emission these regions are determined by
\begin{align} \label{eq:deltaCDgs}
 \overline \Theta_{\rm CD}^{(\gs)} =   \Bigl[
    \overline\Theta_{{\rm SD}_1}^{(\gs)} 
  - \overline\Theta_{{\rm SD}_2}^{(\gs)} \Bigr]
   \,,
\end{align}
where $\overline\Theta_{{\rm SD}_i}^{(\gs)}=1-\Theta_{{\rm SD}_i}^{(\gs)}$ impose the constraints with approximations appropriate for global-soft radiation.

A key difference between the various panels in \Fig{fig:mode1} is whether the two restrictions present for collinear drop (labeled soft dropped and anti-soft dropped) are themselves hierarchically separated or not.  In the upper two panels the choice of parameters makes the constraints hierarchically separated, so we have distinct collinear-soft and global-soft modes on the soft drop and anti-soft drop boundaries. In the lower two panels one or both of the collinear-soft and global-soft modes merge into a single mode because the constraints are no longer fully hierarchical.  In general this distinction will affect the form and results derived from the factorization theorem, however we will see that at NLL order the description is continuous across these cases. 

We will begin by discussing the factorization structure of $P^{\rm CD}_i$ in the hierarchical case in the next section, followed by sections discussing various aspects of this result.  The generalization to non-hierarchical cases is left to \Sec{sec:CDnonhier}, and turns out to be very simple at NLL order.

\subsubsection{Collinear Drop $\Delta m^2$ with Hierarchical Constraints} 
\label{sec:CDhier}

Since the soft drop and anti-soft drop constraints are hierarchically separated, we can factorize the collinear drop constraint such that each boundary condition is individually satisfied by the modes that live on that boundary.  

Generalizing our soft drop discussion, in the hierarchical case we have two sets of global-soft modes, whose scaling is 
\begin{align} \label{eq:pgsCD}
  p_\gsa &\sim (Q z_{{\rm cut}\,1}') \biggl( \frac{R^2}{4\cosh^2\eta_J},1,\frac{R}{2\cosh\eta_J} \biggr)
    \,,
  & p_\gsb &\sim (Q z_{{\rm cut}\,2}') \biggl( \frac{R^2}{4\cosh^2\eta_J},1,\frac{R}{2\cosh\eta_J} \biggr)
    \,,
\end{align}
where
\begin{align}
 & z_{{\rm cut}\, i}'\equiv \tilde z_{{\rm cut}\,i}  \biggl(\frac{R}{\cosh\eta_J}\biggr)^{\beta_i}
  = z_{{\rm cut}\,i} \biggl(\frac{R}{R_0}\biggr)^{\beta_i} 
  \,.
\end{align}
We assume $z_{{\rm cut}\, 1}' \le z_{{\rm cut}\, 2}'$ so the GS$_2$ modes are always more energetic, $E_{\gsb}\ge E_{\gsa}$. 
For later convenience we also define
\begin{align}
 & Q_{{\rm cut}\,i} \equiv 2^{\beta_i}\, \tilde z_{{\rm cut}\,i}\, Q \,.
\end{align}
The corresponding scales where there are no large logarithms for the two global soft modes are $p_\gsa^2 \simeq \mu_\gsa^2$ and $p_\gsb^2 \simeq \mu_\gsb^2$, where
\begin{align}  \label{eq:gsscales}
  \mu_\gsa & = Q_{{\rm cut}1}' \equiv p_T R\,z_{{\rm cut}\,1} \Bigl( \frac{R}{R_0} \Bigr)^{\beta_1} \,,
 & \mu_\gsb & = Q_{{\rm cut}2}' \equiv p_T R\,z_{{\rm cut}\,2} \Bigl( \frac{R}{R_0} \Bigr)^{\beta_2} \,,
\end{align}
and we have $Q_{{\rm cut}i}' = Q_{{\rm cut}i} R^{1+\beta_i}/(2\cosh\eta_J)^{1+\beta_i}$.
Note that we will always have $\mu_\gsa \le \mu_\gsb$. 
Up to one loop the bare global functions for the hierarchical case have the following integral expressions,
\begin{align} \label{eq:SGhier}
 S_{Gj}\bigl( Q_{{\rm cut}\,1}, \beta_1,R,\epsilon\bigr)
   &= 1
     + \frac{4 g^2 C_j\, \mu^{2\epsilon} e^{\epsilon\gamma_E}} 
     {(2\pi)^d (4\pi)^\epsilon }
     \int\! \frac{d^dq\: 2\pi\delta^+(q^2)}{(q^+q^-)}
    \: \Bigl( \overline\Theta_{\rm SD_1}^{(\gs)} -1 \Bigr)
    \: \Theta_{\rm alg}
     \,, \\[5pt]
  \bar S_{Gj}\bigl(Q_{{\rm cut}\,2},\beta_2,R,\epsilon \bigr)
  &= 1
    + \frac{4 g^2 C_j\, \mu^{2\epsilon} e^{\epsilon\gamma_E}} 
     {(2\pi)^d (4\pi)^\epsilon }
    \int\!\! \frac{d^dq\: 2\pi\delta^+(q^2)}{(q^+q^-)}
   \: \Bigl(-\overline\Theta_{\rm SD_2}^{(\gs)} \Bigr)
   \: \Theta_{\rm alg}
   \,, \nn
\end{align}
where $\Theta_{\rm alg}$ is given in \eqn{ThetaSD} and
\begin{align}
    \overline\Theta_{{\rm SD}_i}^{(\gs)} &= 
\Theta\biggl( Q_{{\rm cut}\,i}^{\frac{2}{2+\beta_i}}
      (q^+)^{\frac{\beta_i}{2+\beta_i}} - q^+-q^-\biggr)
   \,.
\end{align}
To derive the form of the constraints in \eqn{SGhier} we use  \eqn{deltaCDgs} and apply the power counting. For $S_{G1}$ we set $\overline\Theta_{\rm SD_2}^{(\gs)}\to 1$ since the energy is always much smaller than the upper bound this constraint imposes. For $S_{G2}$ we set $\overline\Theta_{\rm SD_1}^{(\gs)}\to 0$ since its parametrically larger energy never satisfies this constraint. Performing the calculations gives
\begin{align}
S_{Gj}\bigl( Q_{{\rm cut}\,1}, \beta_1,R,\epsilon\bigr)
   &= 1 +\frac{\alpha_s(\mu)C_j}{\pi(1+\beta_1)} \biggl[
  \frac{1}{2\epsilon^2} + \frac{1}{\epsilon} \ln \frac{\mu}{Q_{{\rm cut}1}'}
    + \ln^2\frac{\mu}{Q_{{\rm cut}1}'} 
  + \ldots
%  - \frac{\pi^2}{24} (3+3\beta_1+\beta_1^2)
  \biggr]
  \,, \nn \\
\bar S_{Gj}\bigl( Q_{{\rm cut}\,2}, \beta_2,R,\epsilon\bigr)
   &= 1 -\frac{\alpha_s(\mu)C_j}{\pi(1+\beta_2)} \biggl[
  \frac{1}{2\epsilon^2} + \frac{1}{\epsilon} \ln \frac{\mu}{Q_{{\rm cut}2}'}
    + \ln^2\frac{\mu}{Q_{{\rm cut}2}'} 
   +\ldots
%  - \frac{\pi^2}{24} (3+3\beta_2+\beta_2^2)
  \biggr] 
  \,,
\end{align}
where the ellipses are terms that can be neglected at NLL order.
This enables us to determine the anomalous dimensions for the renormalized global soft functions
\begin{align} \label{eq:SGCDrge}
  \mu \frac{d}{d\mu} \ln S_{Gj}\bigl( Q_{{\rm cut}\,1}, \beta_1,R,\mu\bigr) 
     & = \frac{2C_j}{1+\beta_1} \Gamma_{\rm cusp}[\alpha_s] \ln\frac{\mu}{Q_{{\rm cut}1}'}  + \gamma_{S_{Gj}}[\alpha_s] 
    \,, \\
  \mu \frac{d}{d\mu} \ln \bar S_{Gj}\bigl( Q_{{\rm cut}\,2},   \beta_2,R,\mu\bigr) 
     & = - \frac{2C_j}{1+\beta_2} \Gamma_{\rm cusp}[\alpha_s] \ln\frac{\mu}{Q_{{\rm cut}2}'}  + \gamma_{\bar S_{Gj}}[\alpha_s] 
   \,,\nn
\end{align}
where $\Gamma_{\rm cusp}$ is given by \Eq{eq:Gcusp} and both $\gamma_{S_{Gj}}[\alpha_s]$ and $\gamma_{\bar S_{Gj}}[\alpha_s]$ vanish at one-loop. Note that it is perfectly consistent to keep the $R$ dependence in these anomalous dimensions. From the point of view of RG consistency this $R$ dependence cancels out in the product $S_{Gj} \bar S_{Gj}$ 

We also now have two sets of collinear-soft modes, as shown in the upper two panels of \Fig{fig:mode1}. They have the following momentum scaling,
\begin{align}  \label{eq:pcsCD}
    p_\csa &\sim \frac{\Delta m^2}{Q\zeta_\csa}
    \Big( \zeta_\csa, \frac{1}{\zeta_\csa},1\Big)\,, 
    &&p_{\csb} \sim \frac{\Delta m^2}{Q\zeta_\csb}
    \Big( \zeta_\csb, \frac{1}{\zeta_\csb},1\Big)\,,
\end{align}
where
\begin{align}
 \zeta_\csi & \equiv \biggl( \frac{\Delta m^2}{QQ_{{\rm cut}\,i}}\biggr)^{\frac{1}{2+\beta_i}} 
 \,.
\end{align}
More explicitly, combining definitions we have
\begin{align}
    p_\csa&\sim \Big(\frac{\Delta m^2}{2E_J},E_J z_{{\rm cut}\,1}\Big(\frac{\Delta m^2}{E_J^2R_0^2z_{{\rm cut}\,1}}\Big)^{\frac{\beta_1}{2+\beta_1}},
\sqrt{\Delta m^2z_{{\rm cut}\,1}}\Big(\frac{\Delta m^2}{E_J^2R_0^2z_{{\rm cut}\,1}}\Big)^{\frac{\beta_1}{2(2+\beta_1)}}\Big)
%\\
%    p_{\overline\csb}&\sim \Big(\frac{\Delta m^2}{E_J},E_J z_{{\rm cut}\,2}\Big(\frac{\Delta m^2}{E_J^2R_0^2z_{{\rm cut}\,2}}\Big)^{\frac{\beta_2}{2+\beta_2}},\sqrt{\Delta m^2z_{{\rm cut}\,2}}\Big(\frac{\Delta m^2}{E_J^2R_0^2z_{{\rm cut}\,2}}\Big)^{\frac{\beta_2}{2(2+\beta_2)}}\Big)
\;,
\end{align}
with an analogous result for $p_{\csb}$. 
The characteristic energy and angular scales are
\begin{align}
    E_\csi = \frac{\Delta m^2}{Q\zeta^2_\csi}=E_J z_{{\rm cut}\,i}\Big(\frac{\Delta m}{E_JR_0\sqrt{z_{{\rm cut}\,i}}}\Big)^{\frac{2\beta_i}{2+\beta_i}}\;,~~~~
    \frac{\theta_\csi}{2} = \zeta_\csi = R_0\Big(\frac{\Delta m}{E_JR_0\sqrt{z_{{\rm cut}\,i}}}\Big)^{\frac{2}{2+\beta_i}}\;.
\end{align}
Note that to have a non-trivial contribution to $\Delta m^2$ requires a non-trivial phase space for collinear-soft modes, which is ensured by the equivalent conditions:
\begin{align}
    E_\csb > E_\csa\;,~~~~\theta_\csb < \theta_\csa \,.
\end{align}
Thus we see that the ${\rm SD}_2$ collinear-soft mode lives at smaller angles.
The corresponding canonical scales for the two collinear-soft scales are
\begin{align}  \label{eq:CDcsscales}
 \mu_\csa & = \bigg( 
    \frac{\Delta m^2}{Q} \bigg)^{\frac{1+\beta_1}{2+\beta_1}} 
    Q_{{\rm cut}\,1}^{\frac{1}{2+\beta_1}} 
    \,, 
 & \mu_\csb & = \bigg( 
    \frac{\Delta m^2}{Q} \bigg)^{\frac{1+\beta_2}{2+\beta_2} }
    Q_{{\rm cut}\,2}^{\frac{1}{2+\beta_2}} 
    \,.
\end{align}
Here we always have $\mu_\csa < \mu_\csb$.
These results can also be written as 
\begin{align}
 \mu_{\csi}
  = \bigg(\frac{\Delta m^2}{p_T R} \bigg)^{\frac{1+\beta_i}{2+\beta_i}} 
    Q_{{\rm cut}\,i}^{\prime\,\frac{1}{2+\beta_i}} 
  = \sqrt{\Delta m^2 z_{{\rm cut}\,i}}\biggl( \frac{\Delta m^2}
    {(p_T R)^2z_{{\rm cut}\,i}}\biggr)^{\frac{\beta_i}{2(2+\beta_i)}}\;,
\end{align}
where the last equality is only true when taking $R=R_0$. The first equality shows that the canonical scale choice for $\mu_\csi$ is independent of $\eta_J$. 

The modes ${\rm SD}_1$ and ${\rm SD}_2$ contribute to the $\Delta m^2$ measurement. For individual soft drop jet masses $m^2_{\rm SD_1}$ and $m^2_{\rm SD_2}$ there are contributions from both collinear modes ($p_c$) and collinear-soft modes ($p_{{\rm cs}i}$), 
%\begin{align}
%    m^2_{\rm SD_1} &= (p_c+p_{\csa} )^2 =p_c^2+ Q\, n\cdot p_{\csa}  +\ldots
%    \,, \\
%    m^2_{\rm SD_2} &= (p_c+p_{\csb} )^2 =p_c^2+ Q\, n\cdot p_{\csb}  +\ldots
%    \,, \nn 
%\end{align}
$m^2_{\rm SD_1} = (p_c+p_{\csa} )^2 =p_c^2+ Q\, n\cdot p_{\csa}  +\ldots$, and $m^2_{\rm SD_2} = (p_c+p_{\csb} )^2 =p_c^2+ Q\, n\cdot p_{\csb}  +\ldots$,
where the ellipses denote terms that are power suppressed. When we take the difference to obtain $\Delta m^2$ the dependence on $p_c^2$ cancels.
Thus the leading power collinear drop measurement is given by
$\Delta m^2 = Q\: (  n\cdot p_{\csa} - n\cdot p_{\csb} )$. Therefore the collinear drop jet mass observable measures a concrete projection of soft radiation within the jet.  To define the momenta $p_{{\rm cs}i}$ we must include the collinear drop phase space constraints as in \eqn{deltaCD}, and implement the power counting for the hierarchical case.
%\begin{align}
%\Delta m^2 &= Q\: (  n\cdot p_{cs_1}
%                 - n\cdot p_{cs_2} )
%    \;.
%\end{align}

The SD$_1$ modes give a collinear-soft function $S_{Ci}$ which is identical to that for soft drop, since these modes have smaller energy and larger angle, and hence have $\Theta_{{\rm SD}_2}=0$ in \eqn{deltaCD}.
The SD$_2$ modes give a dropped collinear-soft function $D_{Ci}$ whose measurement constraint sets $\Theta_{{\rm SD}_1}=1$, which effectively gives the opposite phase space constraint to $S_{Ci}$. Up to one loop the bare functions therefore have the following integral expressions,
\begin{align} \label{eq:SChier}
    S_{Cj}\Bigl(k^+ Q_{{\rm cut}\,1}^{\frac{1}{1+\beta_1}},
       \beta_1,\epsilon\Bigr)
   &= \delta\Bigl(k^+ Q_{{\rm cut}\,1}^{\frac{1}{1+\beta_1}} \Bigr)
     \\
    & + \frac{4 g^2 C_j\, \mu^{2\epsilon} e^{\epsilon\gamma_E}} 
     {(2\pi)^d (4\pi)^\epsilon }
     Q_{{\rm cut}\,1}^{\frac{-1}{1+\beta_1}}\!
     \int\! \frac{d^dq\: 2\pi\delta^+(q^2)}{(q^+q^-)}
     \bigl[ \delta(q^+\!-\!k^+) - \delta(q^+)\bigr] 
    \Theta_{\rm SD_1}
     \,,\nn \\
    D_{Cj} \Bigl(k^+ Q_{{\rm cut}\,2}^{\frac{1}{1+\beta_2}},\beta_2,\epsilon \Bigr)
  &= \delta\Bigl(k^+ Q_{{\rm cut}\,2}^{\frac{1}{1+\beta_2}} \Bigr)
     \nn \\
    & +  \frac{4 g^2 C_j\, \mu^{2\epsilon} e^{\epsilon\gamma_E}} 
     {(2\pi)^d (4\pi)^\epsilon }
    Q_{{\rm cut}\,2}^{\frac{-1}{1+\beta_2}} \!
    \int\!\! \frac{d^dq\: 2\pi\delta^+(q^2)}{(q^+q^-)}
   \bigl[ \delta(q^+ \!-\!k^+) - \delta(q^+)\bigr] 
   \bigl(1-\Theta_{\rm SD_2}\bigr)
   , \nn
\end{align}
where
\begin{align}
    \Theta_{{\rm SD}_i} &= \Theta\bigg(q^-- Q_{{\rm cut}\,i} \Bigl(\frac{q^+}{q^-}\Big)^{\beta_i/2}\biggr)
   \,.
\end{align}
The complement constraint $1-\Theta_{\rm SD_2}$ is effectively equivalent to $-\Theta_{\rm SD_2}$ because of the scaleless integral for the $1$ term, therefore the calculation is the same as the one for soft drop jet mass with an addition minus sign. We find
\begin{align}  \label{eq:SCDCbare}
  S_{Cj}\Bigl(k^+ Q_{{\rm cut}\,1}^{\frac{1}{1+\beta_1}},
       \beta_1,\epsilon\Bigr) 
  &= \delta\Bigl(k^+ Q_{{\rm cut}\,1}^{\frac{1}{1+\beta_1}} \Bigr)
  + \frac{\alpha_s C_j}{\pi} \Biggl\{ 
  \delta\Bigl(k^+ Q_{{\rm cut}\,1}^{\frac{1}{1+\beta_1}} \Bigr) \,
   \frac{2+\beta_1}{1+\beta_1} \Bigl( - \frac{1}{2\epsilon^2} + \frac{\pi^2}{24} \Bigr) 
   \\
  &\qquad + \frac{1}{\epsilon}\, \mu^{\frac{-2-\beta_1}{1+\beta_1}}
     {\cal L}_0\biggl( \frac{k^+ Q_{{\rm cut}\,1}^{\frac{1}{1+\beta_1}} }
     {\mu^{\frac{2+\beta_1}{1+\beta_1}} } \biggr)
    - \frac{2(1+\beta_1)}{2+\beta_1} \mu^{\frac{-2-\beta_1}{1+\beta_1}}
     {\cal L}_1\biggl( \frac{k^+ Q_{{\rm cut}\,1}^{\frac{1}{1+\beta_1}} }
     {\mu^{\frac{2+\beta_1}{1+\beta_1}} } \biggr) 
   \Biggr\} 
   ,\nn\\
  D_{Cj}\Bigl(k^+ Q_{{\rm cut}\,2}^{\frac{1}{1+\beta_2}},
       \beta_2,\epsilon\Bigr) 
  &= \delta\Bigl(k^+ Q_{{\rm cut}\,2}^{\frac{1}{1+\beta_2}} \Bigr)
  - \frac{\alpha_s C_j}{\pi} \Biggl\{ 
  \delta\Bigl(k^+ Q_{{\rm cut}\,2}^{\frac{1}{1+\beta_2}} \Bigr) \,
   \frac{2+\beta_2}{1+\beta_2} \Bigl( - \frac{1}{2\epsilon^2} + \frac{\pi^2}{24} \Bigr) 
   \nn \\
  &\qquad + \frac{1}{\epsilon}\, \mu^{\frac{-2-\beta_2}{1+\beta_2}}
     {\cal L}_0\biggl( \frac{k^+ Q_{{\rm cut}\,2}^{\frac{1}{1+\beta_2}} }
     {\mu^{\frac{2+\beta_2}{1+\beta_2}} } \biggr)
    - \frac{2(1+\beta_2)}{2+\beta_2} \mu^{\frac{-2-\beta_2}{1+\beta_2}}
     {\cal L}_1\biggl( \frac{k^+ Q_{{\rm cut}\,2}^{\frac{1}{1+\beta_2}} }
     {\mu^{\frac{2+\beta_2}{1+\beta_2}} } \biggr) 
   \Biggr\} 
   \nn .
\end{align}
Renormalized $S_{Cj}$ and $D_{Cj}$ functions are obtained in the $\overline{\rm MS}$ scheme by removing all $1/\epsilon^2$ and $1/\epsilon$ terms here by suitable convolutions with counterterms. 

Note that $S_{Cj}$ is the same collinear-soft function as in the soft drop case. In Ref.~\cite{Frye:2016aiz} an all orders argument was given for the dependence of $S_{Cj}$ on only the combination $k^+ Q_{\rm cut}^{\frac{1}{1+\beta}}$. This argument is based on the structure of the soft drop constraint, comparisons made in CA clustering, and boost invariance of the Wilson lines in the operator defining $S_{Cj}$. This same argument applies equally well for the dependence of $D_{Cj}$ on the combination given in its first argument.  Furthermore, just as in soft drop, this implies that there are no non-global logarithms in the $\Delta m^2$ spectrum for this hierarchical case.

Using the Laplace transform of \eqn{SCDCbare}, we find that 
the functions $\tilde S_{Ci}$ and $\tDSC$ satisfy the following multiplicative RG equations,
\begin{align}  \label{eq:SCDCrge}
	\frac{d }{d\ln \mu} \ln \tilde S_{Ci}\Bigl(Q Q_{{\rm cut}\,1}^{\frac{-1}{1+\beta_1}} y,\beta_1,\mu\Bigr)
     	&=  2\Gamma^i_{\rm cusp}(\alpha_s) \ln\frac{Q_{{\rm cut}\,1}^{\frac{1}{1+\beta_1}}}{\mu^{\frac{2+\beta_1}{1+\beta_1}} Q y }
     +\gamma^{S_{Ci}}(\alpha_s)
    \,, \nn\\
     	\frac{d }{d\ln \mu} \ln \tDSC\Bigl(Q Q_{{\rm cut}\,2}^{\frac{-1}{1+\beta_2}} y,\beta_2,\mu\Bigr)
     	&=  -2\Gamma^i_{\rm cusp}(\alpha_s) \ln\frac{Q_{{\rm cut}\,2}^{\frac{1}{1+\beta_2}}}{\mu^{\frac{2+\beta_2}{1+\beta_2}} Q y }
     +\gamma^{\DSC}(\alpha_s)
    \,,
\end{align}
where $\gamma^{S_{Ci}}(\alpha_s)$ and $\gamma^{D_{Ci}}(\alpha_s)$ are zero at one-loop. 

%We see that at the lowest order collinear drop subtracts out the uncorrelated, exclusive $\csb$ phase space region from the $\csa$ phase space region, therefore at higher orders uncorrelated emissions may still contribute to the collinear drop observable. 

Putting the contributions to the $\Delta m^2$ measurement together leads to the following factorized result for $P^{\rm CD}_j$,
\begin{align}
& P^{\rm CD}_j(\Delta m^2,Q,\tilde z_{{\rm cut}\,i}, \beta_i,\mu)
 \\
   & =  Q_{{\rm cut}1}^{\frac{1}{1+\beta_1}}
        Q_{{\rm cut}2}^{\frac{1}{1+\beta_2}}
       \int\!\! dk^+_1dk^+_2\: 
     \delta\big(\Delta m^2\!-Qk^+_1\!-Qk^+_2 \big) 
%     \nn\\
%   &\ \ \times      
      S_{Cj}\Bigl(k^+_1\, Q_{{\rm cut}1}^{\frac{1}{1+\beta_1}},\beta_1,\mu\Bigr)
      D_{Cj}\Bigl(k^+_2\, Q_{{\rm cut}2}^{\frac{1}{1+\beta_2}},\beta_2,\mu\Bigr)
  \nn\\
 &= \int\! dq_1dq_2\: 
     \delta\biggl(\Delta m^2-q_1 Q Q_{{\rm cut}1}^{\frac{-1}{1+\beta_1}}
      -q_2 Q Q_{{\rm cut}2}^{\frac{-1}{1+\beta_2}} \biggr)      
      S_{Cj}\bigl(q_1,\beta_1,\mu\bigr)
      D_{Cj}\bigl(q_2,\beta_2,\mu\bigr)
    \,, \nn
\end{align}
which is a convolution of the collinear-soft function and the dropped collinear-soft function. 
The minus sign for the ${\cal O}(\alpha_s)$ terms in $D_{Cj}$ in \eqn{SCDCbare} can be interpreted as the subtraction of the soft drop distribution contributed from the CS$_2$ collinear-soft mode. In the convolution of collinear-soft functions the CS$_2$ mode subtracts the collinear drop phase space region from the CS$_1$ result, thus implementing the full collinear drop constraint.
Again, it is convenient to study the factorized expression in Laplace space using \eqn{laplace}.
In this case the convolution becomes a product
\begin{align}
    \tilde P_j^{\rm CD}(y,Q,\tilde z_{{\rm cut}\,i},\beta_i,\mu)
   &= \tilde S_{Ci}\Bigl(Q Q_{{\rm cut}1}^{\frac{-1}{1+\beta_1}}y,
    \beta_1,\mu\Bigr)
    \: \tilde D_{Ci} \Bigl(Q Q_{{\rm cut}2}^{\frac{-1}{1+\beta_2}} y,
    \beta_2,\mu\Bigr) 
    \\
   &= 
    \widetilde S_{Ci} \Biggl( \ln\frac{Q_{{\rm cut}1}^{\frac{1}{1+\beta_1}}}
    {yQ\mu^{\frac{2+\beta_1}{1+\beta_1}}}, \beta_1,\alpha_s(\mu) \Biggr)
    \: \widetilde D_{Ci} \Biggl( \ln\frac{Q_{{\rm cut}2}^{\frac{1}{1+\beta_2}}}
    {yQ\mu^{\frac{2+\beta_2}{1+\beta_2}}}, \beta_2,\alpha_s(\mu) \Biggr)
  \,, \nn
\end{align}
where $\tilde P_j^{\rm CD}$, $\tilde S_{Ci}$, and $\tilde D_{Ci}$ are all dimensionless, and in the last line we have defined modified functions that have a logarithms as their first argument.

From \eqn{SCDCrge} the $y$ dependence of the RGE cancels out for the product $\tilde P_j^{\rm CD}=\tilde S_{Ci} \tilde D_{Ci}$, thus properly enabling  its $\mu$ dependence to be canceled by that of $N_j^{\rm CD}$, ensuring that the cross section is $\mu$ independent. 
Thus the RGE for $N_j^{\rm CD}$ is also multiplicative
\begin{align} \label{eq:NjCDrge}
     \frac{d }{d\ln \mu} \ln  N_j^{\rm CD}
  \bigl(\Phi_J,R,\tilde z_{{\rm cut}\,i},\beta_i,\mu\bigr) 
     	&= -2\Gamma^j_{\rm cusp}(\alpha_s) \ln\frac{Q_{{\rm cut}\,1}^{\frac{1}{1+\beta_1}}Q_{{\rm cut}\,2}^{-\frac{1}{1+\beta_2}}}{\mu^{\frac{1}{1+\beta_1}-\frac{1}{1+\beta_2}} }
     + \gamma^{N_j^{\rm CD}}(\alpha_s) \,.
\end{align}
Here $\gamma^{N^{\rm CD}_{j}}(\alpha_s) = \gamma^{S_{Cj}}(\alpha_s)+\gamma^{D_{Cj}}(\alpha_s)$, and also vanishes at ${\cal O}(\alpha_s)$.  At NLL order we observe that the anomalous dimension for $N_j^{\rm CD}$ is fully consistent with the $\mu$ dependent contributions from the two global soft functions, times a $\mu$ independent factor $H_j^{\rm CD}$, 
\begin{align} \label{eq:Njfact}
  N_j^{\rm CD}
  \bigl(\Phi_J,R,\tilde z_{{\rm cut}\,i},\beta_i,\mu\bigr) 
 &= H_j^{\rm CD}\bigl(\Phi_J,R)\: 
   S_{Gj}\bigl( Q_{{\rm cut}\,1}, \beta_1,\mu\bigr) \:
   \bar S_{Gj}\bigl( Q_{{\rm cut}\,2}, \beta_2,\mu\bigr) 
  \,.
\end{align}
In particular, adding the terms in the anomalous dimensions in \eqn{SGCDrge} gives
\begin{align} \label{eq:adimGSsum}
  \frac{2}{1+\beta_1} \ln\frac{\mu}{Q_{{\rm cut}1}'} 
 -\frac{2}{1+\beta_2} \ln\frac{\mu}{Q_{{\rm cut}2}'} 
 &=  -2 \ln \frac{Q_{{\rm cut}\,1}^{\frac{1}{1+\beta_1}}Q_{{\rm cut}\,2}^{-\frac{1}{1+\beta_2}}}{\mu^{\frac{1}{1+\beta_1}-\frac{1}{1+\beta_2}} }  \,,
\end{align}
thus reproducing \eqn{NjCDrge}. Beyond NLL, the $\mu$ independence of $H_j^{\rm CD}$ in \eqn{Njfact} will imply that $\gamma^{N^{\rm CD}_{j}}(\alpha_s) = \gamma^{S_{Gj}}(\alpha_s)+\gamma^{\bar S_{Gj}}(\alpha_s)$ beyond ${\cal O}(\alpha_s)$.

Note how the $\ln R$ contributions in the individual anomalous dimensions cancel when the are summed in \eqn{adimGSsum}. For collinear drop additional contributions to $N_i$ from outside of the jet are not needed to satisfy the RG consistency, unlike the case for soft drop. This occurs because the collinear drop constraint effectively makes the jet behave like an ``unmeasured jet'' (a jet of radius $R$ that is tagged by the jet algorithm, without making further measurements). For example, taking radius $R$ dijets in an $e^+e^-$ collision with a cut $\Lambda$ on energy in the veto region outside the jets, we have
\begin{align}
  H_{j=q}^{\rm CD,e^+e^-}\bigl(\Phi_J,R) = H^{q\bar q}(Q,\mu) 
  J_q^{\rm unmeas}(QR,\mu) J_q^{\rm unmeas}(QR,\mu) 
  S_{q,\rm dijet}^{\rm unmeas}(R,\Lambda,\mu) \,,
\end{align}
where $H^{q\bar q}$ is the standard dijet quark hard function, $J_q^{\rm unmeas}$ is the unmeasured jet function, and $S_{q,\rm dijet}^{\rm unmeas}$ is an unmeasured soft function for the two quark induced dijets. The subscript $j=q$ indicates that we carry out the collinear drop jet mass measurement on one of the quark jets. 
This combination is $\mu$ independent on its own, as can be seen from the perturbative results in Ref.~\cite{Ellis:2010rwa}.  Since we are not interested in summing logarithms of $R$ here, for our purposes the required $H_j^{\rm CD}$ for $pp$ collisions can simply be calculated in fixed order perturbation theory and integrated against the initial state parton distribution functions. 

%So if collinear drop is applied to a jet from a dijet event then
%\begin{align}
%   H_{j}^{\rm CD,dijet}\bigl(\Phi_J,R) = \sum_{a,b,i} f_a(\mu) \otimes f_b(\mu) \otimes \hat H^{dijet}_{a+b\to i+j}(\Phi_J,\mu) 
%  J_i^{\rm unmeas}(QR,\mu) J_j^{\rm unmeas}(QR,\mu) 
%  S_{q,\rm dijet}^{\rm unmeas}(R,\Lambda,\mu) \,,
%\end{align}

For the collinear drop jet mass factorization theorem with resummation we write
\begin{align} \label{eq:CDfactresum}
    \frac{d\sigma}{d \Delta m^2}=\sum_{j=q,g}
 N_j^{\rm CD}(\Phi_J,R,\tilde z_{{\rm cut}\,i}, \beta_i,\mu_\gsa,\mu_\gsb,\mu) 
\:
  P^{\rm CD}_j(\Delta m^2,Q,\tilde z_{{\rm cut}\,i}, \beta_i,\mu_\csa,\mu_\csb,\mu)
  \,.
\end{align}
This notation indicates that in $N_j^{\rm CD}$ we have resummation from $\mu_\gsa$ to $\mu$ for $S_{Gj}$ and from $\mu_\gsb$ to $\mu$ for $\bar S_{Gj}$. And that for $P_j^{\rm CD}$ we have resummation from $\mu_\csa$ to $\mu$ for $S_{Dj}$ and from $\mu_\csb$ to $\mu$ for $C_{Dj}$. The choice of $\mu$ is arbitrary and cancels exactly between the two resummed functions. 
Solving the anomalous dimension equations in \eqn{SCDCrge} the resummed result for $P_j^{\rm CD}$ is
\begin{align} \label{eq:PCDresum}
  &\hspace{-0.2cm} 
  P^{\rm CD}_j(\Delta m^2,Q,\tilde z_{{\rm cut}\,i}, \beta_i,
     \mu_\csa,\mu_\csb,\mu)
   \\
    &= \exp\biggl[-\frac{2(2+\beta_1)}{1+\beta_1}C_j K(\mu_{\csa},\mu)+\frac{2(2+\beta_2)}{(1+\beta_2)}C_j K(\mu_{\csb},\mu) \biggr] 
  \Biggl[\frac{Q_{{\rm cut}1}^{\frac{1}{1+\beta_1}}}
    {Q_{{\rm cut}2}^{\frac{1}{1+\beta_2}}}
    \frac{\mu^{\frac{2+\beta_2}{1+\beta_2}}_{\csb}}{\mu^{\frac{2+\beta_1}{1+\beta_1}}_{\csa}}
    \Biggr]^{2C_j\,\omega(\mu_{\csa},\mu)}\:
   \nn \\
  & \times  \exp\big[\omega_{S_{Cj}}(\mu_{\csa},\mu) + \omega_{D_{Cj}}(\mu_{\csb},\mu)\big]  \: 
  \widetilde D_{Cj}\bigl(\partial_\eta,\beta_2,\alpha_s(\mu_{\csb}) \bigr)
\nn\\
    &\times
     \widetilde S_{Cj}\Biggl(\partial_\eta+
     \ln\frac{Q_{{\rm cut}1}^{\frac{1}{1+\beta_1}}}
     {Q_{{\rm cut}2}^{\frac{1}{1+\beta_2}}}
    \frac{\mu^{\frac{2+\beta_2}{1+\beta_2}}_{\csb}}{\mu^{\frac{2+\beta_1}{1+\beta_1}}_{\csa}} ,\beta_1,\alpha_s(\mu_{\csa}) \Biggr)
    \frac{e^{-\gamma_E\eta}}{\Gamma(\eta)}
   \frac{1}{\Delta m^2}
    \Biggl(\frac{\Delta m^2\, Q_{{\rm cut}2}^{\frac{1}{1+\beta_2}}}
    {\mu^{\frac{2+\beta_2}{1+\beta_2}}_{\csb} Q }\Biggr)^{\eta}
   \:\Bigg|_{\eta=2C_j\, \omega(\mu_{\csa},\mu_{\csb})}
   . \nn
\end{align}
There are no non-global logarithms in the collinear-soft functions $S_{C_j}$ and $D_{C_j}$, so the same holds for $P^{\rm CD}_j$. 
Note that the dependence on the jet rapidity $\eta_J$ cancels in the combinations ${Q_{{\rm cut}1}^{\frac{1}{1+\beta_1}}}{Q_{{\rm cut}2}^{\frac{-1}{1+\beta_2}}}$ and $Q_{{\rm cut}2}^{\frac{-1}{1+\beta_2}}/Q$ which appear in \eqn{PCDresum}.
Solving the anomalous dimension equations in \eqn{SGCDrge} the resummed result for $N_j^{\rm CD}$ is
\begin{align} \label{eq:NCDresum}
  & %\hspace{-1cm} 
N_j^{\rm CD}(\Phi_J,R,\tilde z_{{\rm cut}\,i}, \beta_i,\mu_\gsa,\mu_\gsb,\mu) 
  =   H_j^{\rm CD}(\Phi_J,R)  S_{Gj}(Q_{{\rm cut}1},\beta_1,\mu_\gsa) 
 \bar  S_{Gj}(Q_{{\rm cut}2},\beta_2,\mu_\gsb)
  \nn \\
  &\ \ \times \exp\biggl[\frac{2C_j}{1+\beta_1} K(\mu_{\gsa},\mu)-\frac{2C_j}{1+\beta_2} K(\mu_{\gsb},\mu) \biggr] 
  \exp\bigl[ \omega_{S_{Gj}}(\mu_\gsa,\mu) 
    + \omega_{\bar S_{Gj}}(\mu_\gsb,\mu) \bigr]
 \nn\\
 &\ \ \times 
 \biggl(\frac{\mu_\gsa}{Q_{{\rm cut}1}'}\Bigr)^{
     \frac{2C_j}{1+\beta_1}\,\omega(\mu_\gsa,\mu)} 
 \biggl(\frac{\mu_\gsb}{Q_{{\rm cut}2}'}\Bigr)^{
     \frac{-2C_j}{1+\beta_2}\,\omega(\mu_\gsb,\mu)} 
 \,. 
\end{align}
From the resummed expressions we can see that the canonical scale choices in \eqns{gsscales}{CDcsscales} remove all the logarithms that are not contained in the $K$, $\omega$, or $\omega_F$ evolution kernels.
Thus these solutions sum the desired large logarithms.  To truncate these solutions to NLL order we can set the boundary condition functions $\widetilde D_{Cj}$, $\widetilde S_{Cj}$, $S_{Gj}$ and $\bar S_{Gj}$ to $1$.

\subsubsection{Relaxing Hierarchical Constraints on $\{z_{{\rm cut}\,i},\beta_i\}$}
\label{sec:CDnonhier}

In our analysis so far we have primarily assumed that the two boundaries that define the collinear drop region are hierarchically separated.  However for realistic choices of the $\{z_{{\rm cut}\,i},\beta_i\}$ parameters this is often not the case.  Two examples are shown in \Fig{fig:mode1} in the lower two panels.  In the lower left panel we have the situation where there is a single common collinear-soft mode, and a single common global-soft mode, where both of their phase space is constrained by the two boundaries. In the lower right panel we have the situation where there is a  single common global-soft mode, but we still have two collinear-soft modes that have a hierarchical scaling for their momenta. (The opposite case is also possible, but not shown.)

For the moment we will assume that the collinear-soft modes are well separated from the global-soft modes. There are then two possible ways that the hierarchical situation can be modified. First we may have a single global-soft function when 
\begin{align} \label{eq:samezprime}
z_{ {\rm cut}\,1}' \sim z_{ {\rm cut}\,2}'  \,,
\end{align} 
which for $R_0=R$ is the same as $z_{ {\rm cut}\,1} \sim z_{ {\rm cut}\,2}$. In this case there is a single global soft mode with scaling $p_{gs}\sim p_{gs1}\sim p_{gs2}$ with $p_{gsi}$ from \eqn{pgsCD}. The ${\cal O}(\alpha_s)$ calculation of the global-soft function for this case follows that in \eqn{SGhier}, but with a single phase space constraint given by $\overline\Theta^{(gs)}_{\rm SD_1}-\overline\Theta^{(gs)}_{\rm SD_2}$ for a single emission.  This breaks into two independent pieces, so the result follows immediately from the hierarchical case
\begin{align}
S_{Gj}^{12} \Bigl(Q_{{\rm cut}1},
Q_{{\rm cut}2},\beta_1,\beta_2,\mu\Bigr) 
= S_{Gj}\Bigl(Q_{{\rm cut}1},\beta_1,\mu\Bigr) 
\bar S_{Gj}\Bigl(Q_{{\rm cut}2},\beta_2,\mu\Bigr) 
+{\cal O}(\alpha_s^2)
\,.
\end{align}
The corresponding canonical scale choice is $\mu_{gs} \sim \mu_{\gsa}\sim \mu_{\gsb}$, and its anomalous dimension is
\begin{align} \label{eq:anomdimSG12}
& \mu \frac{d}{d\mu} \ln S_{Gj}^{12}\Bigl(Q_{{\rm cut}1},
Q_{{\rm cut}2},\beta_1,\beta_2,R,\mu\Bigr)
\\
&\quad = \frac{2C_j}{1+\beta_1} \Gamma_{\rm cusp}(\alpha_s) \ln\frac{\mu}{Q_{{\rm cut}1}'} - \frac{2C_j}{1+\beta_2} \Gamma_{\rm cusp}(\alpha_s) \ln\frac{\mu}{Q_{{\rm cut}2}'}  + \gamma_{S^{12}_{Gj}}(\alpha_s) 
\nn  \\
&\quad = - 2C_j \Gamma_{\rm cusp}(\alpha_s) 
\ln\frac{Q_{{\rm cut}\,1}^{\frac{1}{1+\beta_1}} 
	\mu^{\frac{1}{1+\beta_2}} }
{\mu^{\frac{1}{1+\beta_1}} Q_{{\rm cut}\,2}^{\frac{1}{1+\beta_2}} }    + \gamma_{S^{12}_{Gj}}(\alpha_s) 
\,. \nn
\end{align}

Second we could have a single collinear-soft function because
\begin{align} \label{eq:samezeta}
\zeta_{\csa} \sim \zeta_{\csb} \,.
\end{align} 
For this situation we have a single collinear-soft mode with momentum scaling as $p_{cs}\sim p_{cs1}\sim p_{cs2}$, with $p_{csi}$ from \eqn{pcsCD}.  The ${\cal O}(\alpha_s)$ calculation of the corresponding collinear-soft function follows that in \eqn{SChier}, but with a single phase space constraint given by $\Theta_{\rm SD_1}-\Theta_{\rm SD_2}$ for a single emission.  Since this breaks into two independent pieces the result again follows immediately from the hierarchical case
\begin{align}
& S_{Cj}^{12}\Bigl(k^+ Q_{{\rm cut}1}^{\frac{1}{1+\beta_1}},
k^+ Q_{{\rm cut}2}^{\frac{1}{1+\beta_2}},\beta_1,\beta_2,\mu\Bigr)
\\
& =  \int\!\! dk^+_1dk^+_2\: 
\delta\big(k^+ - k^+_1 - k^+_2 \big)       
S_{Cj}\Bigl(k^+_1\, Q_{{\rm cut}1}^{\frac{1}{1+\beta_1}},\beta_1,\mu\Bigr)
D_{Cj}\Bigl(k^+_2\, Q_{{\rm cut}2}^{\frac{1}{1+\beta_2}},\beta_2,\mu\Bigr)
\,.\nn
\end{align}
In this case the corresponding canonical scale choice is $\mu_{cs}\sim \mu_{\csa}\sim \mu_{\csb}$ and the Laplace space anomalous dimension is
\begin{align} \label{eq:anomdimSC12}
& \mu \frac{d}{d\mu} \ln \tilde S_{C_j}^{12}\Bigl(Q Q_{{\rm cut}1}^{\frac{1}{1+\beta_1}} y,
Q Q_{{\rm cut}2}^{\frac{1}{1+\beta_2}} y,\beta_1,\beta_2,\mu\Bigr)
\\ 
&\quad
= 2C_j \Gamma_{\rm cusp}(\alpha_s) \ln\frac{Q_{{\rm cut}\,1}^{\frac{1}{1+\beta_1}}}{\mu^{\frac{2+\beta_1}{1+\beta_1}} Q y }
-2C_j\Gamma_{\rm cusp}(\alpha_s) \ln\frac{Q_{{\rm cut}\,2}^{\frac{1}{1+\beta_2}}}{\mu^{\frac{2+\beta_2}{1+\beta_2}} Q y }
+\gamma_{S_{Cj}^{12}}(\alpha_s)
\nn\\
&\quad
= 2C_j \Gamma_{\rm cusp}(\alpha_s) 
\ln\frac{Q_{{\rm cut}\,1}^{\frac{1}{1+\beta_1}} 
	\mu^{\frac{1}{1+\beta_2}} }
{\mu^{\frac{1}{1+\beta_1}} Q_{{\rm cut}\,2}^{\frac{1}{1+\beta_2}} }   +\gamma_{S_{Cj}^{12}}(\alpha_s)
\,.\nn
\end{align}

The three possible cases with a relaxed hierarchy correspond to $S_{Gj}^{12} \tilde S_{Cj}^{12}$, $S_{Gj}^{12} \tilde S_{Cj} \tilde D_{Cj}$, or $S_{Gj} \bar S_{Gj} \tilde S_{Cj}^{12}$, and in all cases the RGE consistency relations are satisfied. For example, for $S_{Gj}^{12} \tilde S_{Cj}^{12}$ this follows because the anomalous dimensions in \eqns{anomdimSG12}{anomdimSC12} are equal and opposite, with $\gamma_{S_{Cj}^{12}}(\alpha_s) = - \gamma_{S_{Gj}^{12}}(\alpha_s)$.

The simple structure of the phase space constraints at one-loop order has direct implications for obtaining the resummed result for the non-hierarchical cases, where we have one or both of \eqns{samezprime}{samezeta}. The NLL result in all non-hierarchical cases are simply obtained by evaluating \eqns{PCDresum}{NCDresum} at this order with the same scale choices as used in the fully hierarchical case from \Sec{sec:CDhier}. This suffices since the transition to the non-hierarchical cases is fully continuous at this order. However, we do caution that in these non-hierarchical cases that non-global logarithms can appear in $S_{Gj}^{12}$ or $S_{Cj}^{12}$ at ${\cal O}(\alpha_s^2)$. 

It is also interesting to consider the transition between the collinear drop resummed expression, and that for soft drop, by turning off the colliner drop constraint, which could be achieved by taking $z_{{\rm cut}\,2}=1$ and $\beta_2=0$. It is straightforward to see that this reproduces the LL resummed expression for the soft drop jet mass spectrum with a correspondence between anomalous dimensions that has $D_{Cj}\to J_j$ and $\bar S_{G_j}\to H_j$. However beyond LL this correspondence becomes more complicated since the non-cusp anomalous dimensions of $J_j$ and $H_j$ are not obtained by a simple limit from $D_{Cj}$ and $\bar S_{G_j}$. 

\subsubsection{Collinear Drop $\Delta m^2$ with $\beta_1=\beta_2$}

The special case where we take $\beta_1=\beta_2$ is interesting because the result does not contain a leading double logarithmic series. This is analogous to the behavior of soft drop in the $\beta=0$ limit where it reduces to the minimal-mass-drop (MMD) grooming, and there is no double logarithmic series. For MMD the grooming removes the soft $m_J$ dependent logarithm from the series, replacing it by a logarithm of $z_{\rm cut}$.  In the collinear drop case the radiation is always soft, and the leading double logarithmic series is absent for any value of $\beta_1=\beta_2=\beta$. This gives an entire family of observables without a double logarithmic series.

To demonstrate the cancellation of the double logarithms, take $\beta_1=\beta_2=\beta$ in \eqn{PCDresum}, which gives
\begin{align} \label{eq:PCDresumEqualbeta}
  &\hspace{-0.2cm} 
  P^{\rm CD}_j(\Delta m^2,Q,\tilde z_{{\rm cut}\,i}, \beta_i=\beta,
     \mu_\csa,\mu_\csb,\mu)
   \\
    &= \exp\biggl[-\frac{2(2+\beta)}{1+\beta}C_j K(\mu_{\csa},\mu_\csb) \biggr] 
  \Biggl[\frac{z_{{\rm cut}1}}{z_{{\rm cut}2}}
    \frac{\mu^{2+\beta}_{\csb}}{\mu^{2+\beta}_{\csa}}
    \Biggr]^{ \frac{2C_j}{1+\beta}\,\omega(\mu_{\csa},\mu)}\:
   \nn \\
  & \times  \exp\big[\omega_{S_{Ci}}(\mu_{\csa},\mu) + \omega_{D_{Ci}}(\mu_{\csb},\mu)\big]  \: 
  \widetilde D_{Ci}\bigl(\partial_\eta,\beta,\alpha_s(\mu_{\csb}) \bigr)
\nn\\
    &\times
     \widetilde S_{Ci}\Biggl(\partial_\eta+ \frac{1}{1+\beta}
     \ln\frac{z_{{\rm cut}1}}{z_{{\rm cut}2}}
    \frac{\mu^{2+\beta}_{\csb}}{\mu^{2+\beta}_{\csa}} ,\beta,\alpha_s(\mu_{\csa}) \Biggr)
    \frac{e^{-\gamma_E\eta}}{\Gamma(\eta)}
   \frac{1}{\Delta m^2}
    \Biggl(\frac{\Delta m^2\, Q_{{\rm cut}2}^{\frac{1}{1+\beta}}}
    {\mu^{\frac{2+\beta}{1+\beta}}_{\csb} Q }\Biggr)^{\eta}
   \:\Bigg|_{\eta=2C_j\, \omega(\mu_{\csa},\mu_{\csb})}
   . \nn
\end{align}
Furthermore for $\beta_i=\beta$, the canonical values in \eqn{CDcsscales}  give a $\Delta m^2$ independent ratio of scales
\begin{align}
  \frac{\mu_\csb}{\mu_\csa} &= \Big( \frac{z_{{\rm cut}\,2}}{z_{{\rm cut}\,2}} \Big)^{\frac{1}{1+\beta}} \,.
\end{align}
Since at LL only this ratio appears inside $K(\mu_\csa,\mu_\csb)$ and $\omega(\mu_\csa,\mu_\csb)$ in \eqn{PCDresumEqualbeta}, and other $\omega_F$ appear only beyond LL, we see that the LL terms involving double logarithms of $\Delta m^2$ are not present.

\subsection{Transitions with Increasing $\Delta m^2$ for Collinear Drop}
\label{sec:b1isb2}

The above factorization and resummation expressions work for 
$\Delta m^2 \ll (p_T R)^2 z_{{\rm cut}\,1}^\prime $.
In contrast, in the region where 
\begin{align} \label{eq:Deltamtransition}
 \Delta m^2 \ge  (p_T R)^2  z_{{\rm cut}\,1}^\prime 
    \equiv \, \Delta m^2_{{\rm cut}\,1}
  \,,
\end{align}
soft drop ${{\rm SD}_1}$ is now ineffective and we need to match to the effective theory where ${{\rm SD}_1}$ is turned off. When $z_{{\rm cut}\,1}$ is small, such transition can happen at small values of $\Delta m^2$ because the ${{\rm SD}_1}$ constraint can be easily failed by a majority of the jet configurations. This transition is the same as that we discussed for soft drop in \Sec{sec:transition}. Thus we have
\begin{align}  \label{eq:CDtransition}
\mu_{\csa}(\Delta m^2_{{\rm cut}\,1})=\mu_\gsa 
\, .
\end{align}
However unlike the case there, for $\Delta m^2$ the 
${{\rm SD}_2}$ collinear drop constraint is still always at work in the $\Delta m^2 \ge \Delta m^2_{{\rm cut}\,1}$ region. 
In this region the ${{\rm SD}_1}$ collinear-soft mode and GS$_1$ global-soft mode are replaced by a single c-soft mode with 
\begin{align} \label{eq:CDtransition2}
  p_s^\mu \sim \frac{\Delta m^2}{QR^{\prime\,2}} \: \big(R^{\prime\,2},1,R'\big)
\end{align}
where $R'\equiv R/(2\cosh\eta_J)$. Here the characteristic scales are equal to a single soft scale $\mu_s(\Delta m^2)$ as
\begin{align}  \label{eq:CDtransition3}
  \mu_{\csa}(\Delta m^2)=\mu_\gsa(\Delta m^2)= \mu_s(\Delta m^2) = \frac{\Delta m^2}{p_T R}\;,
  \qquad \text{for} \quad
  \Delta m^2 \ge \Delta m^2_{{\rm cut}\,1}
  \ .
\end{align}
In this region $\Delta m^2$ measures the difference between the ${\rm SD}_2$ groomed and ungroomed jet masses. At NLL order our factorization theorem which combines \eqns{PCDresum}{NCDresum} still properly describes the logarithms in this region, simply by implementing the choice of scales in \eqn{CDtransition}. Beyond NLL there will be modifications from the fixed order corrections of the c-soft function for the modes in \eqn{CDtransition}, which will in general differ from the product of fixed order corrections from the $S_{Ci}$ and $S_{Gi}$ functions. Note that once the soft drop grooming is not longer active, that there will be non-global logarithms in the spectrum (through the soft function), like in the ungroomed case.
%\TODO{No longer protected from NGLs, so discuss.} (Beyond NLL there will be modifications from the fixed order corrections of the c-soft function for the modes in \eqn{CDtransition}, which will in general differ from the product of fixed order corrections from the $S_{Ci}$ and $S_{Gi}$ functions.)

Finally we note that there is an upper bound on the $\Delta m^2$ spectrum
\begin{align} \label{eq:Deltamupperlimit}
  \Delta m^2 < (p_T R)^2 \, z_{{\rm cut}\,2}^\prime \,
        \equiv \Delta m^2_{{\rm cut}\,2} \ ,
\end{align}
beyond which the cross section is zero. 
This bound occurs because as $\Delta m^2$ increases the phase space that passes the collinear drop constraint decreases. The available phase space for radiation vanishes  when we reach the bound in \eqn{Deltamupperlimit}. At NLL order the vanishing of our cross section at $\Delta m^2=\Delta m^2_{{\rm cut}\,2}$ occurs because all the scales become equal at this point,
\begin{align} \label{eq:muCDend}
 \mu_s( \Delta m^2_{{\rm cut}\,2})=\mu_{\csb}( \Delta m^2_{{\rm cut}\,2})=\mu_{\gsb}
  \,.
\end{align}

\subsection{Profile Function for $\Delta m^2$}
\label{sec:CDprofile}

\begin{figure}[t!]
\centering
    \includegraphics[height=7cm]{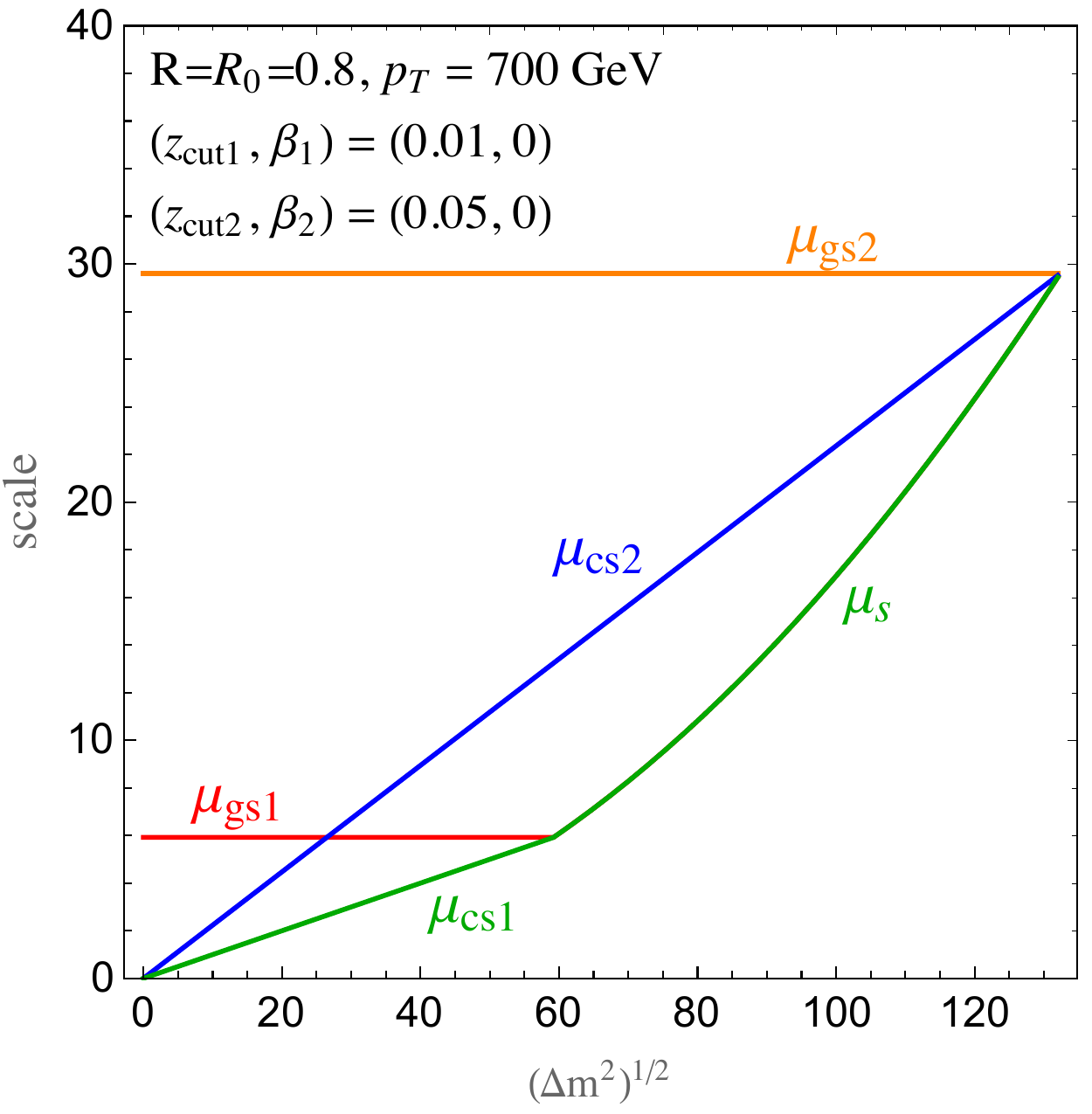}~~~
    \includegraphics[height=7cm]{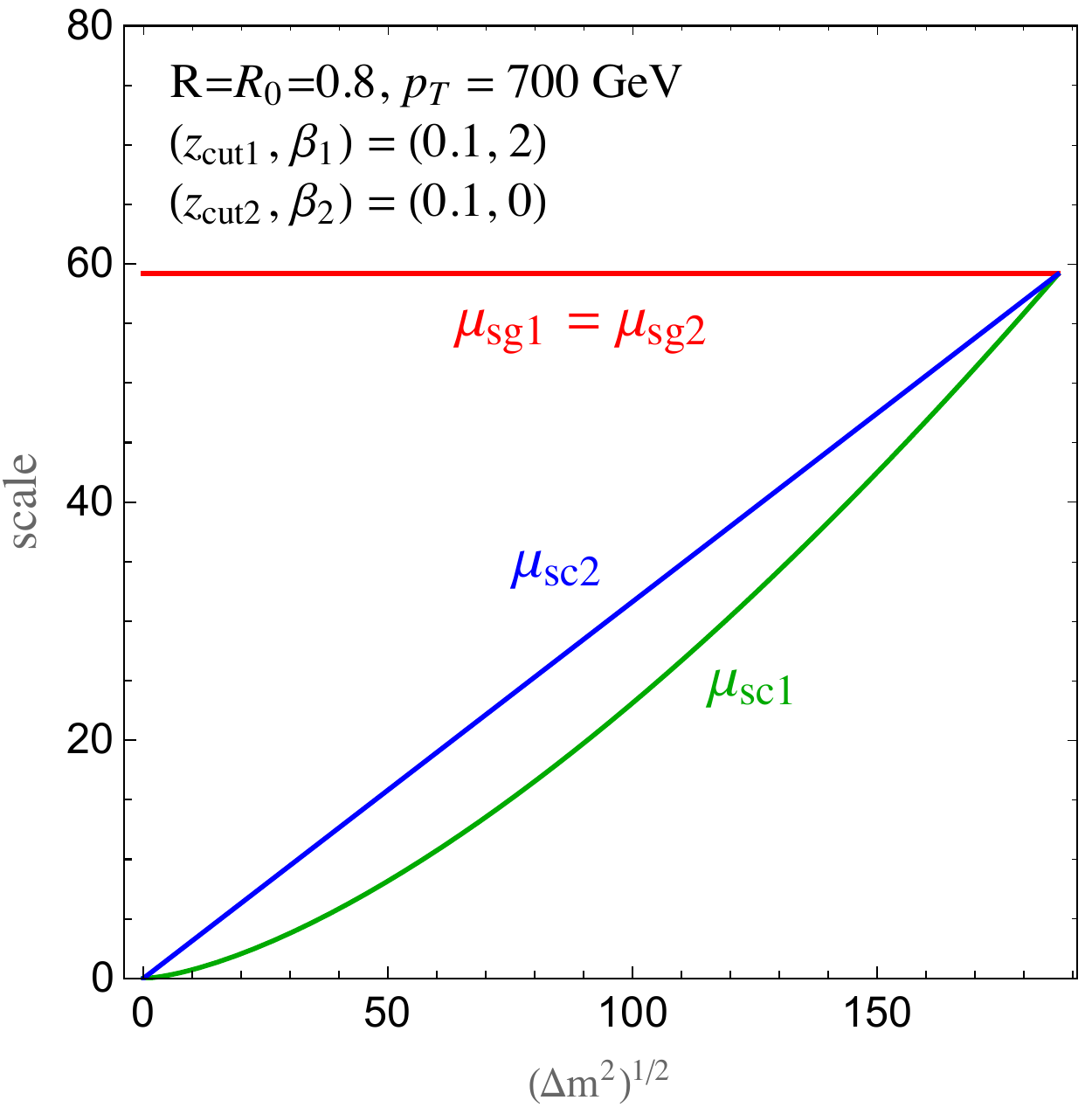}
    \caption{Plots of the canonical scales for collinear drop, including the merging of $\mu_\gsa$ and $\mu_\csa$ at the transition where the ${\rm SD}_1$ grooming becomes ineffective at $(\Delta m^2_{{\rm cut}\,1})^{1/2}$ from \eqn{Deltamtransition}. The upper endpoint of the spectrum occurs at $(\Delta m^2_{{\rm cut}\,2})^{1/2}$, where all the curves meet  at the far right, and depends on the collinear drop parameter $z_{{\rm cut}\,2}$ via \eqn{Deltamupperlimit}.}
\label{fig:scale2}
\end{figure}

We summarize again the canonical scale choices of $\mu_\csa$, $\mu_\csb$, $\mu_\gsa$ and $\mu_\gsb$ in the resummed SCET prediction for $\Delta m^2$,
\begin{align}
\mu_\csa(\Delta m^2)
& =\left\{
\begin{array}{lc}
   \Big( \frac{\Delta m^2}{Q} \Big)^{\frac{1+\beta_1}{2+\beta_1}} Q_{{\rm cut}\; 1}^{\frac{1}{2+\beta_1}}
   & ~~~~~\Delta m^2 < \Delta m^2_{{\rm cut}\,1}\\[5pt]
    \frac{\Delta m^2}{p_TR} & ~~~~~\Delta m^2 \ge \Delta m^2_{{\rm cut}\,1}
\end{array}
\right.\;, 
 \nn\\[5pt]
\mu_\gsa(\Delta m^2)
&=\left\{
\begin{array}{lc}
	p_T R\, z_{{\rm cut}\; 1}' & ~~~~~\Delta m^2 < \Delta m^2_{{\rm cut}\,1}\\[2pt]
    \frac{\Delta m^2}{p_TR} & ~~~~~\Delta m^2 \ge \Delta m^2_{{\rm cut}\,1}
\end{array}
\right.\;,
\nn\\
\mu_\csb(\Delta m^2)
& = \Big( \frac{\Delta m^2}{Q} \Big)^{\frac{1+\beta_2}{2+\beta_2}} Q_{{\rm cut}\; 2}^{\frac{1}{2+\beta_2}}\;,\nn\\
\mu_\gsb(\Delta m^2)
& =  p_T R\, z_{{\rm cut}\; 2}' \;.
\end{align}
The grooming transition  happens at $\Delta m^2 = \Delta m^2_{{\rm cut}\,1}$ in \eqn{Deltamtransition}, at which point we merge $\mu_\csa$ and $\mu_\gsa$ continuously so that they become equal to the single ultrasoft scale $\mu_s= \Delta m^2 / (p_T R)$. Also, as $\Delta m^2$ increases the scales $\mu_\csb$, $\mu_\gsb$ and $\mu_s$ merge at $\Delta m^2 = \Delta m^2_{{\rm cut}\,2}$ in \eqn{Deltamupperlimit}, which is the endpoint of the spectrum. 
In the special case of $z_{{\rm cut}\,1}=z_{{\rm cut}\,2}$, the region $\Delta m^2_{{\rm cut}\,1} < \Delta m^2 < \Delta m^2_{{\rm cut}\,2}$ disappears, so the grooming transition does not happen. 
In all perturbative regions the canonical scales for this collinear-drop jet mass observable obey the relation
\begin{align}
  \mu_\csa^{\frac{2+\beta_1}{1+\beta_1}}  \mu_\gsa^{\frac{-1}{1+\beta_1}}
% = \frac{\Delta m^2}{p_T R}
 = \mu_\csb^{\frac{2+\beta_2}{1+\beta_2}}  \mu_\gsb^{\frac{-1}{1+\beta_2}} 
  \,.
\end{align}

The collinear-soft scales $\mu_\csa$ and $\mu_\csb$ are monotonic functions of $\Delta m^2$. In the $\Delta m^2\rightarrow 0$ limit these two scales can get close to the Landau pole singularity where the strong coupling constant diverges, and the perturbative expressions for the anomalous dimensions break down. In this region there are ${\cal O}(1)$ nonperturbative corrections to the $\Delta m^2$ spectrum. Since $\mu_\csa < \mu_\csb$ it will always be $\mu_\csa$ that gets near to the non-perturbative region first.  These non-perturbative transitions occur for $\mu_\csa$ and $\mu_\csb$ at the values
\begin{align}
  & \Delta m^2 \sim (p_T R) \Lambda_{\rm QCD} \bigg( \frac{\Lambda_{\rm QCD}}{Q_{{\rm cut}\,1}'} \bigg)^{\frac{1}{1+\beta_1}}
  \,,
 &&  \Delta m^2 \sim (p_T R) \Lambda_{\rm QCD} \bigg( \frac{\Lambda_{\rm QCD}}{Q_{{\rm cut}\,2}'} \bigg)^{\frac{1}{1+\beta_2}}
  \,,
\end{align}
respectively, which correspond with $\mu_\csa \sim \Lambda_{\rm QCD}$ and $\mu_\csb \sim \Lambda_{\rm QCD}$. 
These relations have the same form as for the non-perturbative region for the soft-drop jet mass~\cite{Frye:2016aiz}. 
Therefore the running has to be terminated at a low scale $\mu_\csi \sim 1\,{\rm GeV}$, and we do so by modifying the two collinear-soft scales as $\mu_\csi \to f(\mu_\csi)$, using the following profile function,
\begin{align} \label{eq:fmu0}
f(\mu)
=\left\{
\begin{array}{lc}
   \mu& ~~~~~\mu>2\mu_0\\[5pt]
   \mu_0\Big(1+\frac{\mu^2}{4\mu_0^2}\Big) & ~~~~~\mu<2\mu_0
\end{array}
\right.\;.
\end{align}
We take as a default $\mu_0=1$ GeV, which ensures that the collinear-soft scales never go below $\mu_0=1$ GeV. 
Furthermore when $\Delta m^2\to 0$ we have $\mu_\csa=\mu_\csb=\mu_0$, which from \eqn{PCDresum} with $\eta\to 0$ can be seen to force the differential cross section to vanish.
Since the collinear drop spectrum is dominated by smaller values of $\Delta m^2$ than we have for soft drop jet mass or ungroomed jet mass, more of its spectrum is sensitive to non-perturbative effects. 
The choice of $\mu_0$ can modify the partonic cross section in the region where nonperturbative corrections are important, as we discuss in the next section, and hence gives a method for testing the extent of this region.

We will estimate the theoretical uncertainty by varying the scales $\mu_{\gsb}(\Delta m^2)$, $\mu_{\csb}(\Delta m^2)$, $\mu_{\gsa}(\Delta m^2)$ and $\mu_{\csa}(\Delta m^2)$ in the resummation formula, again using profile functions~\cite{Ligeti:2008ac,Abbate:2010xh}.  These scale variations are devised so that they always maintain the conditions in \Eqs{eq:CDtransition}{eq:muCDend}, and the hierarchies between scales so that $\mu_{\gsb}\ge \mu_{\gsa}$ and $\mu_{\csb} \ge \mu_{\csa}$.  In addition, for cases where $z_{{\rm cut}\; 1}=z_{{\rm cut}\; 2}$ so that $\mu_\gsa = \mu_\gsb$, then we retain this equality during the scale variations. For simplicity we quote the variations here taking $R_0=R$. 
For situations with $z_{{\rm cut}\; 1}< z_{{\rm cut}\; 2}$ we consider the following four variations:
\begin{enumerate}
	\item Overall variation of all scales simultaneously up/down by a factor $e_0$, so $\mu_i\to e_0  \mu_i$ with $e_0=1/\sqrt{2}$ or $\sqrt{2}$.

	\item Variation of the $\mu_{\gsa}$ and $\mu_{\csa}$ scales by a multiplicative factor of $e_{sa}=10/9$ or $9/10$ in the region $\Delta m^2\le \Delta m_{{\rm cut}\,1}^2$, while simultaneously multiplying $\mu_s(\Delta m^2)$ for the region $\Delta m^2\ge \Delta m_{{\rm cut}\,1}^2$ by the factor $\big[\Delta m^2/(p_T^2R^2z_{{\rm cut}\,2})\big]^{\ln e_{sa}/\ln(z_{{\rm cut}\,1}/z_{{\rm cut}\,2})}$ to maintain \eqns{CDtransition}{muCDend}.

	\item Variation of the $\mu_{\gsb}$ and $\mu_{\csb}$ scales by a multiplicative factor of $e_{sb}=10/9$ or $9/10$, while simultaneously multiplying $\mu_s(\Delta m^2)$ for the region $\Delta m^2\ge \Delta m_{{\rm cut}\,1}^2$ by the factor $\big[\Delta m^2/(p_T^2R^2z_{{\rm cut}\,1})\big]^{-\ln e_{sb}/\ln(z_{{\rm cut}\,1}/z_{{\rm cut}\,2})}$ to maintain \eqns{CDtransition}{muCDend}.
	
   \item Variation of $\mu_{\csa}$ and $\mu_{\csb}$ simultaneously  by  trumpet factors $\mu_{\csa}\to \mu_{\csa} \big[1+ e_{cs} \big(1- \frac{\Delta m}{\Delta m_{{\rm cut}\,1}} \big)^2 \Theta(\Delta m_{{\rm cut}\,1}-\Delta m)\big]$ and $\mu_{\csb}\to \mu_{\csb} \big[1+ e_{cs} \big(1- \frac{\Delta m}{\Delta m_{{\rm cut}\,2}} \big)^2 \big]$ with $e_{cs}=\pm 1/4$. 
\end{enumerate}
For cases where $z_{{\rm cut}\; 1}= z_{{\rm cut}\; 2}$ we replace the second and third variations by
\begin{enumerate}
	\item[2$^\prime$.] Variation of the $\mu_{\gsa}=\mu_\gsb$ and $\mu_{\csa}$ scales by a common multiplicative factor of $e_{sa}=10/9$ or $9/10$, while simultaneously multiplying $\mu_\csb$ by the factor $\big[\Delta m^2/(p_T^2R^2)\big]^{\ln e_{sa}/\ln(z_{{\rm cut}\,1})}$ to maintain \eqns{CDtransition}{muCDend}.

	\item[3$^\prime$.] Variation of the $\mu_{\gsa}=\mu_\gsb$ and $\mu_{\csb}$ scales by a common multiplicative factor of $e_{sb}=10/9$ or $9/10$, while simultaneously multiplying $\mu_\csa$ by the factor $\big[\Delta m^2/(p_T^2R^2)\big]^{\ln e_{sb}/\ln(z_{{\rm cut}\,1})}$ to maintain \eqns{CDtransition}{muCDend}.
	
\end{enumerate}
We then compute the total uncertainty for collinear drop cross sections at NLL as simply the outer envelope of these four variations. Note that the size of the variation parameters $e_i$ for collinear drop are smaller than in soft drop because the scales tend to be smaller and closer together, and hence smaller variations are required to maintain $\mu_{\csb} \ge \mu_{\csa}$.

\subsection{Partonic SCET Results for $\Delta m^2$}
\label{sec:SCETresults}

We now study the partonic SCET predictions for the $\Delta m^2$ jet mass to gain intuition about these distributions and their dependence on the collinear drop parameters. 

\begin{figure}[t!]
	\hspace{-1.4cm}
	\includegraphics[width=0.55\textwidth]{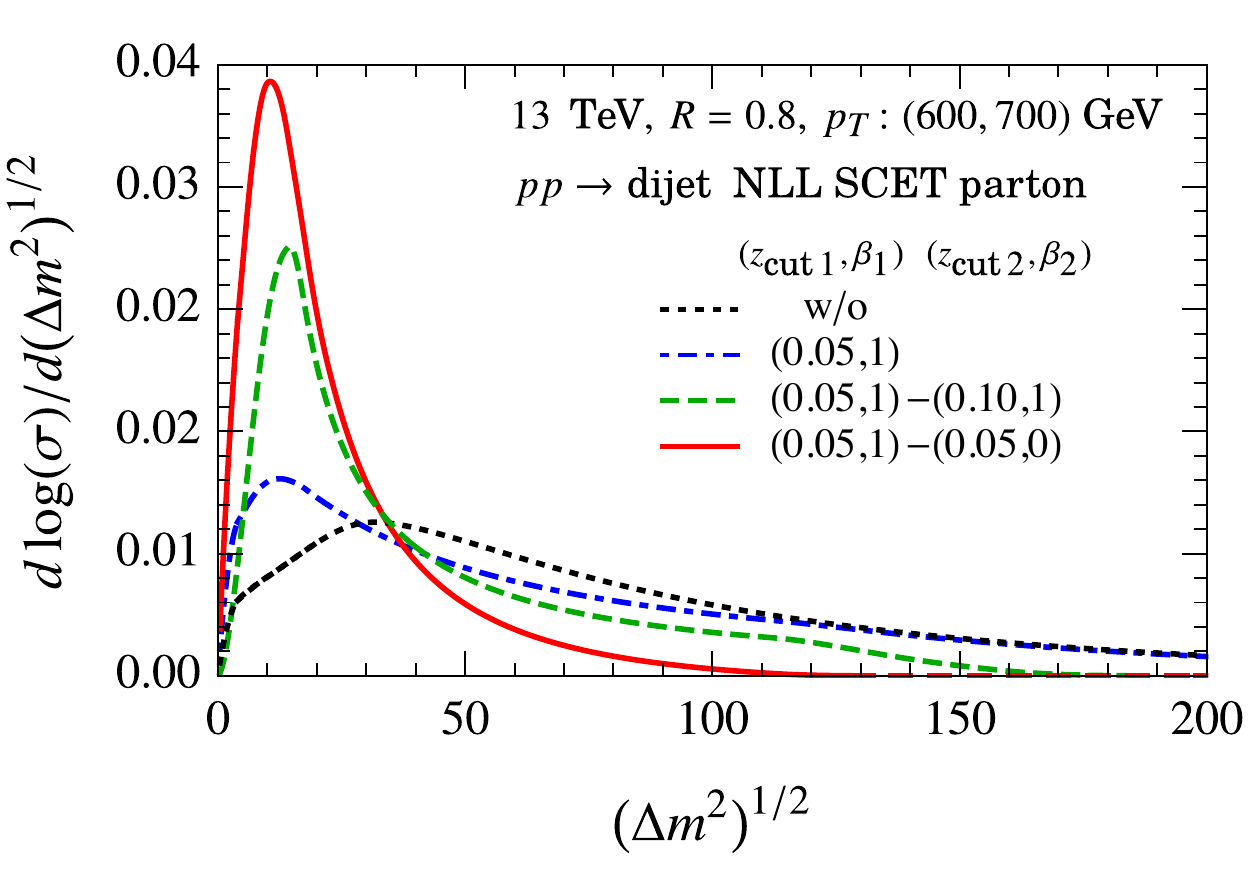}
	\includegraphics[width=0.55\textwidth]{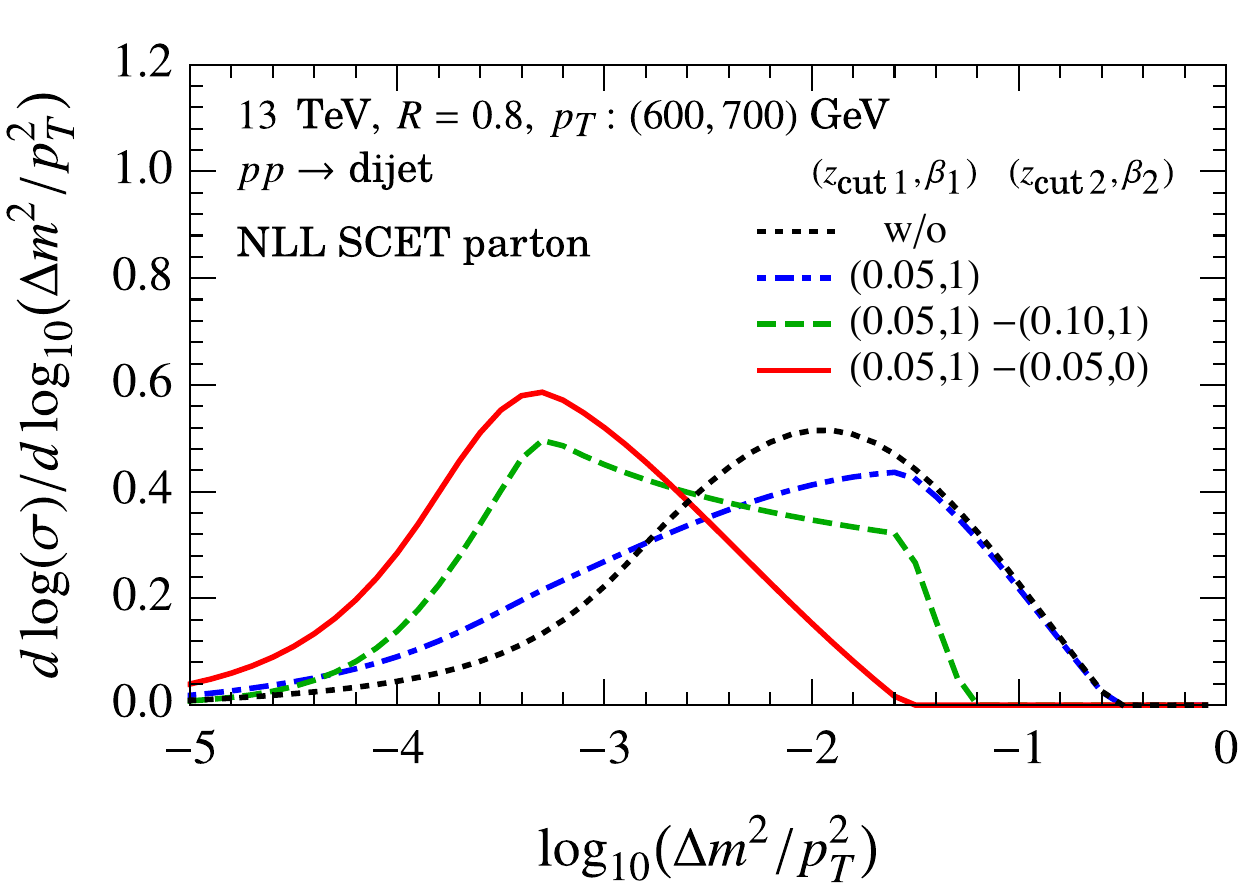}
	\caption{Comparison of collinear drop, soft drop, and ungroomed jet mass spectra calculated using SCET. The soft drop parameters are $z_{{\rm cut}\,1}=0.05$ and $\beta_1=1$ in all cases, while the collinear drop parameters are varied as indicated. The left and right panels show the same spectra but plotted with different axes choices in order to contrast the linear versus logarithmic distributions.
	}
	\label{fig:SCET_soft_collinear_drop}
\end{figure}

Figure \ref{fig:SCET_soft_collinear_drop} shows a comparison between ungroomed (black dotted), soft drop groomed (blue dot-dashed), and collinear drop (green dashed and red solid) jet mass distributions predicted by the SCET formulae. The soft drop and collinear drop curves are at NLL accuracy, while the ungroomed curve is shown for illustration and only includes so-called NLL global logarithms, while neglecting non-global terms. Note that although the horizontal axis has been labeled with $\Delta m^2$, the ungroomed and soft-drop cases have $\Delta m^2 = m_J^2$. The soft-drop parameters are chosen to be $(z_{{\rm cut}\;1},\beta_1)=(0.05,1)$ and the two collinear-drop observables are constructed by varying the value of $z_{\rm cut}$ with a fixed $\beta$: $(z_{{\rm cut}\;2},\beta_2)=(0.10,1)$ (green), or varying $\beta$ with a fixed $z_{\rm cut}$: $(z_{{\rm cut}\;2},\beta_2)=(0.05,0)$ (red). The left panel shows the distributions linearly with $\sqrt{\Delta m^2}$ while the right panel shows the same distributions, but plotted with the variable $\log_{10}(\Delta m^2/p_T^2)$. From the left panel of \Fig{fig:SCET_soft_collinear_drop} we observe that collinear drop distribution significantly softens the jet mass distribution, and makes it narrower, as expected for the removal of energetic collinear radiation. The same softening of the spectrum is even more clearly visible in the right panel, where the peaks of the two collinear drop distributions are significantly to the left of both the ungroomed and soft drop distributions.

A noticeable feature of the collinear drop distributions, seen most clearly in the right panel of \Fig{fig:SCET_soft_collinear_drop}, is that their upper boundary occurs earlier than that of the non-collinear drop spectra. We recall that at the order we are working it occurs at $\Delta m^2 = p_T^2 R^2 z_{{\rm cut}\;2}$, which corresponds to $\sqrt{\Delta m^2}\approx 125$ GeV and $\log_{10}(\Delta m^2/p_T^2)\approx -1.5$ for $z_{{\rm cut}\;2}=0.05$, and $\sqrt{\Delta m^2}\approx 177$ GeV and $\log_{10}(\Delta m^2/p_T^2)\approx -1.2$ for $z_{{\rm cut}\;2}=0.10$, taking $p_T\approx 650$ GeV and $R=0.8$. The green collinear drop curve with $z_{{\rm cut}\,2}=0.1$ also exhibits the  same transition as the blue soft drop curve for the groomed to ungroomed transition point, which is at $\Delta m^2 = p_T^2 R^2 z_{{\rm cut}\;1}$ corresponding to $\sqrt{\Delta m^2}\approx 125$ GeV and $\log_{10}(\Delta m^2/p_T^2)\approx -1.5$. In contrast the red collinear drop curve with  $z_{{\rm cut}\,1}=z_{{\rm cut}\,2}$ has no such transition. Note that the red and green collinear drop distributions have quite distinct shapes.

\begin{figure}[t!]
    \hspace{-1.4cm}
	\includegraphics[width=0.55\textwidth]{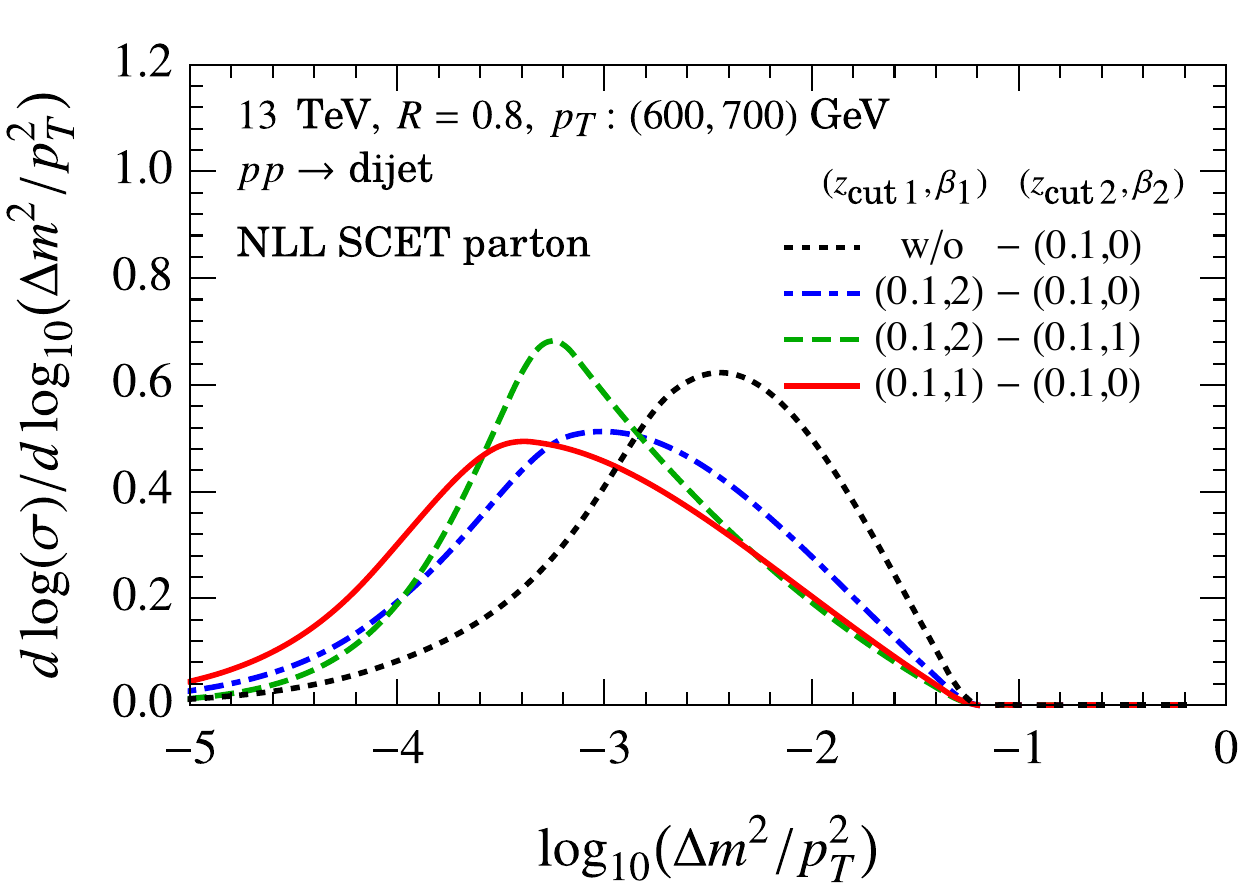}
	\includegraphics[width=0.55\textwidth]{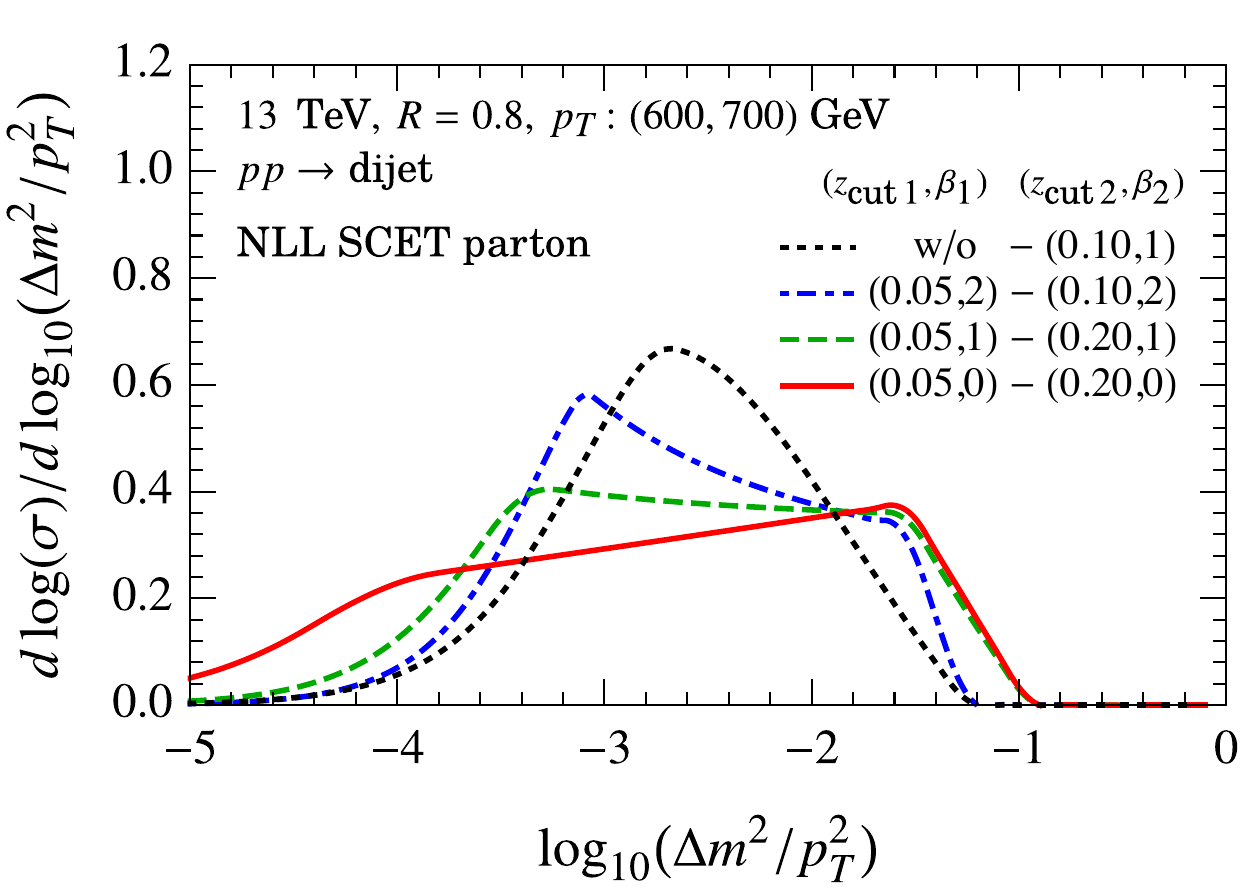}
	\caption{Comparison of collinear drop distributions calculated using SCET. The left panel shows examples from varying the angular parameters $\beta_i$ while holding $z_{{\rm cut}\,1}=z_{{\rm cut}\,2}$ fixed, whereas the right panel gives examples with varied $z_{{\rm cut}\,i}$ with fixed $\beta_1=\beta_2$.  
	}
	\label{fig:SCET_collinear_drop}
\end{figure}

In \Fig{fig:SCET_collinear_drop} we contrast 8 different collinear drop observables, which probe different parts of the soft phase space. The left panel has a fixed value of $z_{{\rm cut}\,1}=z_{{\rm cut}\,2}=0.1$ and varies $\beta_1$ and $\beta_2$ within the values $0,1,2$ for the blue dot-dashed, green dashed, and red solid curves as indicated. These sets of parameters are the ones used in the ATLAS soft drop jet mass measurements. (The CMS~\cite{Sirunyan:2018xdh} soft drop jet mass uses $\beta=0$.) This makes it straightforward to carry out new measurements for these observables based on the same ATLAS data set. Since $z_{{\rm cut}\;1}=z_{{\rm cut}\;2}$ these results are groomed throughout the full spectrum. The choice of $\beta_i$ values mostly effects the shape and location of the peak. 
We also show with the black curve a comparison of a collinear-drop observable that does not include the soft drop grooming, and hence retains the soft wide-angle radiation.  It peaks further to the right, though still to the left of the curves without collinear drop from \Fig{fig:SCET_soft_collinear_drop}.   

In the right panel of \Fig{fig:SCET_collinear_drop} we show a different type of collinear drop observables, holding $\beta_1=\beta_2$ fixed, taking
$z_{{\rm cut}\,1}=0.05$ and varying $z_{{\rm cut}\,2}= 0.1, 0.2$. Recall that although this is simply a special case of the generic NLL formula, that for $\beta_1=\beta_2$ only the LL $\Delta m^2$ dependent logarithms are summed at the order we are working.   For fixed $\beta_1=\beta_2$, varying $z_{{\rm cut}\,2}$ does not lead to large differences, so we choose to use $\beta_1=\beta_2=0,1,2$ for the red solid, green dashed, and blue dot-dashed curves respectively. The most notable feature in the comparison of these collinear drop distributions is the slope in the central region, which varies in each case. We show with the black curve a different collinear-drop observable that again does not have soft drop grooming. This curve has double logarithmic $\Delta m^2$ dependence, and a different shape. 
Note again that both the grooming transition and the upper bound are determined by the values of $z_{{\rm cut}\;1}$ and $z_{{\rm cut}\;2}$, respectively, which are clear features one can exploit. 

\begin{figure}[t!]
	\hspace{-1.4cm}
	\includegraphics[width=0.55\textwidth]{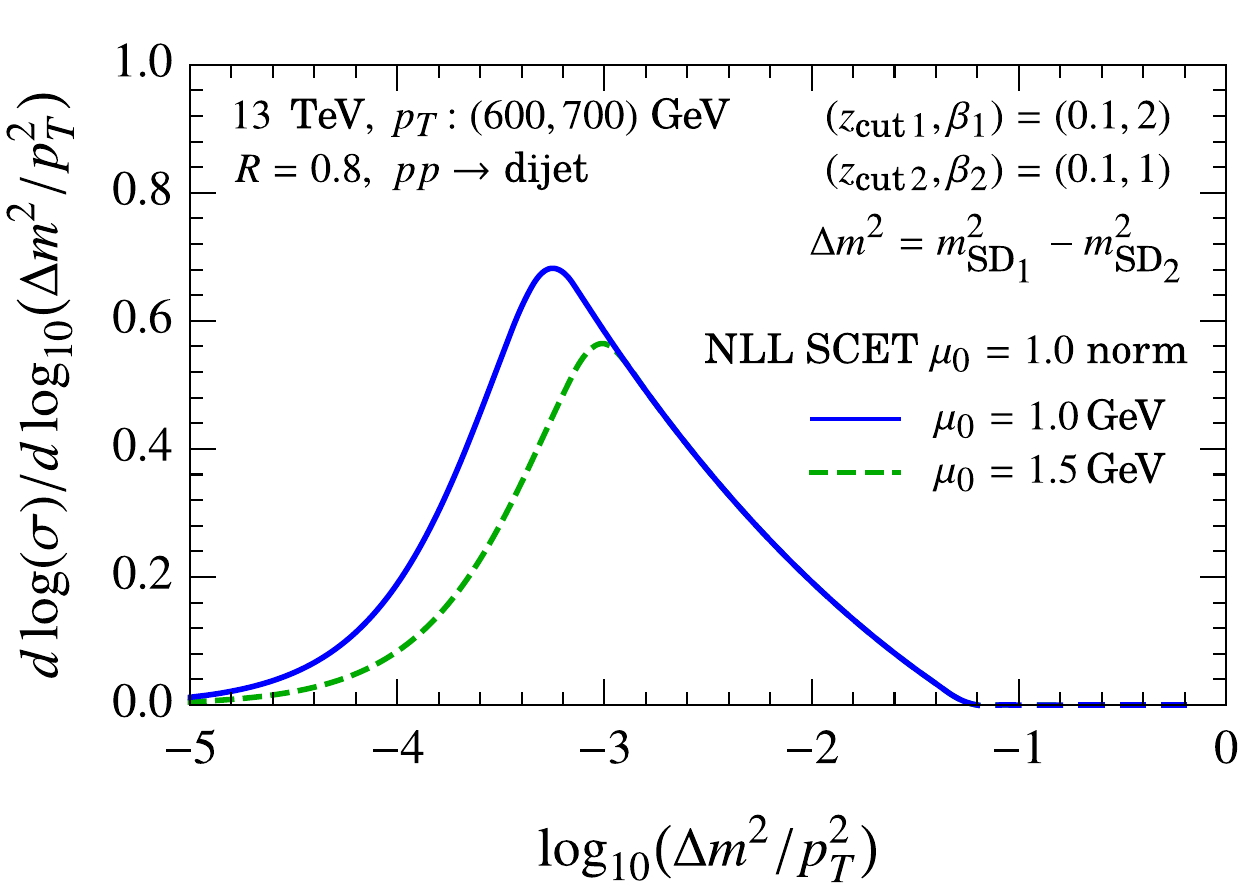}
	\includegraphics[width=0.55\textwidth]{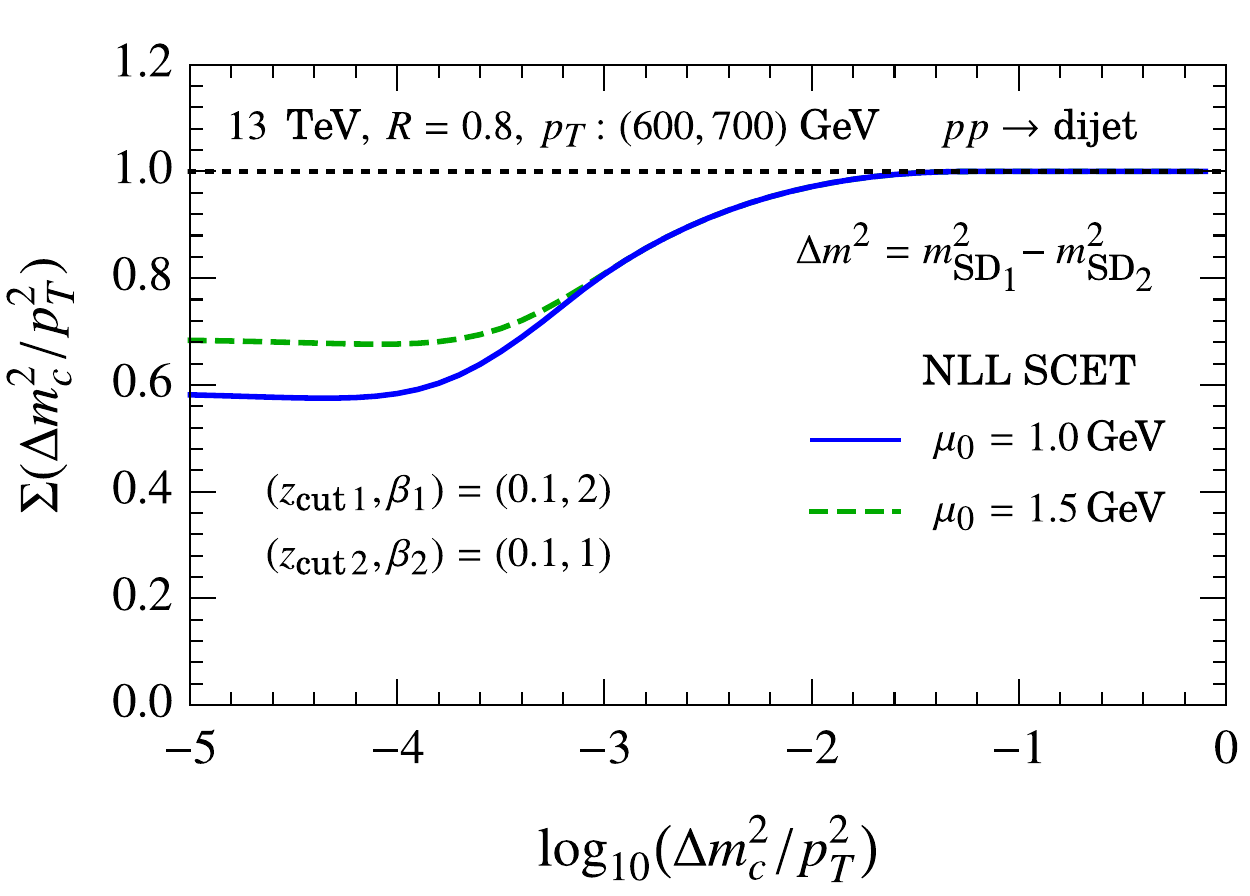}
	\caption{Comparison of Collinear Drop distributions calculated using SCET with $(z_{{\rm cut}\,1},\beta_1)=(0.1,2)$ and $(z_{{\rm cut}\,2},\beta_2)=(0.1,1)$ and different cutoff scale $\mu_0$ of the running of strong coupling constant. The left panel shows the differential distributions while the right panel shows the cumulant ones.
	}
	\label{fig:MC_SCET_collinear_drop_zcut0p1_x0}
\end{figure}

Another interesting feature to examine is the sensitivity to hadronization corrections. From the purely partonic SCET calculation we can get an idea about this sensitivity by varying the parameter $\mu_0$ in \Eq{eq:fmu0}, which is the scale where we choose to freeze the running of the strong coupling constant to ensure that it does not enter the nonperturbative region for the evaluation of perturbative anomalous dimensions. 
In \Fig{fig:MC_SCET_collinear_drop_zcut0p1_x0} we take an example collinear drop distribution, and vary $\mu_0$ from its default of $1\,{\rm GeV}$ (green dashed curve) up to $1.5\,{\rm GeV}$ (blue solid curve). 
In the left panel we show the differential distributions, using the same normalization from the $\mu_0=1\,{\rm GeV}$ result for both curves so as to not obscure differences in the spectrum. We clearly see that the change to $\mu_0$ only modifies the results below some value of $\Delta m^2$, and the region where these curves differ provides a rough indicator for the region where we can expect larger corrections from hadronization. 
In the right panel of \Fig{fig:MC_SCET_collinear_drop_zcut0p1_x0} we show the analogous results for the cumulative collinear drop cross section
\begin{align}
  \Sigma(\Delta m_c^2/p_T^2) 
    = \int_0^{\Delta m_c^2}\!\! d(\Delta m^2)\
      \frac{1}{\sigma} \frac{d\sigma}{d\Delta m^2}
     \,.
\end{align}
To obtain NLL SCET results for $\Sigma$ we integrate \eqn{PCDresum} which replaces $(\Delta m^2)^{-1+\eta}/\Gamma(\eta) \to (\Delta m_c^2)^{-\eta}/\Gamma(1+\eta)$, and we use $\Delta m_c^2$ in place of $\Delta m^2$ for all the scales $\mu_i$. From \Fig{fig:MC_SCET_collinear_drop_zcut0p1_x0} we see that the results asymptotes to $1$ at large $\Delta m_c^2$ as expected. 
Again we see that for large enough $\Delta m_c^2$ that the curves with two different values for $\mu_0$ agree, but start to deviate at smaller $\Delta m_c^2$ in the region where nonperturbative corrections are more relevant. 
Figure~\ref{fig:MC_SCET_collinear_drop_zcut0p1_x0} also exhibits an important feature of the collinear drop cross section, namely that $\Sigma$ goes to a non-trivial constant as $\Delta m_c^2\to 0$. This differs from the ungroomed or soft drop groomed observables where this constant would be $\approx 0$. The reason for this behavior is that due to the collinear drop constraint, we are always removing perturbative radiation, even as $\Delta m^2\to 0$. Hence, rather than being dominated by a Sudakov suppression for the radiation, we instead find an interesting constant that corresponds to the fraction of radiation that is retained by collinear drop in this limit.  Although not shown in the figure, we find that this constant exhibits strong dependence to the choice of collinear drop parameters, and hence is an interesting event fraction observable in its own right.  We will leave further dedicated study of these collinear drop event fractions to future work. We will return to the study of hadronization corrections in \Sec{sec:MC}, where we use Monte Carlo simulations to examine these effects for collinear drop. 

\begin{figure}[t!]
   \hspace{-1.4cm}
	\includegraphics[width=0.55\textwidth]{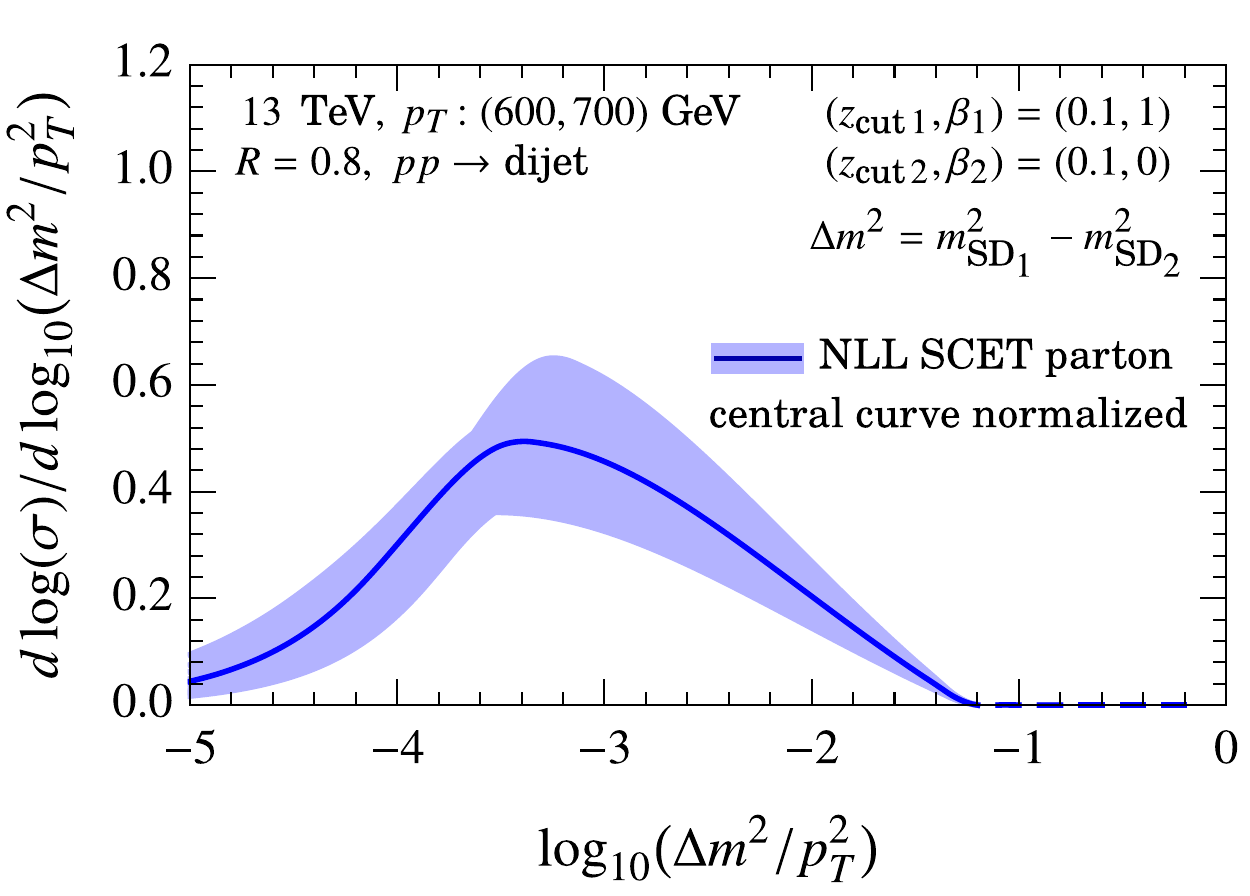}
	\includegraphics[width=0.55\textwidth]{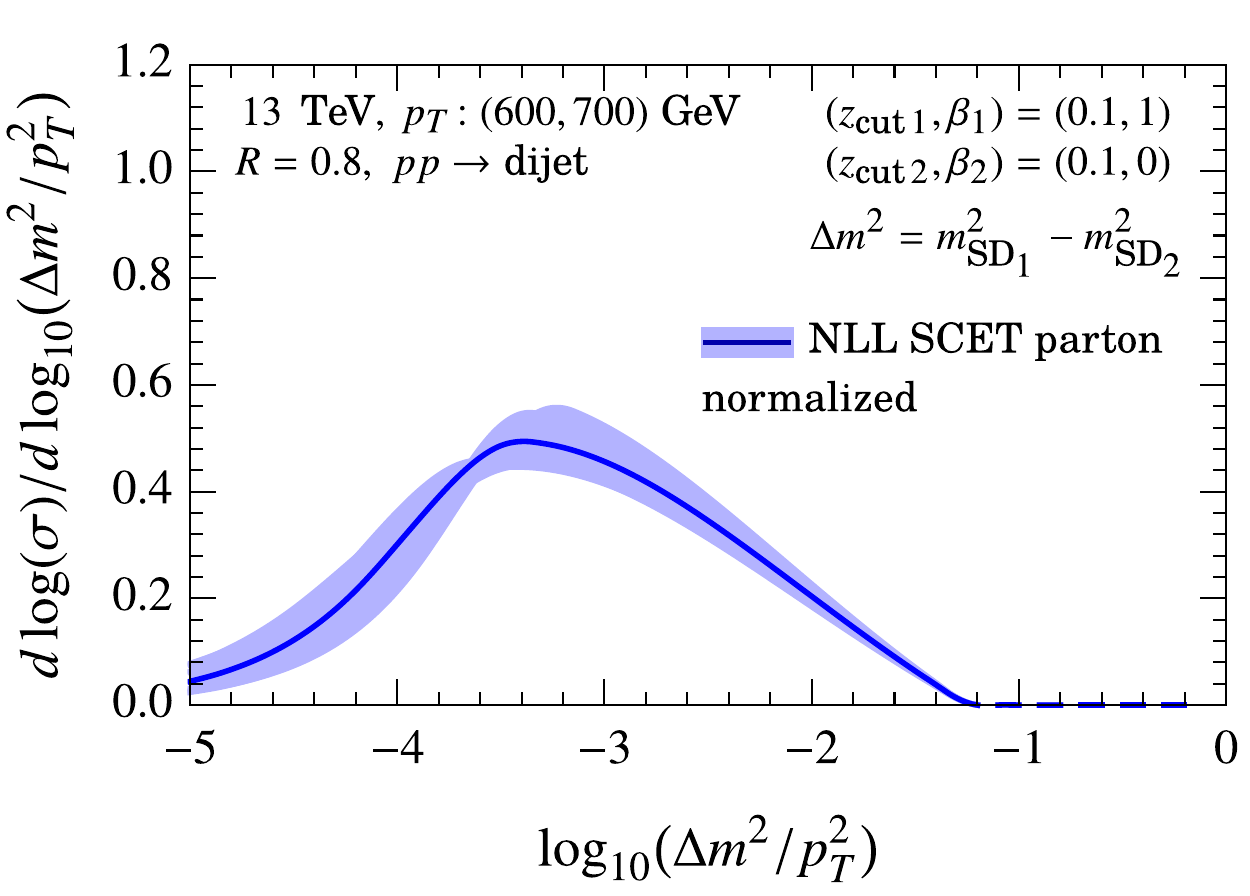}
	\caption{An example of collinear drop distributions showing uncertainty bands at NLL order. The left panel includes the normalization uncertainty, while the right panel only includes shape uncertainty. 
	}
	\label{fig:MC_collinear_drop_norm}
\end{figure}

So far our collinear drop results have been based on central values obtained with canonical profile scales. In \Fig{fig:MC_collinear_drop_norm} we include uncertainty bands from varying the scales $\mu_i$ following the prescription outlined in \Sec{sec:CDprofile}, with the overall uncertainty determined by the envelope of individual variations.  In the left panel the central curve is normalized over the range shown, while the individual variations are not further normalized, implying that this result includes an estimate for the (relative) normalization uncertainty. In the right panel the individual profile variations are themselves normalized, so the band only estimates shape uncertainty, and hence is not as large. This pattern echos what we observed already for soft drop in \Sec{sec:SDcurves}.  Examining \Fig{fig:MC_collinear_drop_norm}, and its analog for a wide range of other collinear drop parameters, we conclude that the uncertainty results obtained by our proposed scale variations are a reasonable estimate for the uncertainties at NLL order.

\section{Monte Carlo Analysis and Comparison to Analytic Predictions}
\label{sec:MC}

In the previous section we derived analytic factorization based predictions for the partonic collinear drop jet mass observable $\Delta m^2$, and examined the resulting partonic SCET distributions at NLL order. In this section we carry out further analysis of these collinear drop spectra using  Monte Carlo simulations. In particular we compare simulation results between \Pythia and \Vincia, and our SCET based factorization results. We also examine the impact on collinear drop observables of final state hadronization and of the multi-parton interaction model for underlying event effects.
When observables have different quark and gluon compositions for a jet sample this can also significantly affect jet substructure distributions, and it is interesting to see how accurate these channels are or whether their discription can be improved. Therefore we also perform separate comparisons for these two components, since . For both the simulations and factorization based results, the identity of a jet as being quark or gluon induced is determined at the stage of the initial hard scattering.

Since this work focuses on analytic predictions at the parton level, comparing our results with simulations generated using different parton showers will provide useful information about the impact of parton shower accuracy on jet substructure observables. For the Monte Carlo analysis, we use \Pythia 8.223 and \Vincia 2.0.01 to generate jet samples from dijet events. Here jets are reconstructed using the anti-$k_t$ jet algorithm with radius $R=0.8$. We study the leading two jets in inclusive jet events in 13 TeV proton-proton collisions, and we impose the following kinematic selection:  $600~{\rm GeV} < p_T < 700~{\rm GeV}$ and $|y|<2.0$.  The main difference between the MC simulations is that \Pythia uses a dipole shower where we can talk about the radiator for individual branches, whereas \Vincia uses a antennae shower with radiation produced by color correlated pairs. Both of these MCs use a string fragmentation model to implement hadronization.
 
As was mentioned previously, ATLAS recently measured soft-drop jet mass with parameters $z_{\rm cut}=0.1$ and $\beta=0,1,2$ (and CMS with $\beta=0$). Since the same data can be readily used to construct the collinear drop observables $\Delta m^2$, we will include this parameter choice in our collinear drop analysis.

\subsection{Monte Carlo Partonic Results for $\Delta m^2$}
\label{subsec:MC}

\begin{figure}[t!]
    \centering
	\includegraphics[width=0.49\textwidth]{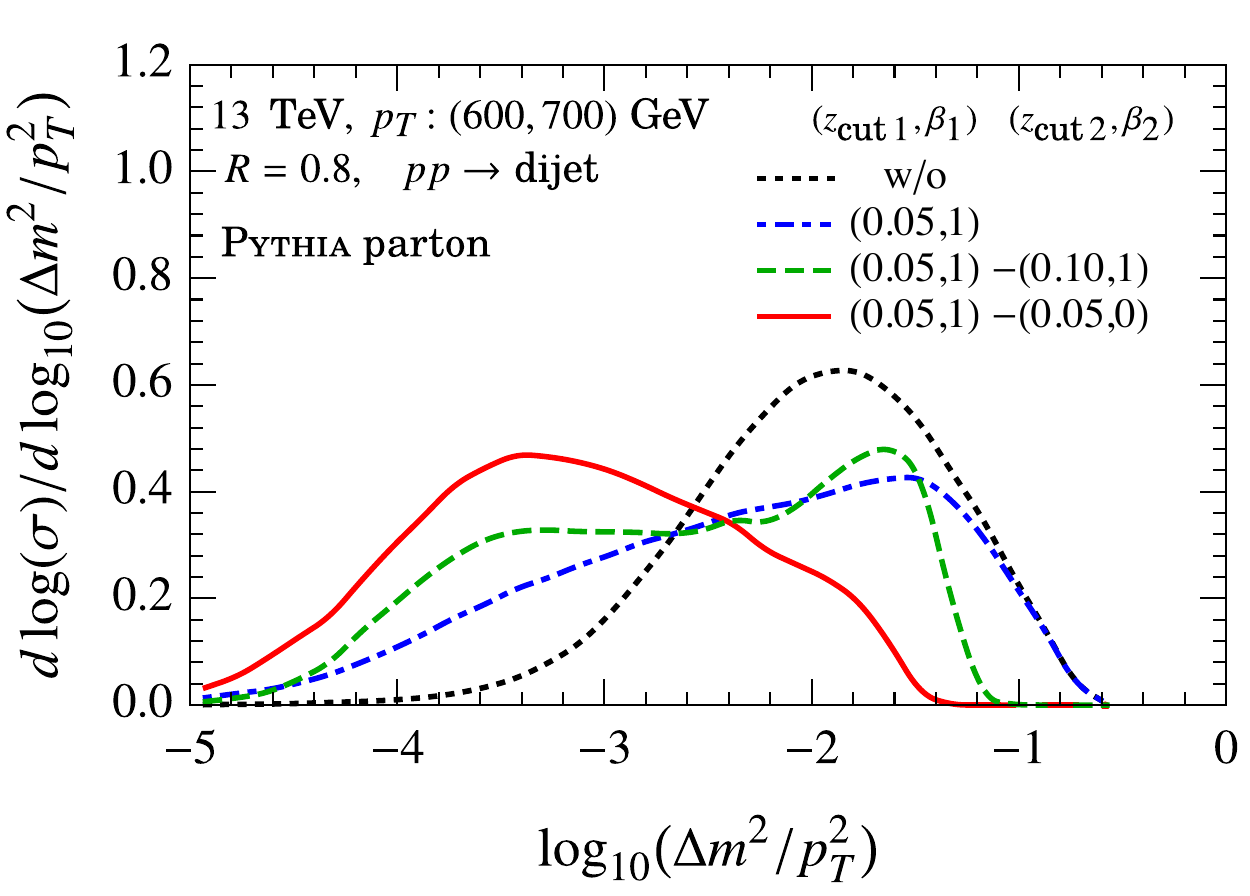}
	\includegraphics[width=0.49\textwidth]{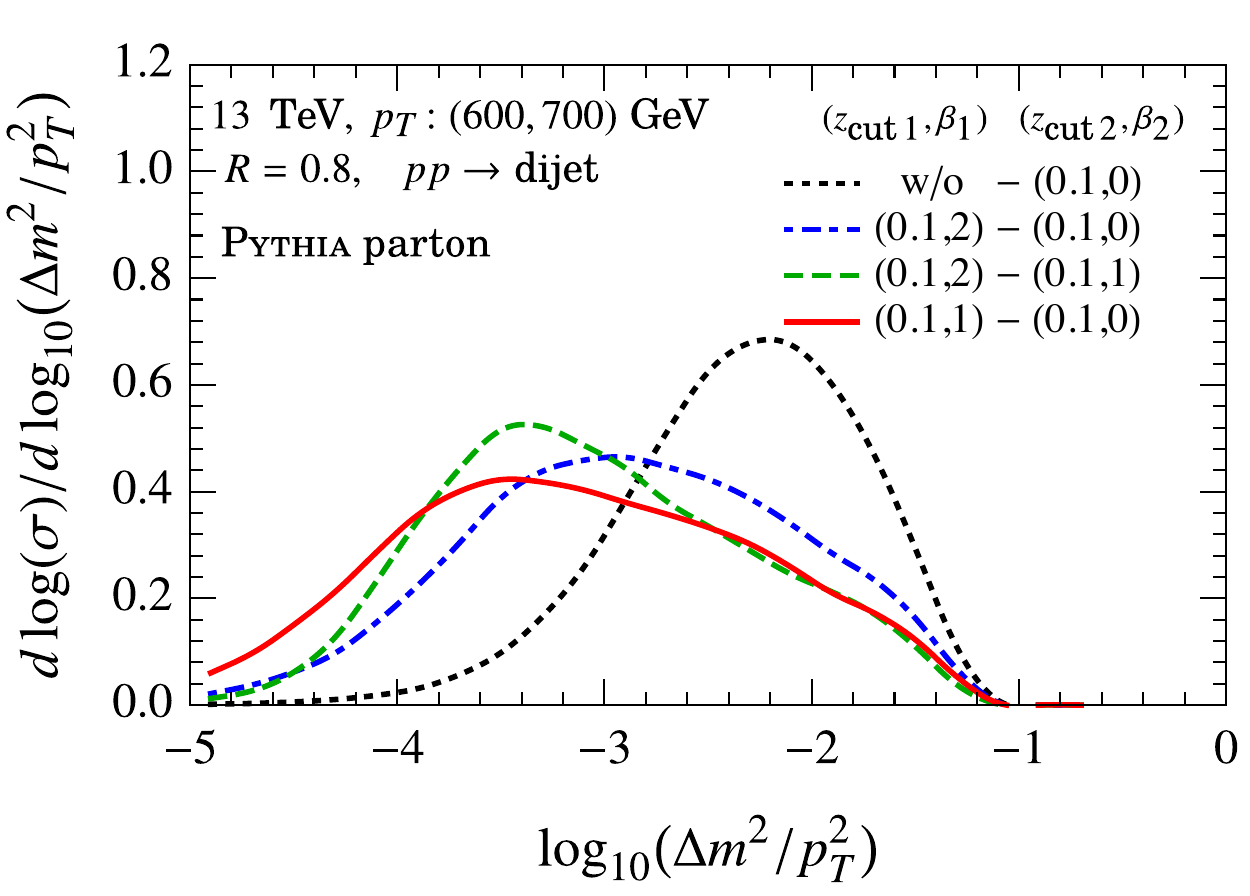}
	\includegraphics[width=0.49\textwidth]{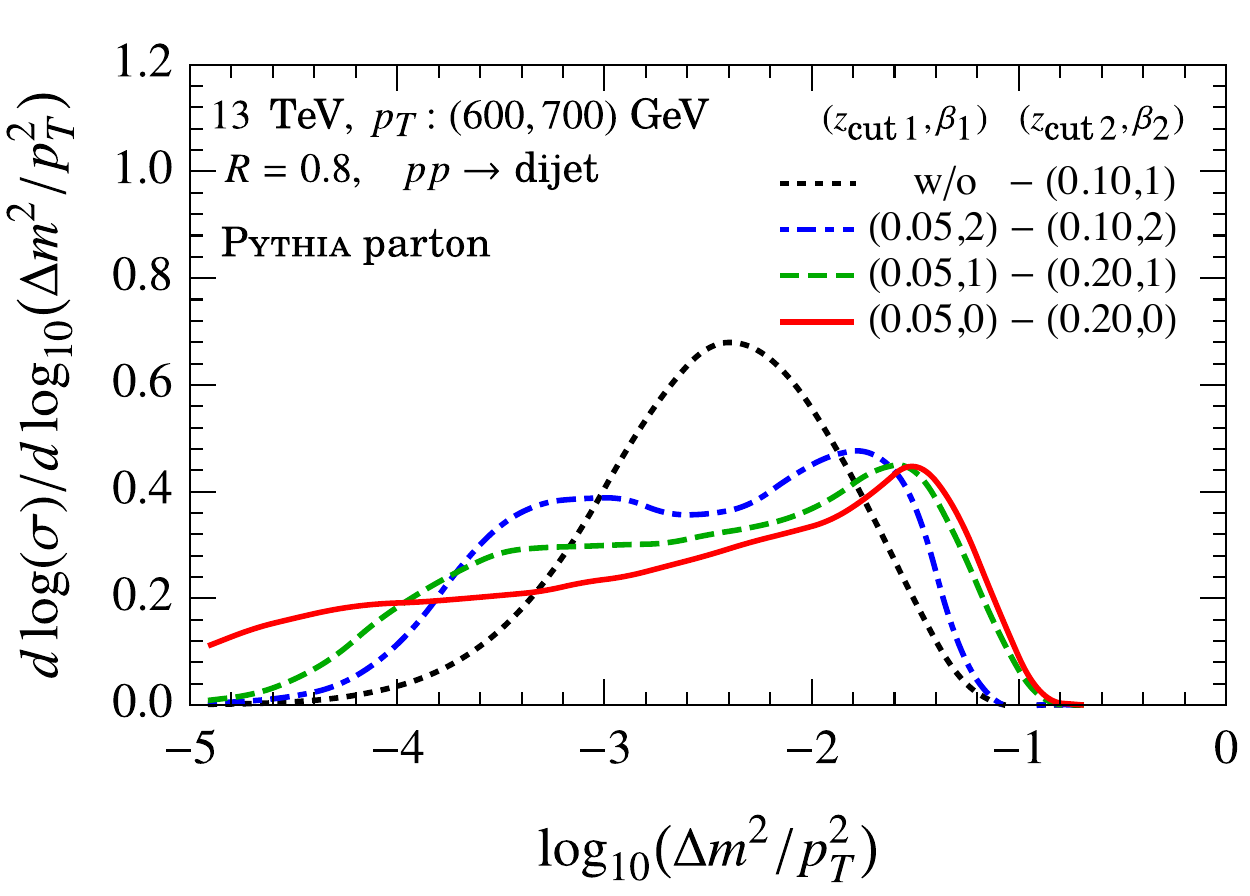}
	\caption{\Pythia collinear drop results for various parameter choices.  
     The top left panel compares collinear drop to the soft drop and ungroomed partonic jet mass spectra analogous to the NLL SCET results in the right panel of \Fig{fig:SCET_soft_collinear_drop}, while the remaining two figures show partonic results for 8 choices of collinear drop parameters analogous to the SCET results in \Fig{fig:SCET_collinear_drop}.
	}
	\label{fig:MC_soft_collinear_drop}
\end{figure}

We begin in \Fig{fig:MC_soft_collinear_drop} by reproducing with \Pythia at the parton level some results that were obtained using factorization in \Figs{fig:SCET_soft_collinear_drop}{fig:SCET_collinear_drop} of \Sec{sec:SCETresults}. The top left panel of \Fig{fig:MC_soft_collinear_drop} compares distributions for ungroomed (black dotted), soft-drop (blue dot-dashed) and collinear-drop (green dashed and red solid) jet mass, to be compared with partonic SCET results in the right panel of \Fig{fig:SCET_soft_collinear_drop}.  As already discussed earlier, the soft drop curves are quite close. The collinear drop curves also exhibit the same hierarchies in different regions and the same endpoints, but the precise shape does show some differences, in particular for the green curves.  More collinear drop results are shown in the right most panel and bottom panel of \Fig{fig:MC_soft_collinear_drop}, which can be directly compared to the two panels in \Fig{fig:SCET_collinear_drop}. Again the pattern of curves is similar, but there are noticeable differences in the precise shape, particularly for the bottom panel of \Fig{fig:MC_soft_collinear_drop}. This motivates carrying out a more detailed comparison, including the NLL uncertainties, to which we now turn.

\subsection{Comparison to Partonic SCET Results for $\Delta m^2$}

We now consider a more detailed comparison between the partonic NLL SCET predictions with Monte Carlo simulations generated with both \Pythia and \Vincia at the parton level, pointing out places where they differences in their predictions for collinear drop observables. Such comparisons can point the way to methods for improving both parton shower and analytic predictions. All curves are normalized over the displayed range, unless otherwise indicated.

\begin{figure}[t!]
	\centering
	\includegraphics[width=0.49\textwidth]{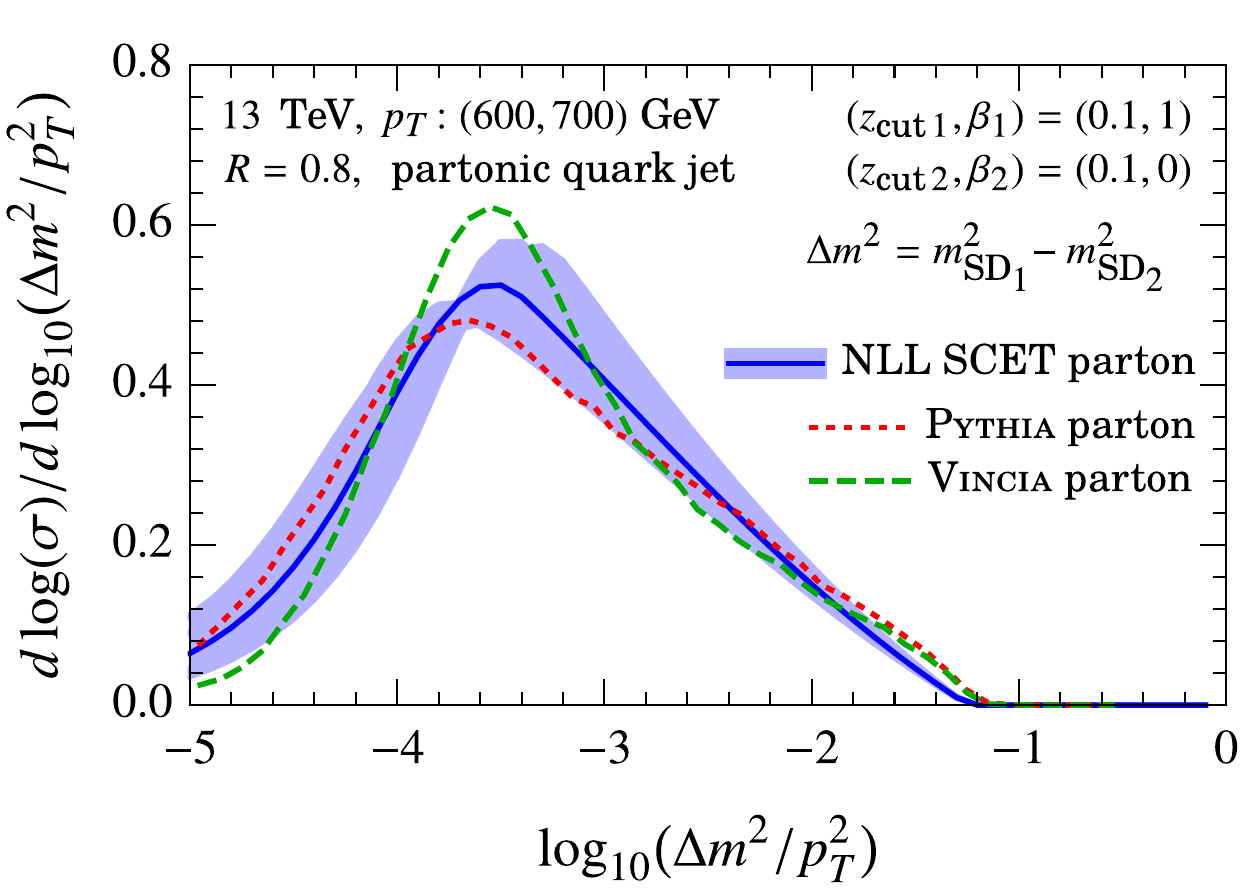}
	\includegraphics[width=0.49\textwidth]{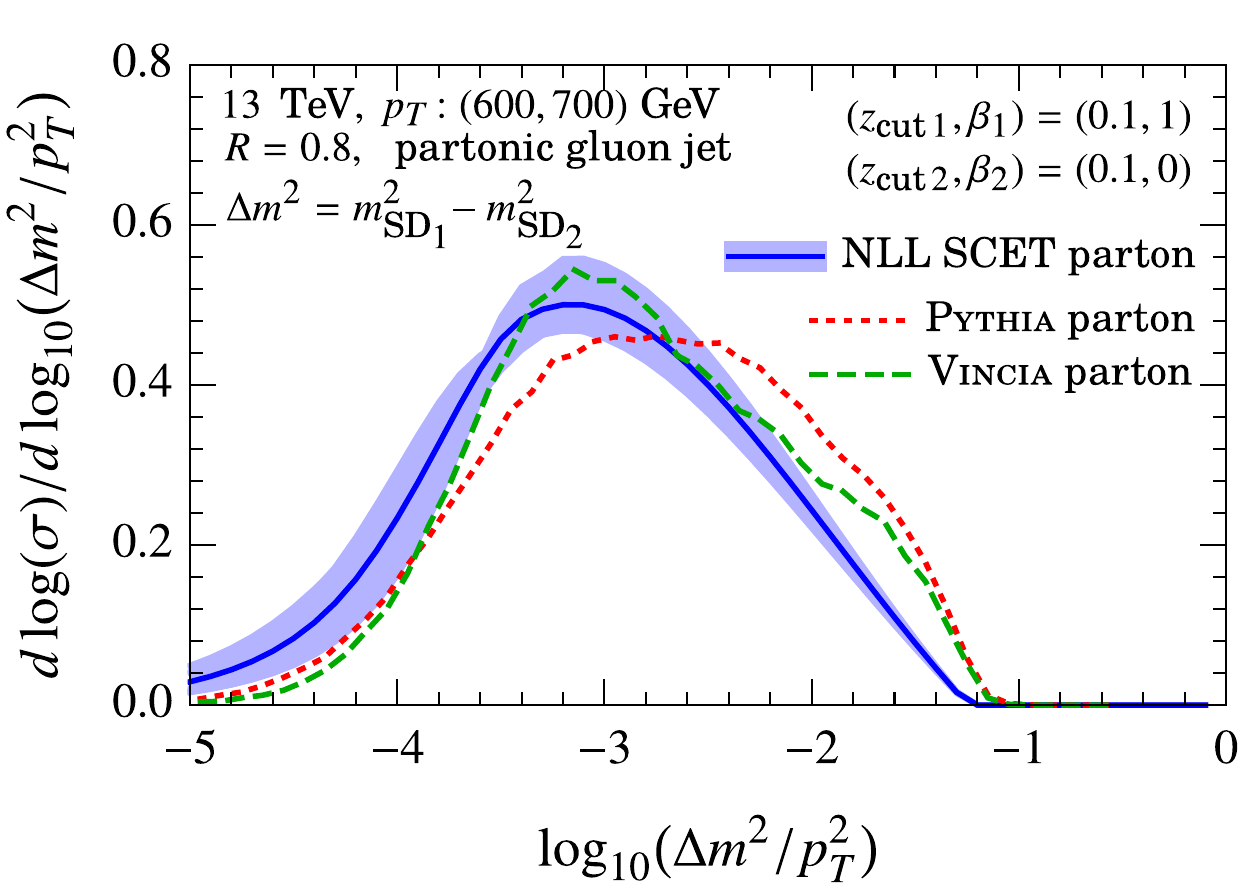}
    \\
	\includegraphics[width=0.49\textwidth]{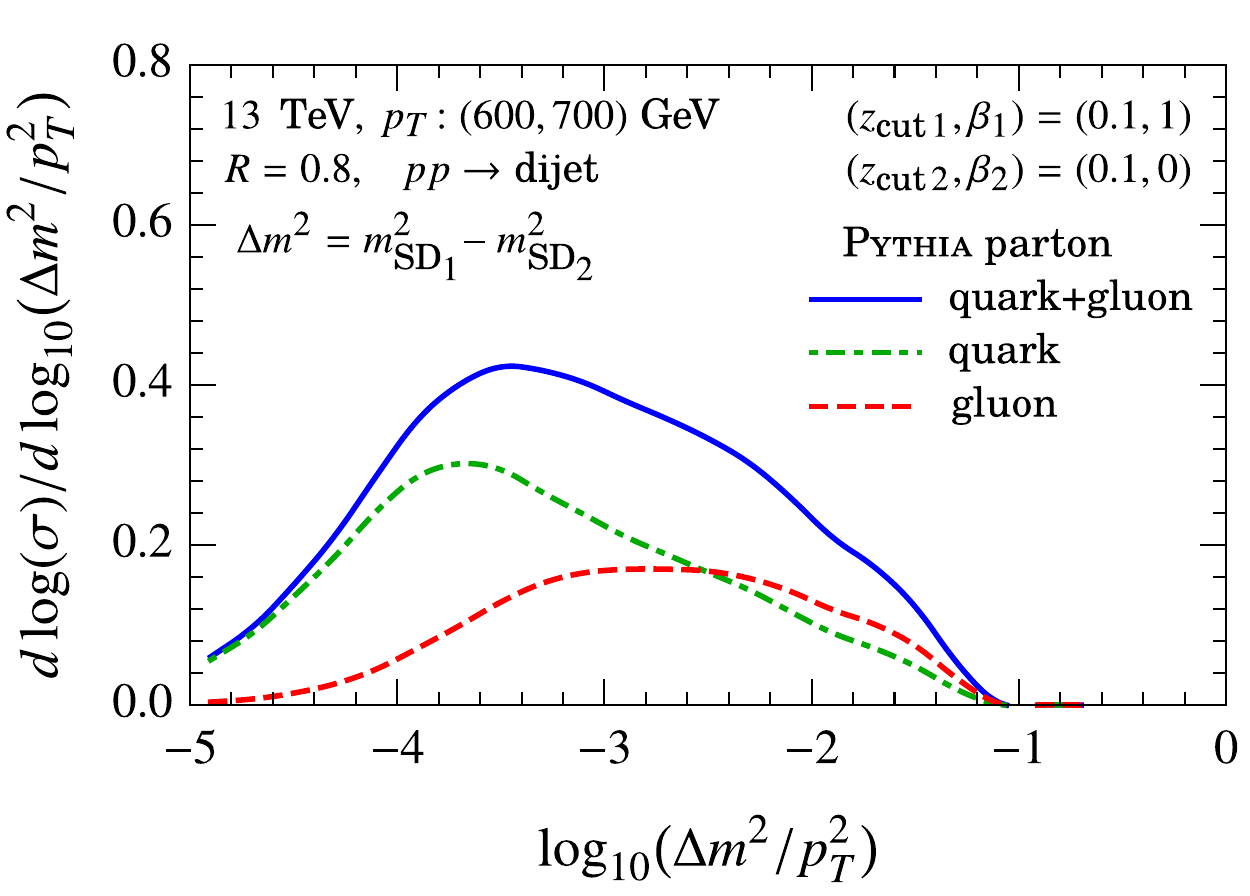}
	\includegraphics[width=0.49\textwidth]{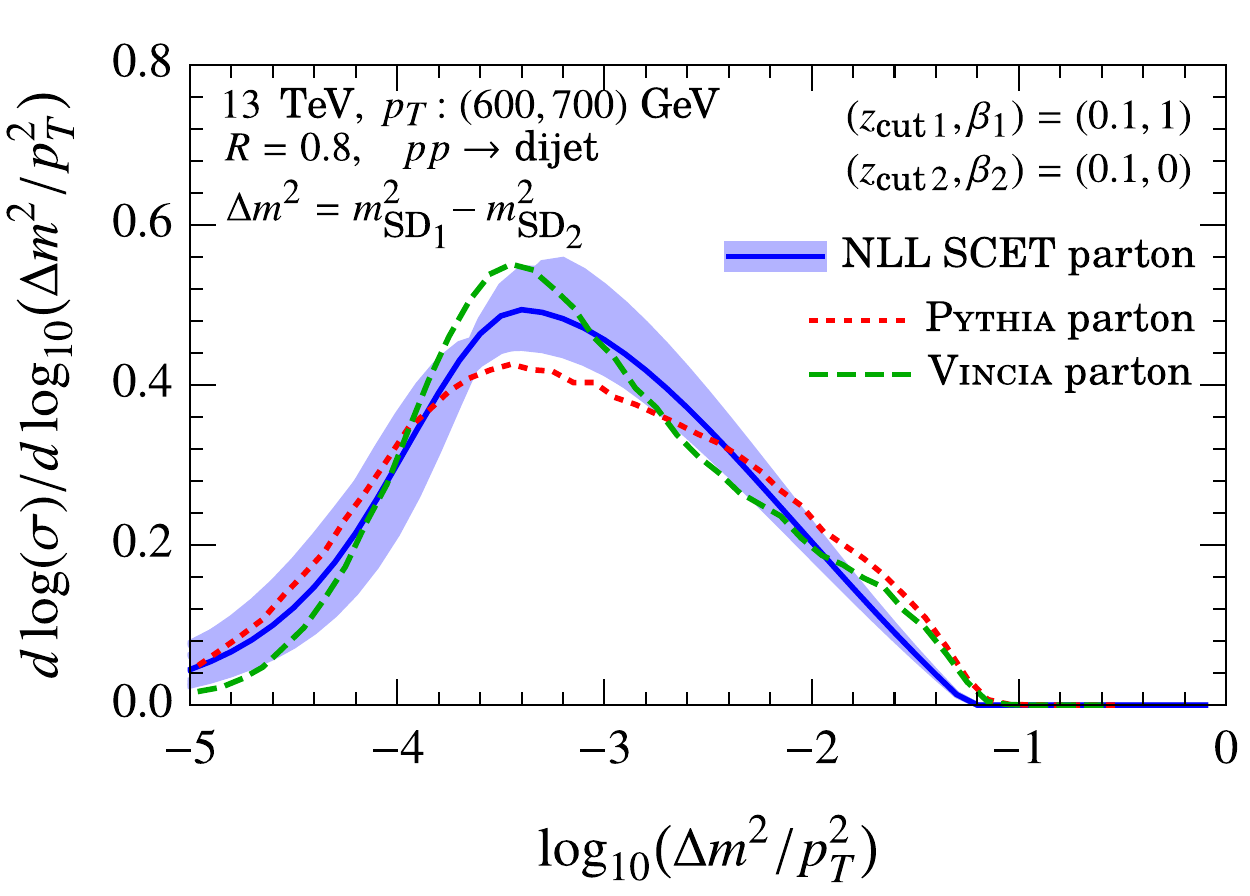}
	\caption{Collinear Drop distributions with $(z_{{\rm cut}\,1},\beta_1)=(0.1,1)$ and $(z_{{\rm cut}\,2},\beta_2)=(0.1,0)$ from \Pythia and \Vincia simulations as well as SCET calculations with theoretical uncertainty estimation (blue bands). The top two panels show the distributions for quark-initiated jets (left panel) and gluon-initiated jets (right panel), and the bottom right panel gives the dijet distributions. The bottom left panel gives a decomposition of the Collinear Drop distribution from \textsc{Pythia} simulations into quark and gluon components.
	}
	\label{fig:MC_SCET_collinear_drop_zcut0p1_qg}
\end{figure}

\begin{figure}[t!]
	\hspace{-1.4cm}
	\includegraphics[width=0.55\textwidth]{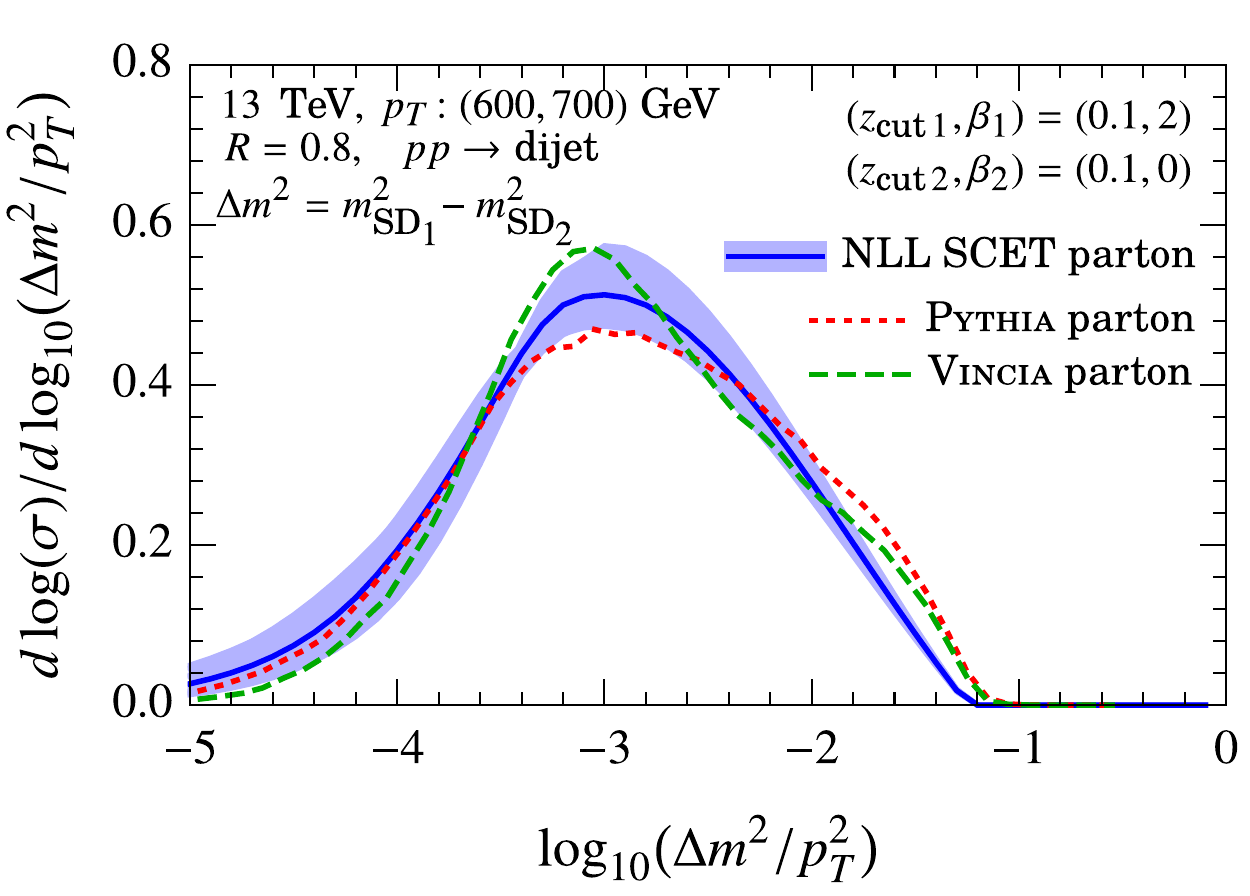}
	\includegraphics[width=0.55\textwidth]{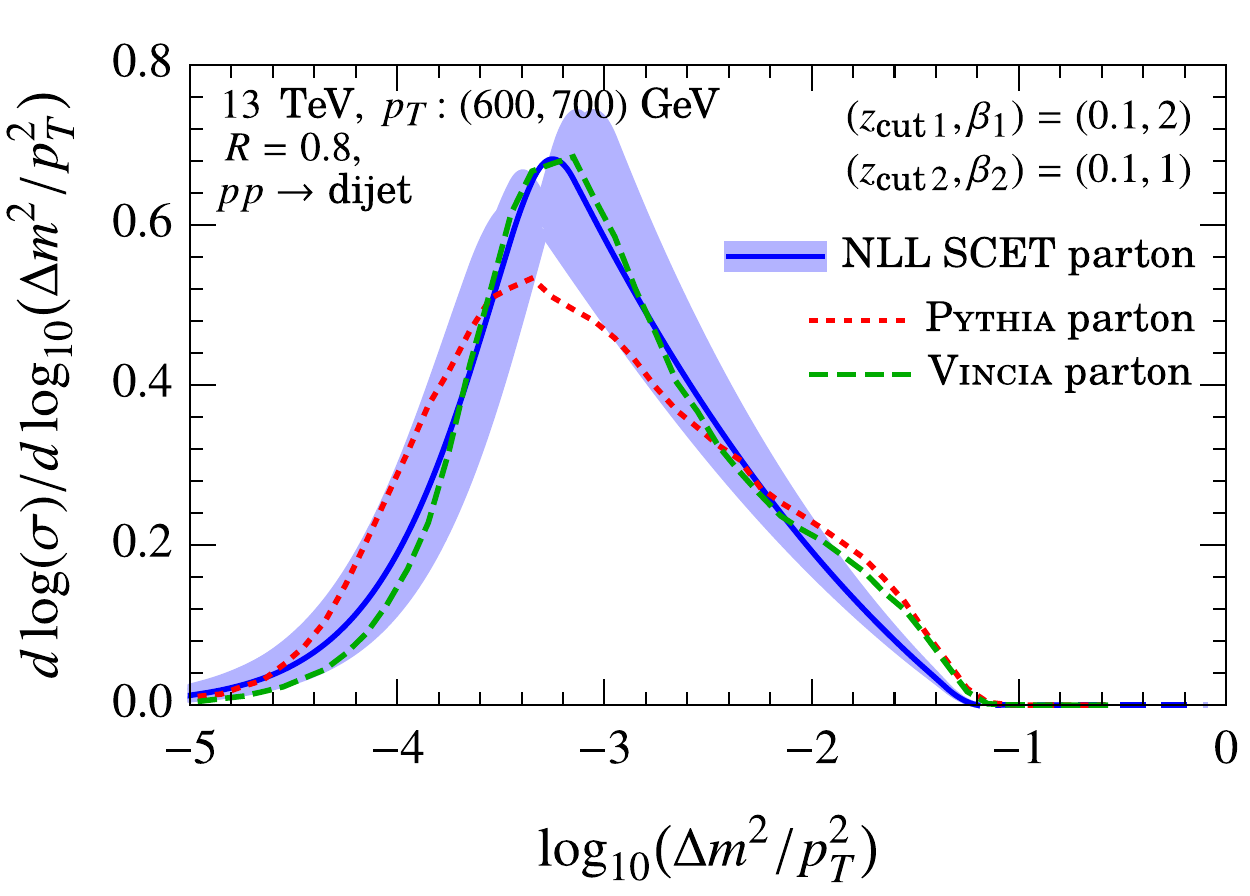}
	\caption{More Collinear Drop distributions with $(z_{{\rm cut}\,1},\beta_1)=(0.1,2)$ and $(z_{{\rm cut}\,2},\beta_2)=(0.1,0)$ (left panel) and $(z_{{\rm cut}\,1},\beta_1)=(0.1,2)$ and $(z_{{\rm cut}\,2},\beta_2)=(0.1,1)$ (right panel) from \Pythia and \Vincia simulations as well as SCET calculations. 
	}
	\label{fig:MC_SCET_collinear_drop_zcut0p1}
\end{figure}

In \Fig{fig:MC_SCET_collinear_drop_zcut0p1_qg} we consider the collinear-drop observable $\Delta m^2$ with fixed $z_{{\rm cut}\;1} = z_{{\rm cut}\;2} = 0.1$, with $\beta_1=1$ and $\beta_2=0$. 
The partonic results from SCET at NLL accuracy are shown by the solid blue lines, \Pythia by dotted red lines, and \Vincia by dashed green lines.  The blue band corresponds to theoretical uncertainty estimated by scale variation following the method described in \Sec{sec:CDprofile}, and studied in \Sec{sec:SCETresults}. 
The top left panel of \Fig{fig:MC_SCET_collinear_drop_zcut0p1_qg} shows the results for quark-initiated jets while the right panel corresponds to gluon-initiated jets. For quark initiated jets the \Vincia results are significantly more peaked than for \Pythia, but both simulations agree with the SCET results within the uncertainty band. For gluon initiated jets the SCET results are closer to those of \Vincia, while \Pythia is broader and peaks at larger $\Delta m^2$ values. In the $\log_{10}(\Delta m^2/p_T^2)\sim -1.5$ region, the analytic and simulation results for gluons differ. In general this region is the most sensitive to fixed order corrections and corrections from beyond leading power in the collinear drop expansions. 
In the lower left panel of \Fig{fig:MC_SCET_collinear_drop_zcut0p1_qg} we show the breakdown of quark and gluon contributions for dijets in \Pythia. Here only the total quark+gluon curve is normalized, while the individual quark and gluon curves add to this total.  As expected the quarks dominate for smaller $\Delta m^2$, whereas the gluon contributions are broader and peak at larger values. In the lower right panel of \Fig{fig:MC_SCET_collinear_drop_zcut0p1_qg} we consider the collinear observable for $pp\to $ dijets, again comparing the partonic collinear drop predictions. Clear differences are still evident in this figure between \Pythia and \Vincia, particularly in the peak region. This motivates both the corresponding experimental measurement, as well as carrying out more precise SCET calculations beyond NLL, to shed light on these differences.

In \Fig{fig:MC_SCET_collinear_drop_zcut0p1} we extend the comparison of dijet predictions to two other collinear drop observables in the same class, still fixing $z_{{\rm cut}\;1} = z_{{\rm cut}\;2} = 0.1$, but using other values of $\beta_1$ and/or $\beta_2$.  The results for $\beta_1=2$, $\beta_2=0$ in the the left panel are similar to those of \Fig{fig:MC_SCET_collinear_drop_zcut0p1_qg}, with somewhat smaller differences between the predictions.  On the other hand, the results for $\beta_1=1$, $\beta_2=2$ in the the right panel of \Fig{fig:MC_SCET_collinear_drop_zcut0p1}  exhibit even clearer differences between \Pythia and \Vincia.  In this case the SCET NLL results appear to clearly favor the \Vincia result.

%The disagreement in the large $\log_{10}(\Delta m^2/p_T^2)$ region again shows up because of the significant fraction of gluon-initiated jets in the dijet sample. 

\begin{figure}[t!]
	\centering
	\includegraphics[width=0.49\textwidth]{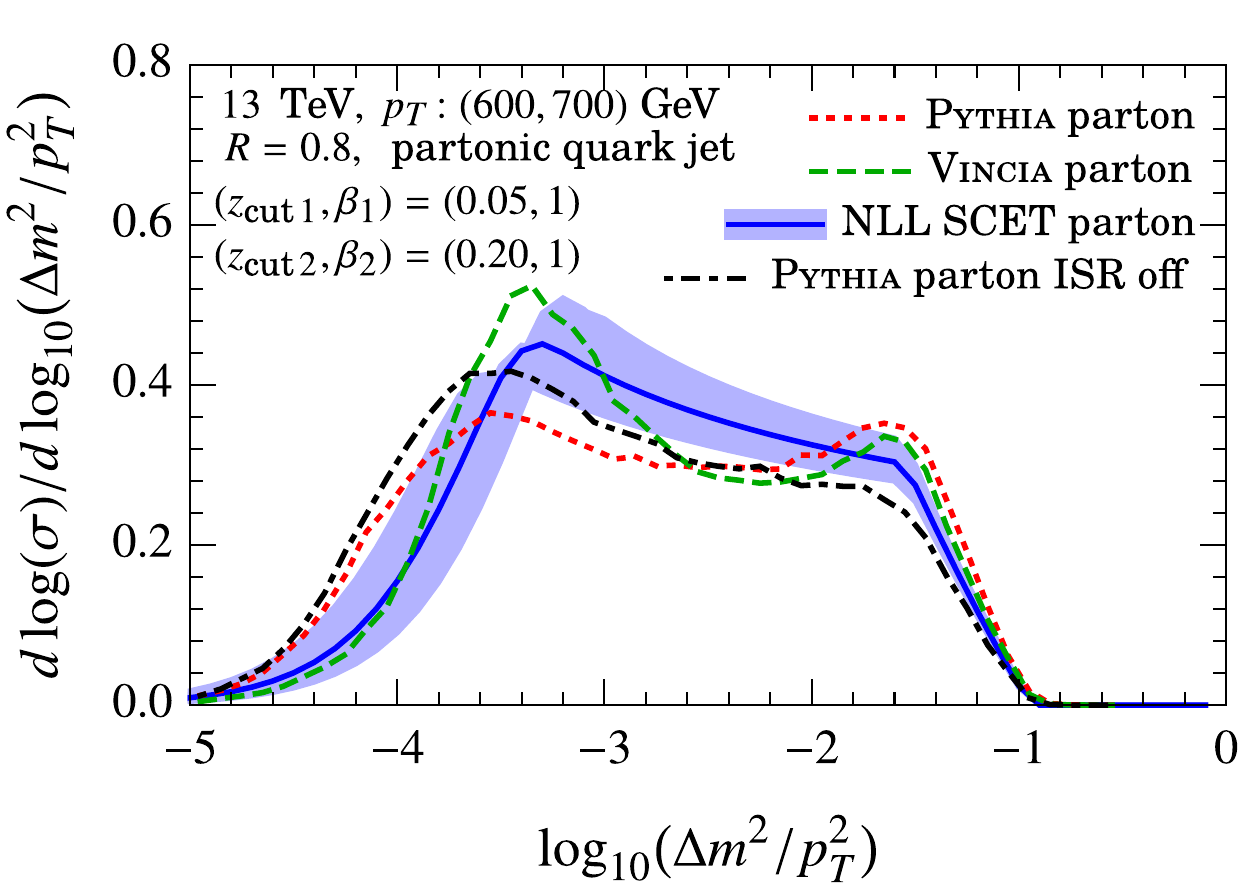}
	\includegraphics[width=0.49\textwidth]{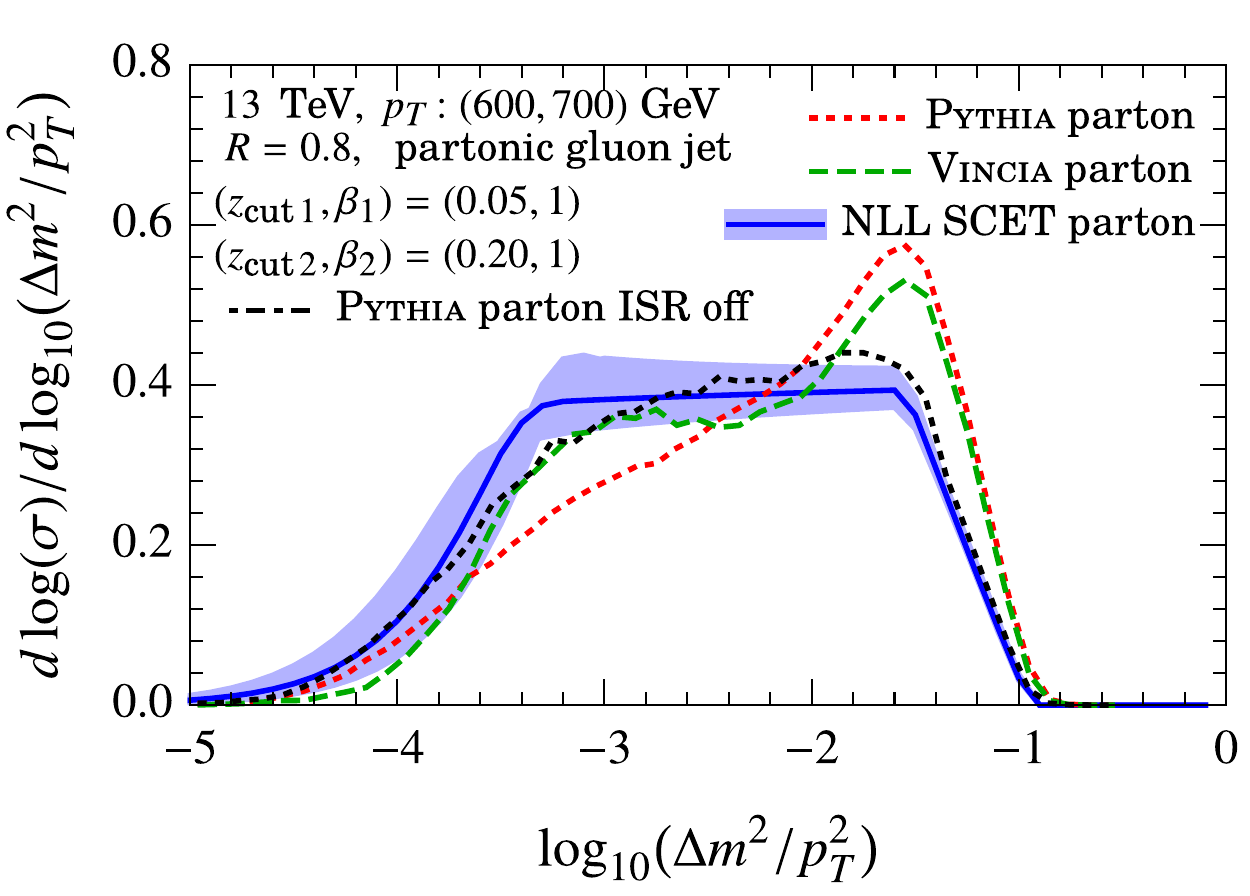} 
    \\
	\includegraphics[width=0.49\textwidth]{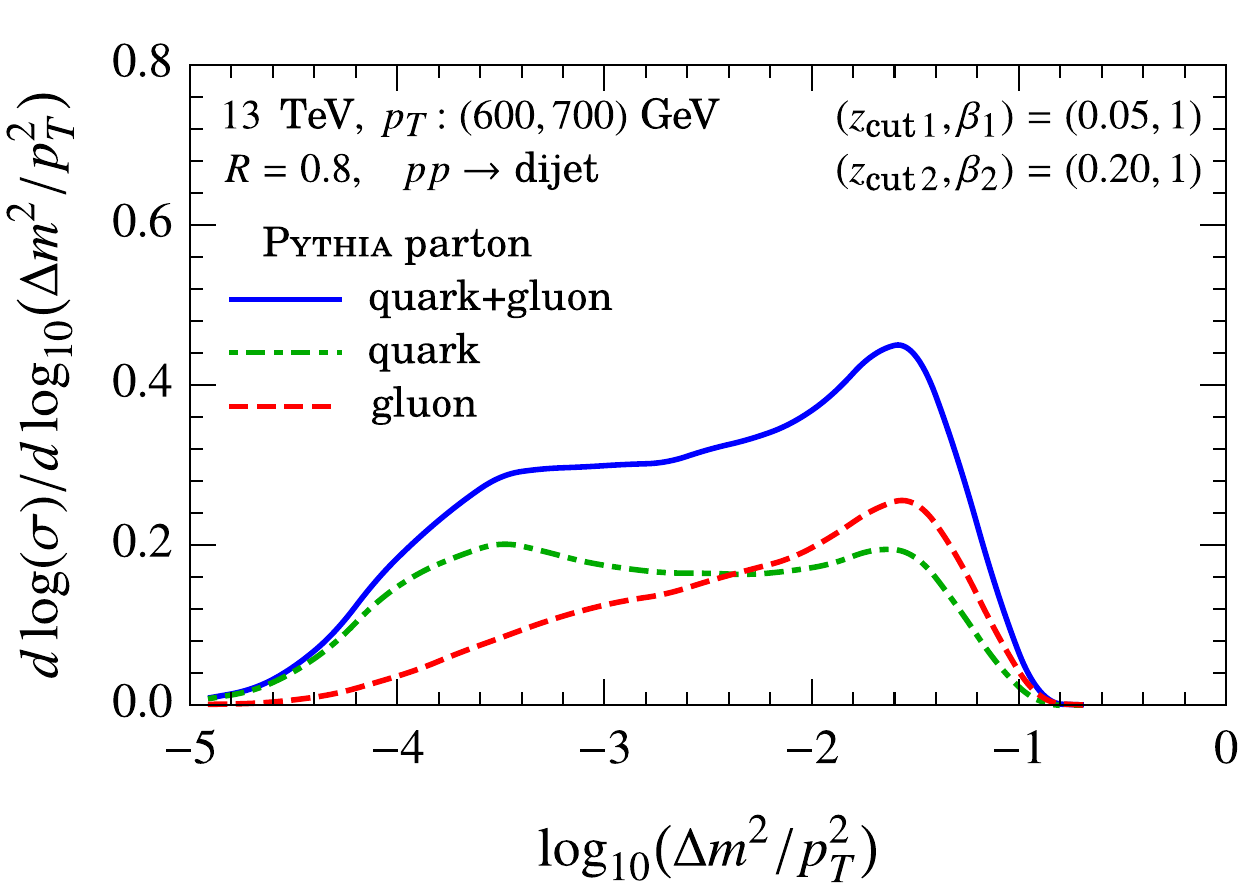}
	\includegraphics[width=0.49\textwidth]{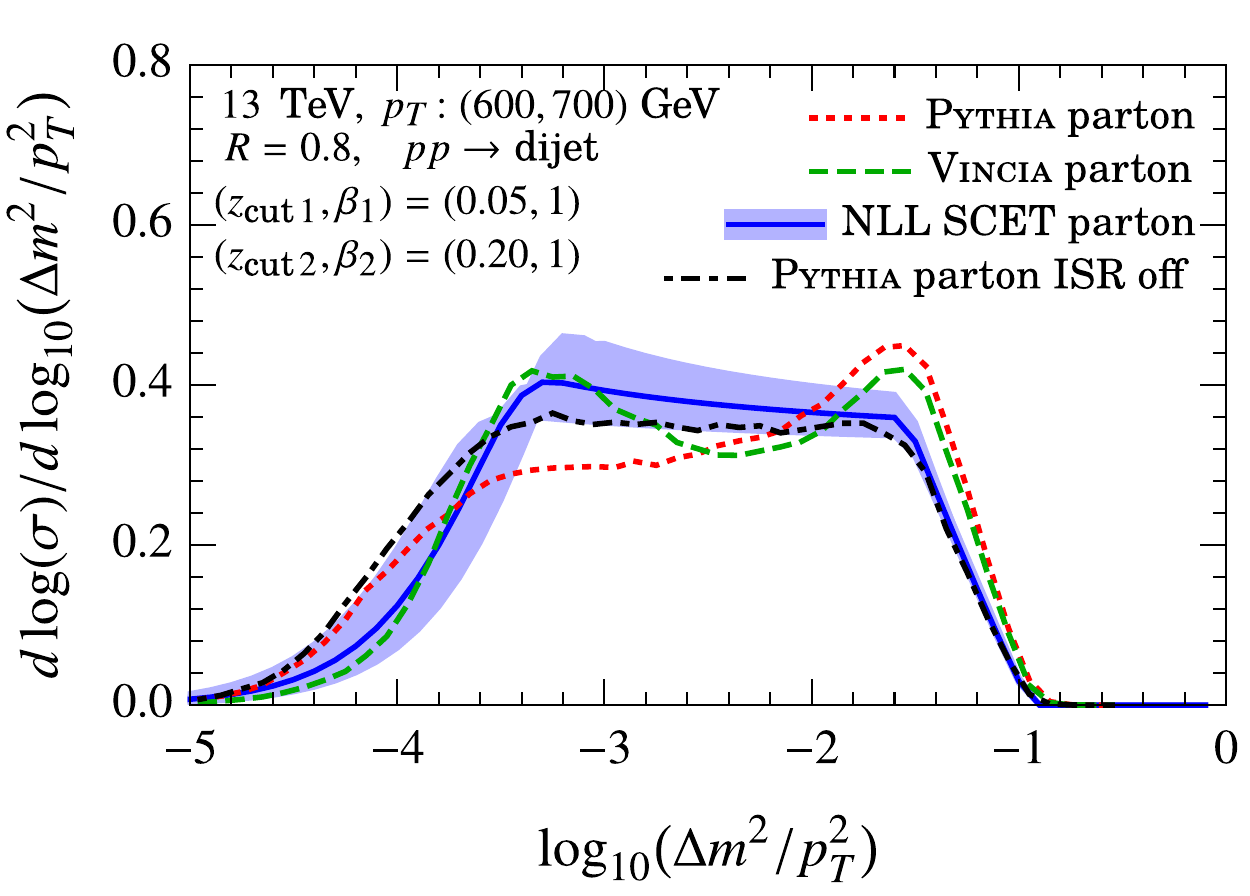}
	\caption{Collinear Drop distributions with $(z_{{\rm cut}\,1},\beta_1)=(0.05,1)$ and $(z_{{\rm cut}\,2},\beta_2)=(0.2,1)$ from \Pythia and \Vincia simulations as well as SCET calculations with theoretical uncertainty estimation (blue bands). Results from \textsc{Pythia} simulations with initial state radiation turned off are also provided. The top two panels show the distributions for quark-initiated jets (left panel) and gluon-initiated jets (right panel), and the bottom right panel gives the dijet distributions. The bottom left panel gives a decomposition of the Collinear Drop distribution from \textsc{Pythia} simulations into quark and gluon components.
	}
	\label{fig:MC_SCET_collinear_drop_beta1_qg}
\end{figure}

We now turn to the analysis of a different class of collinear drop observables, where we have $z_{{\rm cut}\;1}< z_{{\rm cut}\;2}$ and $\beta_1=\beta_2$. In \Fig{fig:MC_SCET_collinear_drop_beta1_qg} we take $\beta_1=\beta_2=1$, with $z_{{\rm cut}\;1}=0.05$ and $z_{{\rm cut}\;2}=0.2$.  Again we compare partonic results, and the blue curves are NLL SCET, dotted red are \Pythia, and dashed green are \Vincia.  In addition we show black dot-dashed curves which correspond to \Pythia results with initial state radiation (ISR) turned off. The top left panel of \Fig{fig:MC_SCET_collinear_drop_beta1_qg} shows quark initiated jets, while the right panel shows the result for gluon initiated jets. Interestingly, there are again quite significant differences between the \Pythia and \Vincia curves, which in this case are evident for quark jets in the region $\log_{10}(\Delta m^2/p_T^2)< -3$ where nonperturbative corrections are expected to become more significant. The lower panels \Fig{fig:MC_SCET_collinear_drop_beta1_qg} again show the breakdown of quark and gluon contributions (left panel) and the predictions for dijets (right panel).  In general the SCET results at NLL exhibit a less peaky structure than the MC simulations, and are in general closer to the \Vincia results. 

For gluon jets near  $\log_{10}(\Delta m^2/p_T^2)\simeq -1.5$ in \Fig{fig:MC_SCET_collinear_drop_beta1_qg} there is a clear difference between the simulation and SCET results, since there is a significant peak in both MC simulation results that does not appear in our NLL theory curve. This corresponds to the value where the groomed to ungroomed transition occurs, where it is known that fixed order corrections become more important. Since soft drop grooming is no longer being effective in this region, there can also now be significant corrections from wide angle soft radiation that are not included in our NLL calculations here.  To test the importance of such radiation, we have included \Pythia results with ISR radiation turned off (black dot-dashed curves). In this case the peak structure near the groomed to ungroomed transition is removed and the spectrum from \Pythia simulations with ISR off  agrees better with our partonic NLL results. 
It would therefore be interesting to increase the perturbative precision of the SCET calculation in this transition region, by including both ISR effects and higher order matching corrections. We leave this for future work. 

\begin{figure}[t!]
	\hspace{-1.4cm}
	\includegraphics[width=0.55\textwidth]{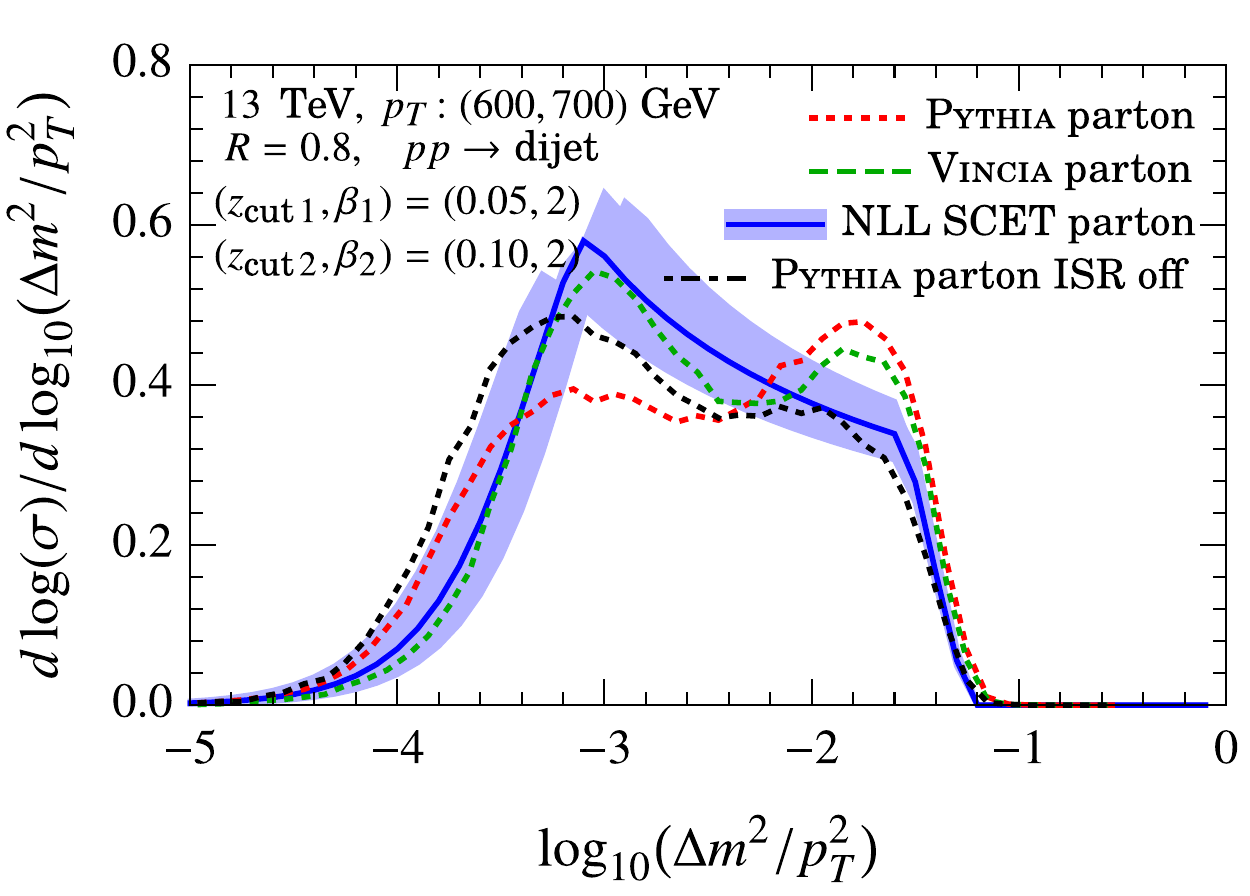}
	\includegraphics[width=0.55\textwidth]{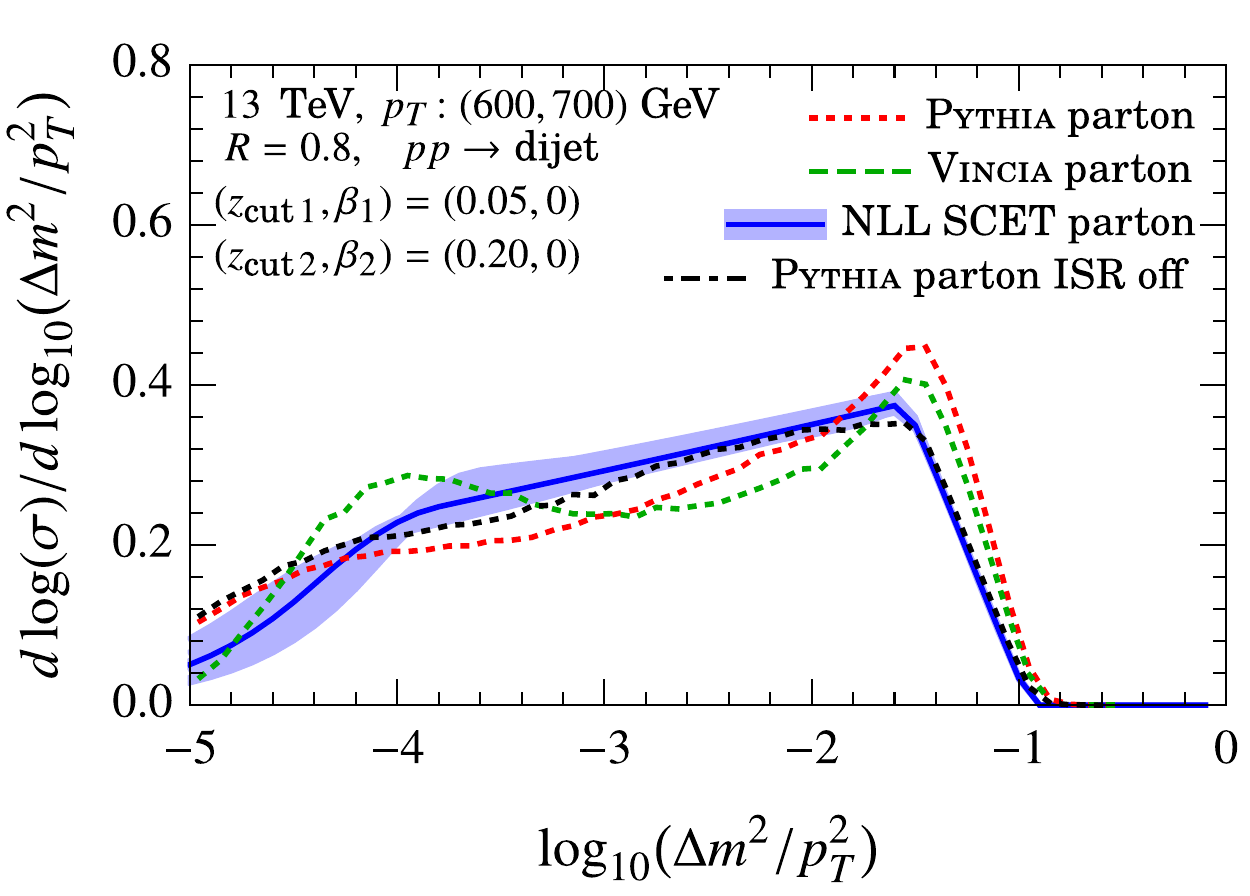}
	\caption{More Collinear Drop distributions with $(z_{{\rm cut}\,1},\beta_1)=(0.05,2)$ and $(z_{{\rm cut}\,2},\beta_2)=(0.1,2)$ (left panel) and $(z_{{\rm cut}\,1},\beta_1)=(0.05,0)$ and $(z_{{\rm cut}\,2},\beta_2)=(0.2,0)$ (right panel) from \Pythia and \Vincia simulations as well as SCET calculations. Results from \textsc{Pythia} simulations with initial state radiation turned off are also provided.
	}
	\label{fig:MC_SCET_collinear_drop_beta1}
\end{figure}

In \Fig{fig:MC_SCET_collinear_drop_beta1} we compare dijet results for two different collinear drop observables which also have $\beta_1=\beta_2$ and $z_{{\rm cut}\;1}< z_{{\rm cut}\;2}$. The left panel which is more peaked takes $\beta_1=\beta_2=2$, while the right panel which is wider and flatter uses $\beta_1=\beta_2=0$. Again we see significant differences between the \Pythia and \Vincia results in both cases, and the presence of significant wide angle soft radiation contributions near the groomed to ungroomed transition point. Away from that point the SCET results agree more closely with \Vincia for the left panel, and do not clearly favor either MC in the right panel. 

%The difference between the distributions of quark-initiated and gluon-initiated jets is again predicted to be larger in MC simulations. \TODO{If we want to say this we should quote numbers with uncertainties for the comparison.}

In general we conclude that there are noticeable and interesting differences between \Pythia and \Vincia simulation predictions for collinear drop observables. The NLL SCET calculations performed here show somewhat of a preference for the \Vincia results, though higher order calculations should be carried out with reduced uncertainties to more clearly pin this down. Such studies should be carried out independently for quarks and gluons, with the combinations giving dijets then compared to experimental data. Also prominent is the advantage to studying the $z_{{\rm cut}\;1}= z_{{\rm cut}\;2}$ class of collinear drop observables, where soft wide angle radiation is always more suppressed due to the lack of a groomed to ungroomed transition region. On the other hand if the goal is to study this radiation, then the prominent peaks in this region provided by the $\beta_1=\beta_2$ class of collinear drop observables, provide a means to do so.

\subsection{Hadronization and MPI for $\Delta m^2$}

In this section we carry out a study of the sensitivity of collinear-drop observables to hadronization and MPI effects as implemented in MC simulations. 

\begin{figure}[t!]
    \hspace{-1.4cm}
	\includegraphics[width=0.55\textwidth]{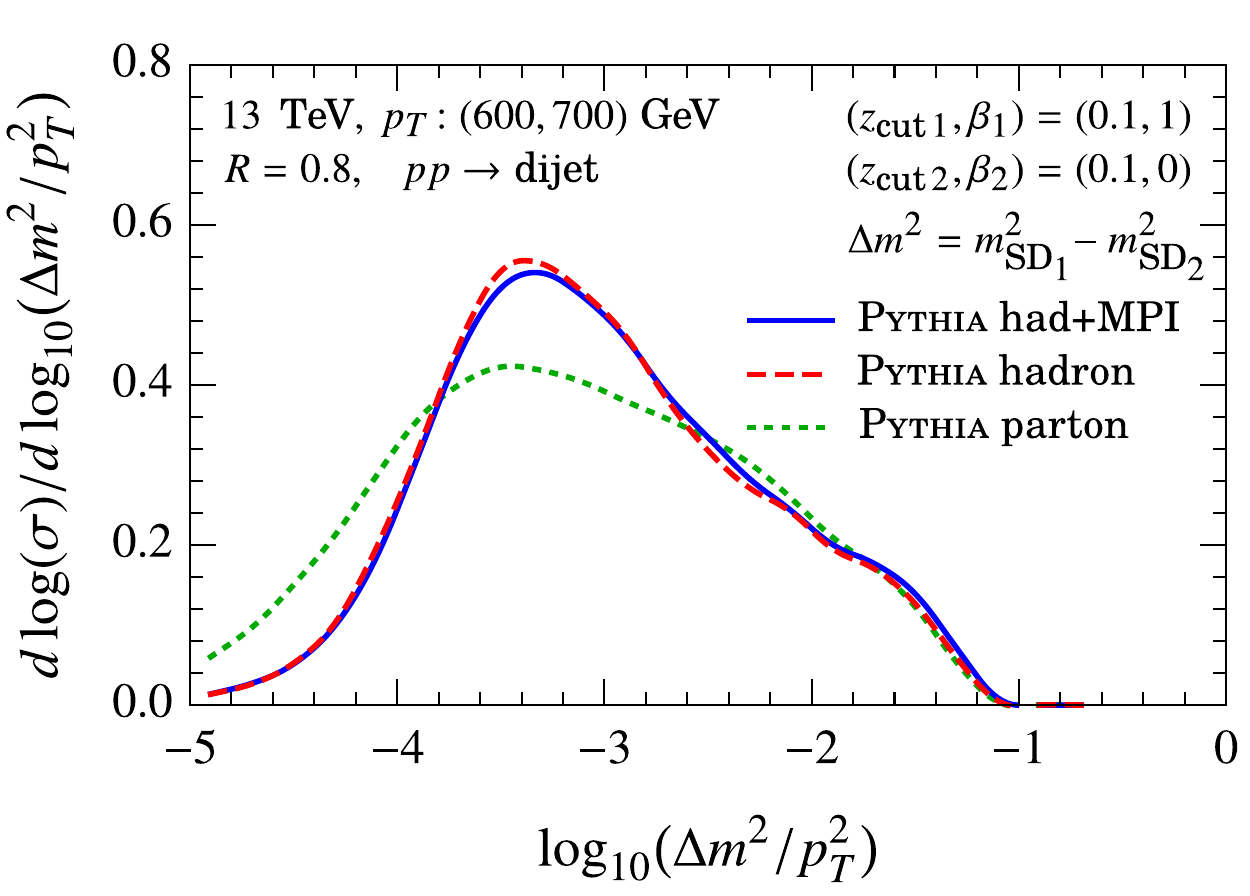}
	\includegraphics[width=0.55\textwidth]{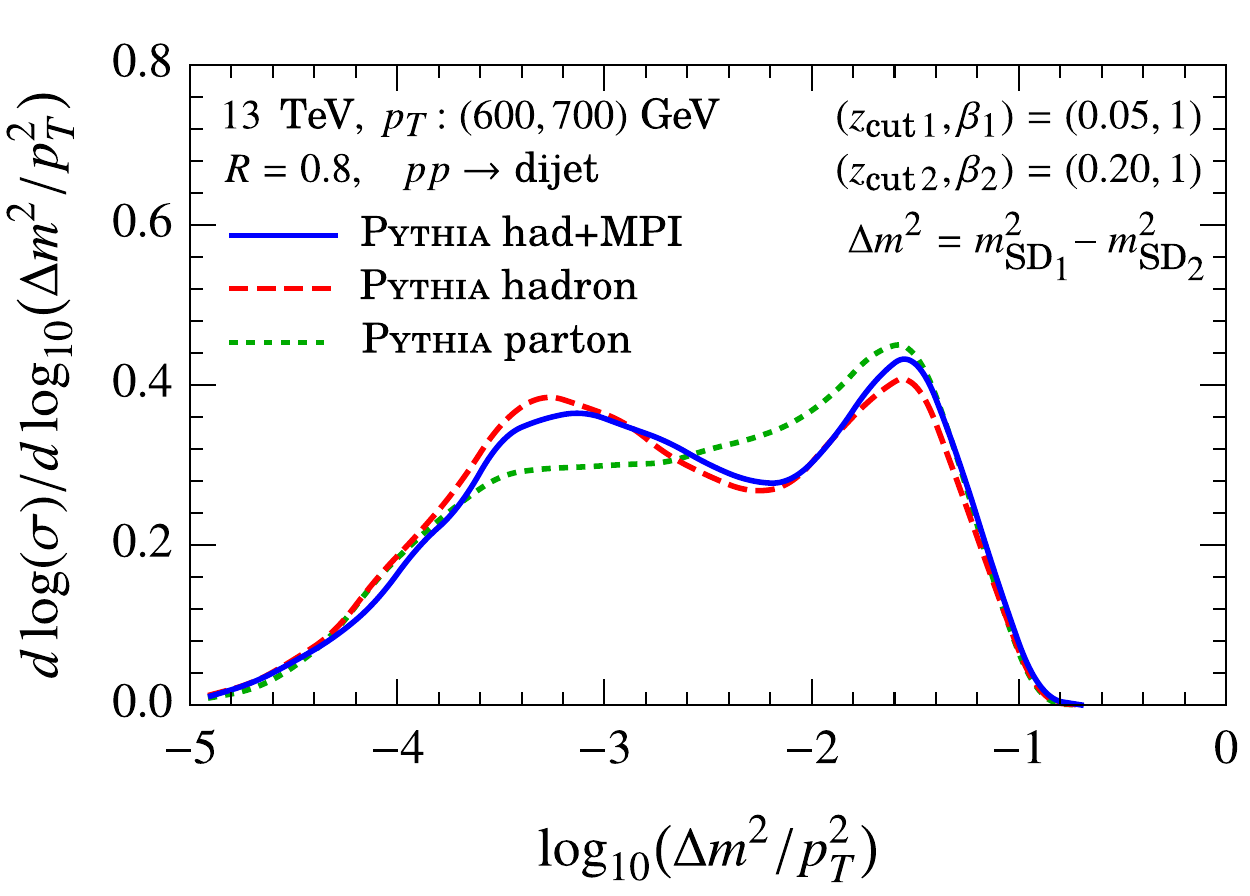}
	\caption{Collinear Drop distributions with $(z_{{\rm cut}\,1},\beta_1)=(0.1,1)$ and $(z_{{\rm cut}\,2},\beta_2)=(0.1,0)$ (left panel) and $(z_{{\rm cut}\,1},\beta_1)=(0.05,1)$ and $(z_{{\rm cut}\,2},\beta_2)=(0.2,1)$ (right panel) from \Pythia simulations. Results at parton and hadron (with and without multi-parton interactions) levels are provided.}
	\label{fig:MC_collinear_drop_had_MPI}
\end{figure}

\begin{figure}[t!]
	\hspace{-1.4cm}
	\includegraphics[width=0.55\textwidth]{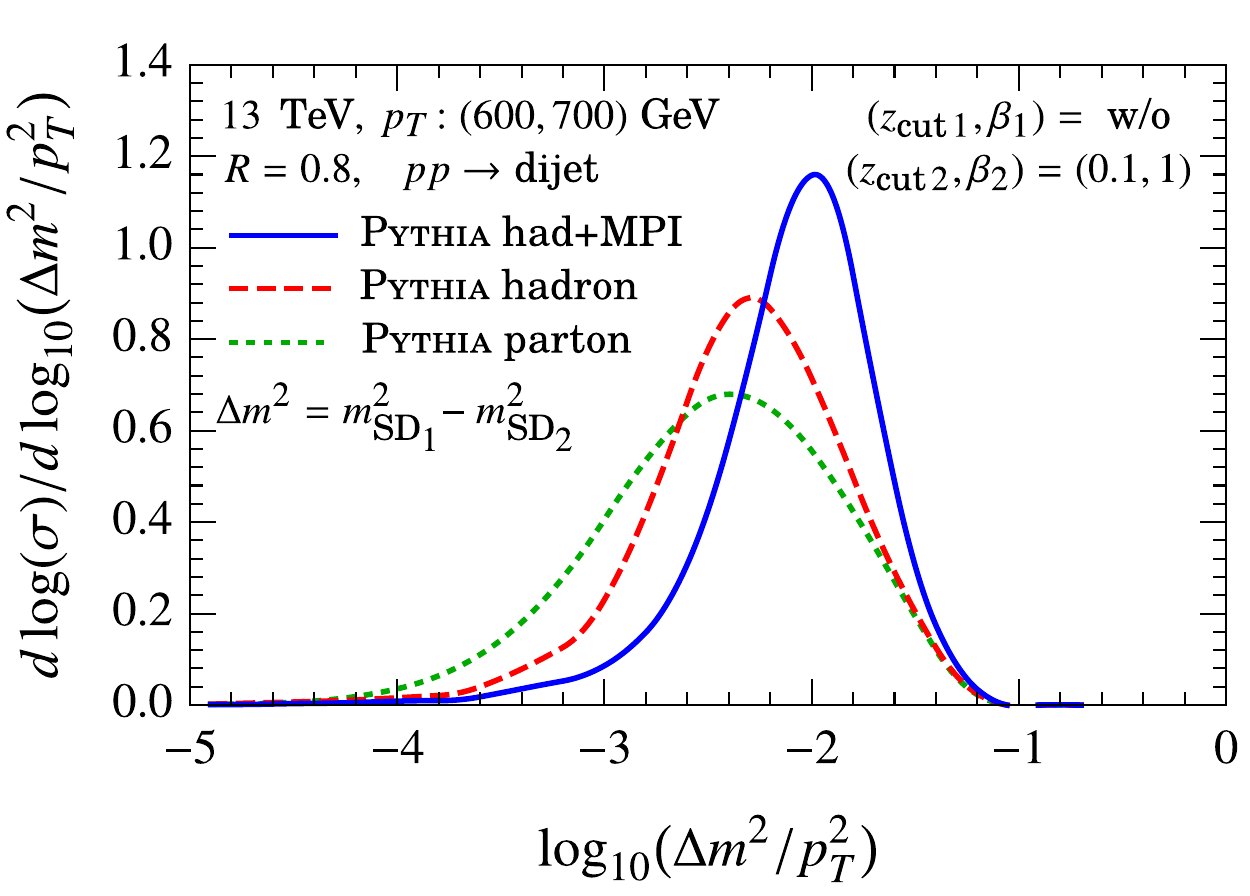}
	\includegraphics[width=0.55\textwidth]{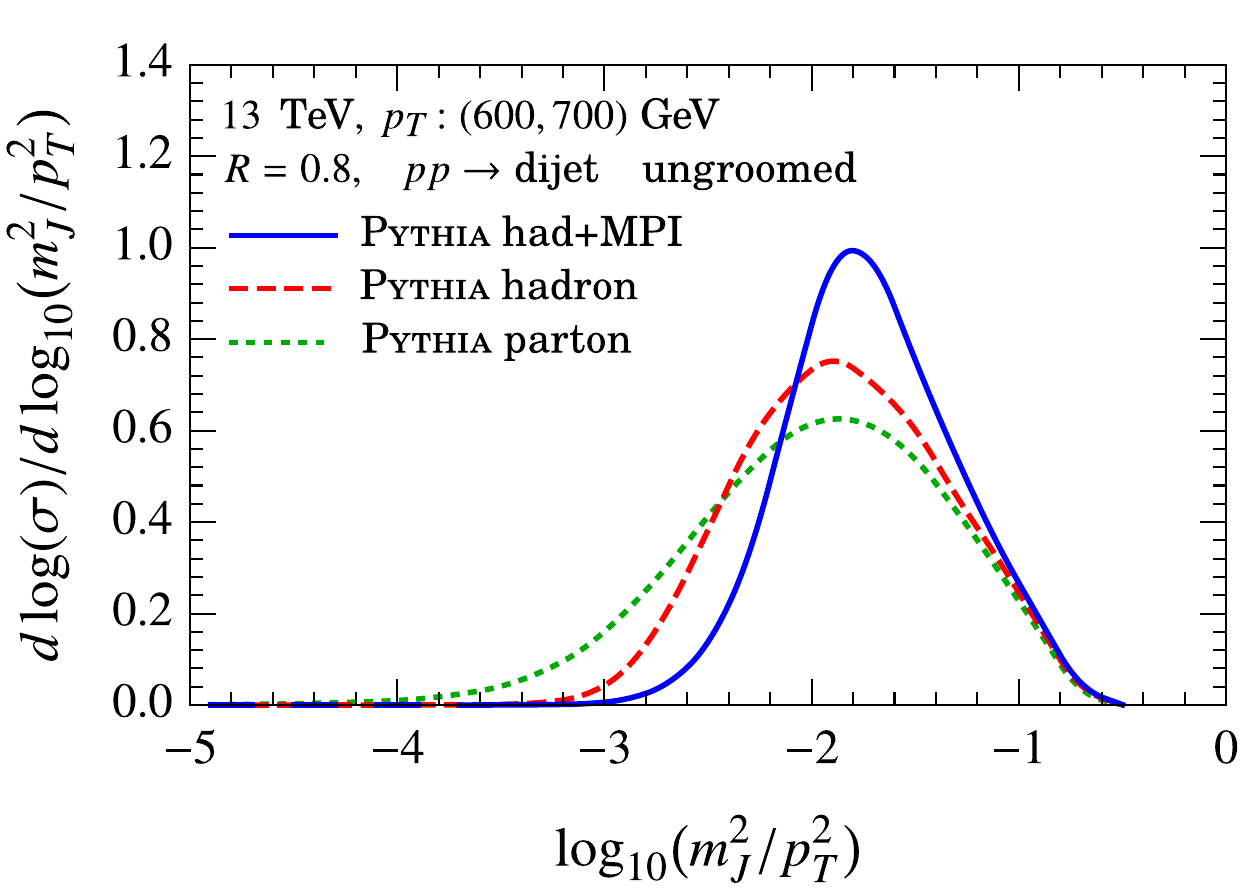}
	\caption{Collinear Drop with $(z_{{\rm cut}\,1},\beta_1)$ turned off and $(z_{{\rm cut}\,2},\beta_2)=(0.1,1)$ as well as ungroomed jet mass distributions from \Pythia simulations.  Results at parton and hadron (with and without multi-parton interactions) levels are provided.
	}
	\label{fig:MC_collinear_drop_ungroomed_had_MPI}
\end{figure}

Figure \ref{fig:MC_collinear_drop_had_MPI} shows the $\log_{10}(m^2/p_T^2)$ distributions with two sets of collinear-drop parameters: $z_{{\rm cut}\;1}=z_{{\rm cut}\;2}=0.1$ and $\beta_1=1$, $\beta_2=0$ (left panel), as well as $\beta_1=\beta_2=1$ and $z_{{\rm cut}\;1}=0.05$, $z_{{\rm cut}\;2}=0.2$ (right panel).  The curves include parton level (dotted green), hadron level without MPI effects (dashed red) and hadron level with MPI effects (solid blue). For both of these results we see  by comparing the red and green curves that there are, as expected, significant hadronization corrections.  For the left panel these  predominantly occur for $\log_{10}(m^2/p_T^2)<-2.8$, whereas in the right panel the hadronization corrections cause the distribution to become more peaked in both the groomed to ungroomed transition region, and for small masses. In both cases the comparison of green and blue curves shows that the MPI effects are  suppressed.  For this choice of collinear drop observables the soft drop cut has protected us from MPI effects, while still providing interesting observables for studying hadronization.

Collinear drop observables can also be designed to have more sensitivity to MPI.  To demonstrate this we consider in \Fig{fig:MC_collinear_drop_ungroomed_had_MPI} (left panel) the collinear drop observable that takes $z_{{\rm cut}\,2} =0.1$, $\beta_2=1$, but does not include soft drop grooming with $z_{{\rm cut}\,1},\beta_1$. Here there is a significant difference between the (dashed red) hadron level MC curve, and the (solid blue) curve including both hadronization and MPI.  Due to the collinear drop this observable is sensitive to soft MPI radiation, and is now not protected from large effects due to the absence of soft drop grooming. Indeed, the effect of MPI is even larger for this observable than for ungroomed jets, which are shown in the right panel. This makes it an interesting observable for testing the accuracy of the modeling of MPI effects in MC, through comparison with experimental data.

Every jet sample is a mixture of quark-initiated jets and gluon-initiated jets, and it is worth noting that hadronization effects can also differ for jets with different partonic origins.  It should also be possible to apply the formalism for studying nonperturbative corrections to soft drop observables developed in Ref.~\cite{Hoang:2019ceu} to the collinear drop observables proposed here. We leave further studies of hadronization and MPI effects to future work.

\subsection{Annulus Energy Fraction}

Having discussed in detail the example of collinear drop observable $\Delta m^2$, in this section we study one other example of a collinear drop observable with MC simultations, namely the annulus energy fraction $x=\tau_{\theta_a}$ defined with \eqns{CDshape1}{theta-annulus}.

\begin{figure}[t!]
    \hspace{-1.5cm}
	\includegraphics[width=0.55\textwidth]{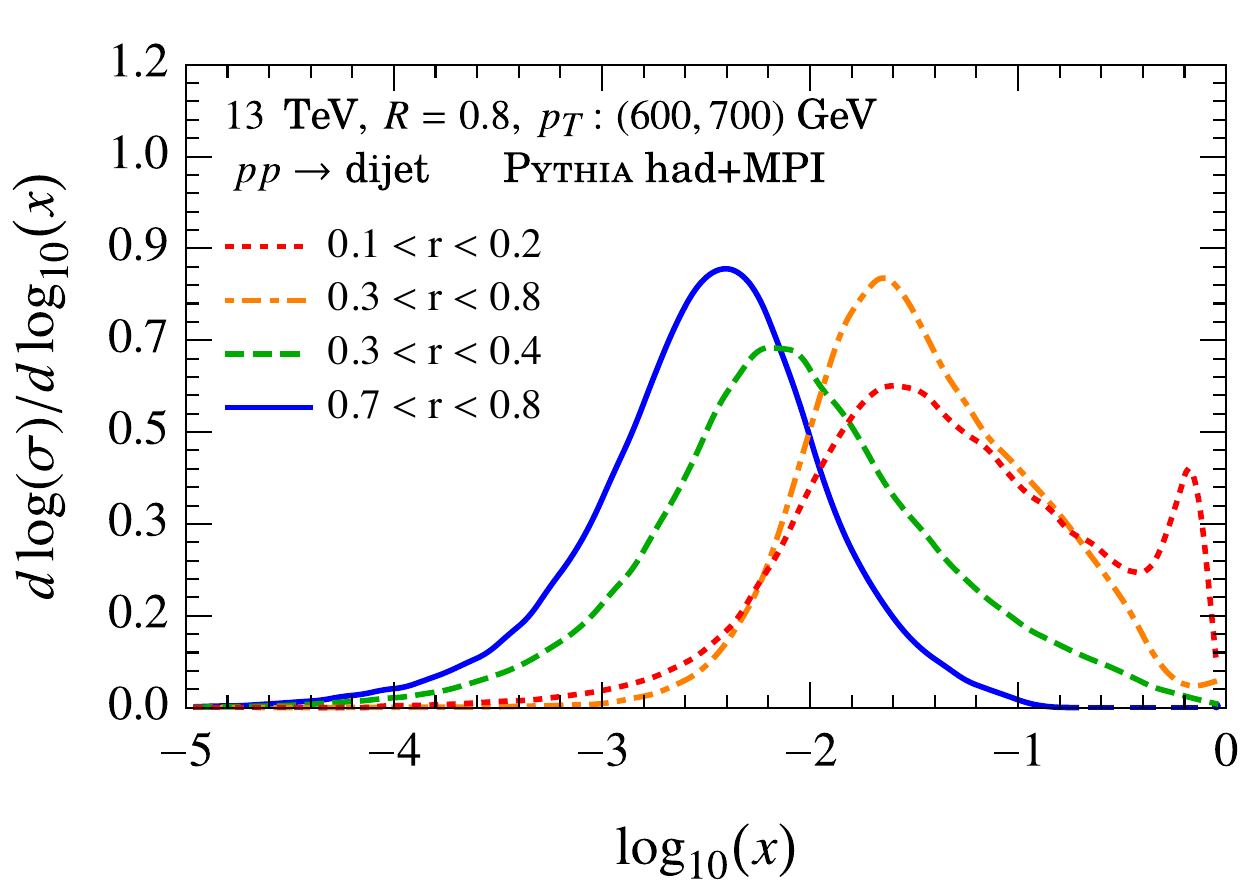}
	\includegraphics[width=0.55\textwidth]{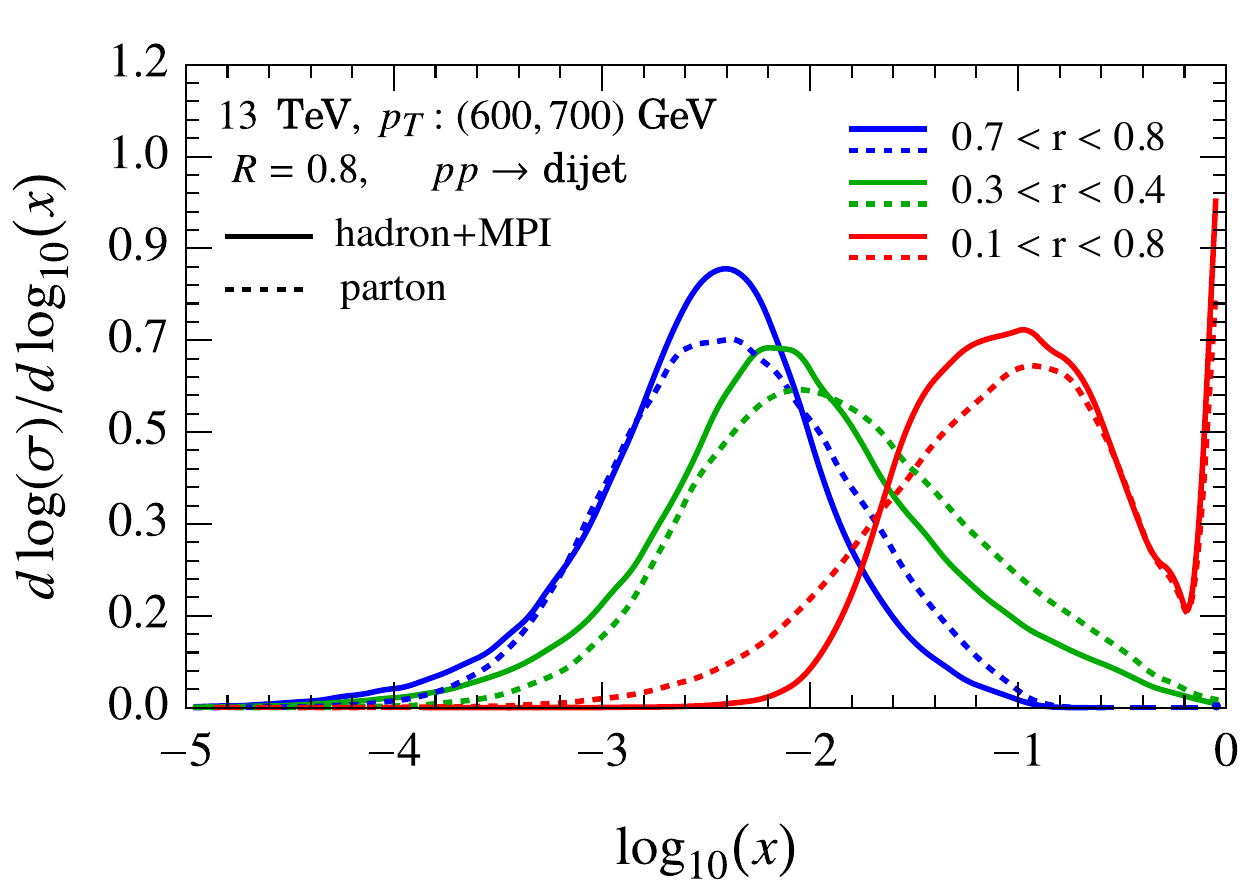}
	\caption{Annulus Energy Fraction distributions of various ring regions. The left panel shows the results with hadronization and MPI effects, while the right panel shows the partonic results as well as the hadronic results with multi-parton interactions. }
	\label{fig:MC_AEF_1}
\end{figure}

The left panel of Figure \ref{fig:MC_AEF_1} shows the $x$ distributions with hadronization and MPI effects for different ring regions: $0.1<r<0.2$ (dotted red), $0.3<r<0.8$ (dot-dashed orange), $0.3<r<0.4$ (dashed green) and $0.7<r<0.8$ (solid blue). We can see that a significant fraction of the jet energy is contained within a ring away from the core of the jet. The $0.1<r<0.2$ region is even capable of capturing energetic, collinear splittings and develops a peak at large values of $\log_{10} x$ (and is not itself a collinear drop observable). As one moves away from the jet axis, the fraction of jet energy decreases quickly. Less than $10\%$ of the jet energy is contained in the $0.7<r<0.8$ region. The right panel of Figure \ref{fig:MC_AEF_1} shows the hadronization and MPI effects to the annulus energy fraction. We provide the distributions for  $0.7<r<0.8$ (blue), $0.3<r<0.4$ (green) and $0.1<r<0.8$ (red). Unlike many other jet substructure distributions where hadronization and MPI effects tend to increase the values of the observables, these effects can give a qualitatively different trend and cause a depletion of the annulus energy fraction in certain ring regions so that the peak position moves to smaller values. Analytic calculations using SCET for such observables will be discussed in future work. 

\section{Conclusions and Outlook}
\label{sec:conclusion}

Systematic improvements of the understanding of collider events and searches for new physics require an efficient probe of the Standard Model phase space, especially in regions with more complicated soft dynamics or hadronic activity. 
In this paper we introduce a new class of jet substructure observables called collinear drop, which allows us to optimize the sensitivity to 
%any particular region
soft regions of QCD phase space from 
%the highest collision energy scale
higher energy perturbative scales
down to the confinement scale. 
We used techniques of jet grooming and jet shapes to give concrete examples of constructing collinear drop observables. 
In particular, we used multiple soft-drop jet grooming algorithms to select controlled internal jet regions by removing energetic, collinear particles as well as soft, wide angle particles. We work out the analytic description of collinear drop observables using the soft-collinear effective theory, and we provide theoretical predictions at next-to-leading logarithmic (NLL) accuracy. We also developed scale variation methods to estimate perturbative uncertainties for these observables that are compatible with transition regions.

As a validation of the theoretical framework, we provide comparisons of soft-drop jet mass distributions between our theoretical predictions to \textsc{Pythia} simulations, which agree well at parton level, and were also contrasted with ATLAS data.
We then compare analytic results of partonic collinear drop distributions to different Monte Carlo simulations generated by \textsc{Pythia} and \textsc{Vincia}. We observe interesting differences between Pythia and Vincia with collinear drop observables, indicating that they are useful experimental observables for testing and improving MC simulations. While in general the NLL SCET results are closer to the Vincia results, the reduced theoretical uncertainties expected at one higher order (next-to-next-to leading logarithmic accuracy) will be needed in order to utilize SCET to truly distinguish features of different parton shower event generators. We also demonstrated that collinear drop observables can be utilized to study hadronization in jets in a manner independent from underlying event contamination, and with different settings, can be also used as a sensitive probe of underlying event effects themselves. Thus predictions for collinear drop observables provide key probes of soft phase space that are useful both for systematic improvements of Monte Carlo event generators and for rigorous study of underlying event and non-perturbative hadronization, paving the road toward higher precision QCD results for hadron-hadron, electron-ion, and heavy-ion collisions.

There are many other potential applications of collinear drop observables, especially for probing the color coherence of soft particles which can allow us to distinguish quark, gluon and color neutral particle initiated jets. For hadronically decaying boosted electroweak bosons, standard tagging methods exploit the two-prong structure inherent from the boson masses and kinematics \cite{Thaler:2010tr,Larkoski:2013eya,Moult:2016cvt}, local color flow information due to color connection \cite{Gallicchio:2010sw,Hook:2011cq,Curtin:2012rm}, or by visualizing jets in the Lund jet plane~\cite{Dreyer:2018nbf,Andrews:2018jcm}. On the other hand, collinear drop observables can be used to perform color-singlet jet isolation \cite{Chien:2013kca,Chien:2017xrb,Chien:2018dfn} which has been seen to improve the $W/Z$ and top tagging efficiency. Recently, an observable ${\cal O}_2$ that is efficient for quark gluon discrimination was studied in Ref.~\cite{Komiske:2018cqr}, which also suppresses collinear radiation. The analytic calculation of collinear drop observables for hadronic electroweak boson jets, and their prospects for improving tagging methods, will be discussed in a separate paper.

\section*{Acknowledgments}

The authors thank Yi Chen, Raghav Elayavalli, Edmond Iancu, Yen-Jie Lee, Christopher McGinn, Bernhard Mistlberger, George Sterman and Jesse Thaler for helpful discussions. 
This work was supported by the U.S. Department of Energy, Office of Nuclear Physics, from DE-SC0011090.
I.S. was also supported in part by the Simons Foundation through the Investigator grant 327942, and Y.-T.C. was supported in part by the LHC Theory Initiative Postdoctoral Fellowship under the National Science Foundation grant PHY-1419008.

\bibliography{delta_m2}

\end{document}